\documentclass[fleqn,usenatbib]{mnras}
\usepackage{newtxtext,newtxmath}
\usepackage[T1]{fontenc}
\usepackage{textgreek}
\DeclareRobustCommand{\VAN}[3]{#2}
\let\VANthebibliography\thebibliography
\def\thebibliography{\DeclareRobustCommand{\VAN}[3]{##3}\VANthebibliography}

\usepackage{graphicx}	
\usepackage{amsmath}	
\usepackage[normalem]{ulem}
\usepackage{xcolor}

\newcommand{\hmpc}{h^{-1}{\rm Mpc}}

\title[Galaxy environments in CAMELS]{CAMELS Environments: The Impact of Local Neighbours on Galaxy Evolution across the SIMBA, IllustrisTNG, ASTRID, and Swift-EAGLE Simulations}

\author[Sims et al.]{
Xavier Sims,$^{1}$\thanks{E-mail: xavier.sims@uconn.edu}
Daniel Anglés-Alcázar,$^{1}$
Boon-Kiat Oh,$^{1,2}$
Daisuke Nagai,$^{3,4}$
Jonathan Mercedes-Feliz,$^{1}$ \newauthor
Isabel Medlock,$^{4}$
Yueying Ni,$^5$
Christopher C. Lovell,$^6$
Francisco Villaescusa-Navarro,$^{7,8}$ 
\\
$^{1}$Department of Physics, University of Connecticut, 196 Auditorium Road, U-3046, Storrs, CT, 06269, USA\\
$^2$School of Physics, Korea Institute for Advanced Study, 85 Hoegiro, Dongdaemun-gu, Seoul 02455, Republic of Korea\\
$^{3}$Department of Physics, Yale University, New Haven, CT 06520, USA \\
$^{4}$Department of Astronomy, Yale University, New Haven, CT 06520, USA\\
$^5$Michigan Institute for Data and AI in Society, University of Michigan, Ann Arbor, MI, 48109, USA\\
$^6$Kavli Institute for Cosmology, University of Cambridge, Cambridge CB3 0HA, UK\\
$^7$Center for Computational Astrophysics, Flatiron Institute, New York, NY 10010, USA\\
$^8$Department of Astrophysical Sciences, Princeton University, Princeton, NJ 08544, USA
}

\date{Accepted XXX. Received YYY; in original form ZZZ}

\pubyear{2026}

\begin{document}
\label{firstpage}
\pagerange{\pageref{firstpage}--\pageref{lastpage}}
\maketitle

\def\SIMBA{{\it SIMBA}\;}
\def\EAGLE{{\it EAGLE}\;}
\def\TNG{{\it IllustrisTNG}\;}
\def\Astrid{{\it Astrid}\;}
% Abstract of the paper
\begin{abstract}
\label{sec:abstract}
Internal feedback from massive stars and active galactic nuclei (AGN) play a key role in galaxy evolution, but external environmental effects can also strongly influence
galaxies. We investigate the impact of environment on galaxy evolution, and its dependence on baryonic physics implementation, using Cosmology and Astrophysics with MachinE Learning Simulations (CAMELS) spanning a wide range of stellar and AGN feedback implementations in the SIMBA, IllustrisTNG, ASTRID, and Swift-EAGLE galaxy formation models. We show that satellite galaxies are significantly affected by the environment in all simulation models, with their gas fraction and star formation rate (SFR) suppressed in overdense regions compared to similar mass satellites in underdense environments at $z=0$. Central galaxies are less sensitive to environment but tend to show lower gas fraction and SFR in overdense regions at low stellar mass, transitioning to higher gas fraction and SFR for massive galaxies in higher-density environments. 
Halo baryon fraction ($f_{\rm B}$) and circumgalactic medium mass fraction ($f_{\rm CGM}$) at $z=0$ show clear environmental effects. In SIMBA, low-mass haloes in overdense regions have systematically lower $f_{\rm B}$ and $f_{\rm CGM}$ at fixed halo mass, while Swift-EAGLE haloes in overdense regions have systematically higher $f_{\rm B}$ and $f_{\rm CGM}$ across the full halo mass range, and IllustrisTNG and ASTRID show opposite trends at the low and high mass ends. Environmental effects can flip at higher redshift, with SFR and $f_{\rm B}$ increasing with local density in low-mass haloes before quenching at an increasing overdensity threshold. Our results demonstrate that the impact of environment on galaxy evolution depends significantly on galaxy formation model, and higher-density environments can either suppress or enhance star formation depending on galaxy mass and cosmic epoch.

\end{abstract}


%%%%%%%%%%%%%%%%%%%%%%%%%%%%%%%%%%%%%%%%%%%%%%%%%%

%%%%%%%%%%%%%%%%% BODY OF PAPER %%%%%%%%%%%%%%%%%%

\section{Introduction}

\label{sec:introduction}

Galaxy evolution is influenced by a variety of astrophysical processes, from the gravitational collapse and growth of dark matter haloes, radiative cooling of gas, and star formation to feedback from massive stars and active galactic nuclei (AGN) powered by accreting supermassive black holes (SMBHs) \citep{somerville2015, Crain&voort2023}. Feedback from massive stars in the form of stellar winds, radiation, and supernovae (SNe) power large-scale galactic winds that can unbind gas in low-mass galaxies and significantly suppress global star formation \citep{Hopkins2014, Angles-Alcazar2017b, Schneider2020, Steinwandel2024}. For galaxies more massive than the Milky Way, AGN feedback is believed to become a dominant internal feedback process, which may also have an impact in low-mass galaxies \citep{Koudmani2022, Wellons2023}. As SMBHs at the centers of galaxies grow larger, they drive energetic AGN feedback in the form of large-scale outflows and jets that can evacuate gas from galaxies and suppress further inflows from the circumgalactic medium (CGM) to the central regions \citep{Fabian2012, Choi2015, Harrison2017, Hlavacek_Larrondo_2022, Mercedes-feliz2023, Vayner2024}. Stellar and AGN feedback can both affect the amount of gas within the galaxy and how efficiently gas can cool to form stars. However, in addition to these internal feedback processes, external environmental effects can also strongly influence galaxy evolution \citep{Kauffman2004}.

These external processes encompass a wide range of physical mechanisms that can shape how galaxies evolve. Galaxy mergers can redistribute angular momentum and trigger bursts of star formation \citep{Hopkins2008, Ellison2019, Moreno2019, Moreno2022}, while competition among neighbouring galaxies can limit gas accretion from the surrounding medium \citep{Aragon-calvo2019}. Shock heating of gas in the deep potential wells of galaxy groups and clusters can suppress further cooling and inflow \citep{Birnboim2003}, and even AGN feedback from nearby galaxies can heat or expel gas across intergalactic distances \citep{Gilli2019, Martin2019, Martin2021, gonzalez2025, Wang2025}. The relative importance of these various mechanisms depends strongly on environment. This is supported in part by the existence of massive, red, quiescent galaxies, which are predominately found in overdense regions, while low-density environments tend to host smaller, blue, star-forming galaxies \citep{Blanton2009}. Large-scale surveys such as the Sloan Digital Sky Surveys (SDSS) have enabled detailed studies of the connection between environment and galaxy evolution \citep{Gomez2003, Hogg2004, Rojas2004, Blanton2005, Kuutma2017}. Specific impacts of environment on galaxy properties such as lower star formation rate (SFR) in overdense regions at a given stellar mass have been shown in previous works \citep{Dressler1980, Kauffman2004, Boselli2006, Blanton2009}.  

The extent to which environment affects galaxy properties is further impacted by whether the galaxy is a central or a satellite \citep{Faltenbacher2010, Peng2012}. Satellite galaxies are prone to astrophysical processes that do not affect the evolution of central galaxies, such as ram pressure and tidal stripping. These processes can strongly suppress the gas content and star formation activity of satellite galaxies \citep{Gunn&Gott1972, Abadi1999, Boselli2022,Pathak2025}. As satellites move through host halo gas with sufficiently high velocity, they experience ram-pressure stripping \citep{Tonnesen2009, Roy2024}. Further, tidal stripping can occur when the central galaxy's gravity pulls on the satellite asymmetrically, stripping away stars and gas from the satellite and thus suppressing growth and star formation \citep{dekel2003}.

Cosmological hydrodynamic simulations provide a powerful framework for investigating how environment influences galaxy evolution. Their ability to track galaxies and their neighbours in three dimensions allows for precise characterization of the environments in which galaxies reside. Large-volume cosmological simulations have provided valuable insights linking environment to key processes such as the efficiency of star formation \citep{Hasan2023, bulichi24}, the quenching of satellite galaxies \citep{Wright22,Rohr2023}, or how environmental effects vary with redshift \citep{Ghodsi2024}.
The definition of “environment” can vary depending on the physical question being explored. Common approaches include using the local number density of galaxies \citep{2017MNRAS.466.3460V}, the ratio between the distance from and size of a neighbouring halo \citep{Haas2012}, or distance to nearest cosmic web filament \citep{kraljic2020, bulichi24}. Each metric captures different aspects of environmental influence. However, disentangling the intrinsic effects of environment is complicated by the strong correlation between environment and halo mass, which can bias trends inferred from environmental studies. In addition, it has been shown that the mass of a galaxy affects how its local environment impacts its properties such as alignment with cosmic web filaments \citep{kraljic2020}. Efforts to decouple halo mass from environmental measures are therefore critical to isolating the true role of environment in shaping galaxy evolution \citep{Haas2012}. 

Previous simulation-based studies of environmental effects on galaxy evolution have typically relied on a single galaxy formation model, each implementing a specific set of sub-grid prescriptions to capture unresolved physical processes such as star formation, SMBH growth, and associated feedback processes. Recent large-volume cosmological hydrodynamic simulations are converging on the set of physical processes that govern galaxy evolution and are achieving increasingly better agreement with observations \citep{somerville2015, Habouzit2021, Habouzit2022, Crain2023}. However, the success of most current models still depends on extensive tuning of free parameters that control the efficiency and parameterize the uncertainty of the underlying physical processes. It is therefore desirable to investigate the role of environment on galaxy evolution across a variety of galaxy formation models in order to identify robust trends that are independent of specific model assumptions.

In this work, we use simulations from the Cosmology and Astrophysics with Machine Learning Simulations (CAMELS) project \citep{CAMELS_PAPER,Villaescusa-Navarro2023} to investigate the impact of environment on galaxy evolution. CAMELS contains thousands of hydrodynamic simulations varying cosmological and astrophysical parameters across different galaxy formation models, including SIMBA \citep{SIMBA.PAPER}, IllustrisTNG \citep{Weinberger2017, Pillepich2018}, ASTRID \citep{Bird2022,Ni2022}, and Swift-EAGLE (Lovell et al. {\it in prep}). Originally designed to train machine learning algorithms for applications ranging from fast emulation of simulations \citep{Hassan2022, Delgado2023, Lee2024} to inference of cosmological parameters \citep{Villaescusa-Navarro2021arx, Nicola2022, Villaescusa-Navarro2022, DeSanti2023, Lovell2025}, CAMELS provides a powerful framework to investigate the physical mechanisms driving galaxy evolution as a function of model implementation details \citep{Ni2023, Tillman2023, Gebhardt2024, Medlock2024, Medlock2025b}. Additionally, with the vast amount of CAMELS data, we can combine tens of thousands of galaxies simulated under different modeling assumptions to robustly investigate the impact of environment on galaxy evolution.

This paper is organized as follows: In Section~\ref{sec:Methods}, we describe the CAMELS project and the various definitions of environment that we use in this work. In Section~\ref{sec: Overview of Simulations}, we provide an overview of the impact of environment across four different galaxy formation models in CAMELS and quantify the range of environments present in the CAMELS simulations. In Section~\ref{sec: Impact of Environment}, we describe the impact of environment as measured by galaxy overdensity on multiple halo and galaxy properties including halo baryon fraction, CGM fraction, galaxy gas fraction, and galaxy SFR. In Section~\ref{sec:Redshift}, we describe how the impact of environment changes with redshift. In Section~\ref{sec:Definition of Environment}, we explore the impact of environment on halo and galaxy properties as quantified by an alternative definition of environment. In Section~\ref{sec:Discussion}, we discuss our results in the context of the current state of the field, and we summarize the conclusions of our work in Section~\ref{sec:Conclusions}.

\section{Methods}
\label{sec:Methods}

For this work, we utilize the publicly available CAMELS dataset\footnote{https://camels.readthedocs.io/en/latest/}, comparing results obtained using the SIMBA, IllustrisTNG, ASTRID, and Swift-EAGLE suites. We briefly describe the simulations and environment analysis techniques below.

\subsection{CAMELS}

The CAMELS project\footnote{https://www.camel-simulations.org} \citep{CAMELS_PAPER} contains >9000 hydrodynamic simulations and >7000 N-body simulations. In this work, we use hydrodynamic simulations produced with four different galaxy formation models: SIMBA, IllustrisTNG, ASTRID, and Swift-EAGLE. Every simulation has a (25\,$h^{-1}$Mpc)$^3$ volume and follows the evolution of \begin{math} 256^3 \end{math} dark matter particles. The hydrodynamic simulations follow the evolution of an additional \begin{math} 256^3 \end{math} gas resolution elements. Physical processes implemented in all models include gravitational and hydrodynamic forces, gas cooling, star formation, stellar feedback, SMBH growth, and SMBH feedback. Haloes are defined as friends-of-friends (FOF) groups with linking length $b=0.2$ and galaxies are identified using the SUBFIND algorithm \citep{Springel2001}. Each of the CAMELS suites contains four sets of simulations; the cosmic variance (CV) set, the 1 parameter (1P) set, the Latin Hypercube (LH) set, and the Extreme (EX) set. In this work, we utilize only the CV sets. Each CV set contains 27 simulations run with different initial conditions but with the same subgrid parameter values corresponding to each fiducial model. The 27 different initial conditions in the CV sets are shared across simulation suites, which enables a direct comparison between galaxy formation models.

\subsubsection{SIMBA}
The SIMBA galaxy formation model \citep{SIMBA.PAPER} is built from its predecessor, MUFASA \citep{Dave2016}. SIMBA uses the GIZMO \citep{Hopkins2015} meshless finite mass hydrodynamics code. The gravity equations are solved with a tree/particle-mesh method adapted from GADGET-III code \citep{Springel2005}. Stellar feedback is modeled with two-phase galactic winds with 30\% of the wind particles being ejected with their temperature set by the difference between SNe energy and wind kinetic energy. The mass-loading factor and velocity of galactic winds driven by stellar feedback scale with stellar mass as per the FIRE zoom-in simulations \citep{Muratov2015,Angles-Alcazar2017b}. SMBH particles are seeded in galaxies with $M_\star \gtrsim 10^{9.5}$\,M$_\odot$. SMBH growth in SIMBA is implemented with the gravitational torque accretion model \citep{Hopkins2011, Angles-Alcazar2017a} and the Bondi accretion model \citep{Bondi1952} for cold and hot gas respectively. Additionally, SIMBA implements kinetic AGN feedback \citep{Angles-Alcazar2017a} for two different modes depending on Eddington ratio, high speed collimated jets and quasar-mode winds, as well as X-ray radiative feedback following \citet{Choi2012}. 

\subsubsection{IllustrisTNG}
The IllustrisTNG galaxy formation model \citep{Weinberger2017, Pillepich2018} is built from its predecessor, Illustris \citep{Genel2014, Vogelsberger2014}, and uses the Arepo hydrodynamics code \citep{Springel2010}. The Arepo code uses a hybrid tree/particle-mesh method to solve the equations of gravity, and a moving Voronoi mesh to solve for ideal magnetohydrodynamics. IllustrisTNG includes stellar feedback driven galactic-scale outflows and three-mode AGN feedback \citep{Springel2018}. Galactic winds are modeled with stochastic and isotropic ejection of particles with energy and mass loading factors based on local velocity dispersion and metallicity. SMBH particles are seeded in haloes with $M_{\rm halo}>5 \times 10^5$\,$h^{-1}$M$_\odot$. The thermal mode AGN feedback (high accretion state) injects thermal energy into the gas around the SMBH, and the kinetic mode (low accretion state) injects kinetic energy. The AGN radiative feedback mode adds the radiation flux of the SMBH to the cosmic ionizing background. Unlike the first two modes, the radiative mode is always active.

\subsubsection{ASTRID}
The ASTRID galaxy formation model \citep{Bird2022, Ni2022} uses the MP-gadget code, an optimized version of the Gadget-2 code \citep{Springel2005} using a hybrid tree/particle-mesh method to solve for gravity and an entropy-conserving formulation of smoothed particle hydrodynamics. ASTRID is adapted for CAMELS from the original ASTRID simulations; see \citet{Ni2023} for a detailed description of model updates. ASTRID is built with kinetically implemented galactic winds with wind particles sourced from newly formed star particles and mass loading and velocity scaling with the local dark matter velocity dispersion following the Illustris model \citep{Vogelsberger2014}. SMBH particles are seeded in haloes with $M_{\rm halo} = 5\times10^{10}$\,$h^{-1}$M$_\odot$ and grow by Bondi accretion and mergers. CAMELS-ASTRID also utilizes a two-mode SMBH feedback model. SMBH feeback uses either thermal or kinetic energy injection depending on the Eddington ratio of the SMBH accretion rate. 

\subsubsection{Swift-EAGLE}
The CAMELS Swift-EAGLE galaxy formation model (Lovell et al. {\it in prep.}) is a modified version of the original EAGLE simulation \citep{Crain2015, Schaye2015} implemented in the Swift code \citep{SWIFT} to solve the equations of gravity and the SPHENIX formulation of smoothed particle hydrodynamics \citep{Borrow2022}. The CAMELS version of Swift-EAGLE utilizes the same physics as the model discussed in \citet{SWIFT}, but is recalibrated at a lower resolution to maintain consistency with the other CAMELS galaxy formation models (the new CAMELS Swift-EAGLE calibration is presented in Lovell et al. {\it in prep.}). Stellar feedback from massive stars follows the stochastic thermal approach described in \citet{Dalla_Vecchia&Schaye}. SMBH particles are seeded in haloes more massive than $10^{10}\,{\rm M_\odot}$, and SMBH growth follows a modified Bondi accretion prescription. AGN feedback is based on stochastic injection of thermal energy, analagous to the implementation of stellar feedback \citep{Booth&Schaye09}.

\subsection{Measures of environment}\label{sec:env_def}

\begin{figure*}
    \includegraphics[width=0.45\textwidth]{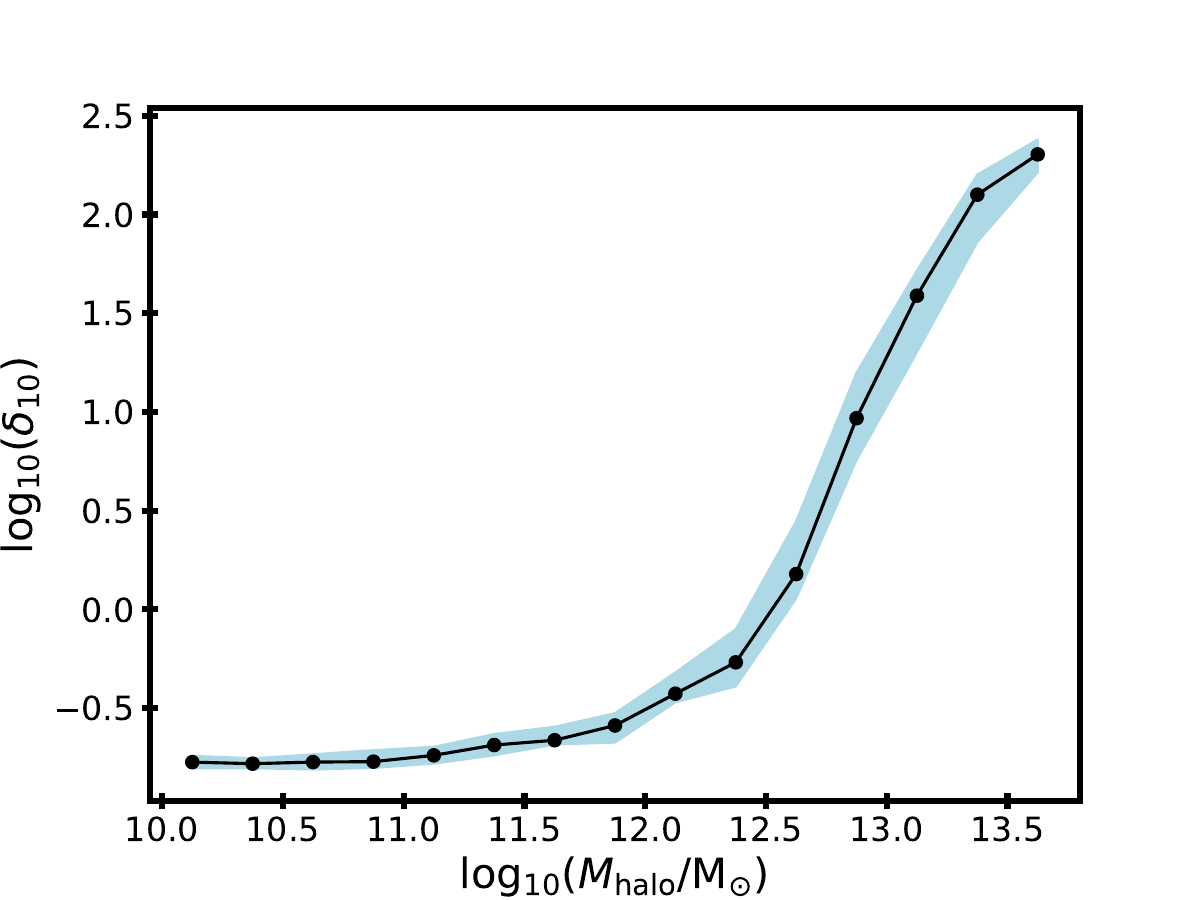}
    \includegraphics[width=0.45\textwidth]{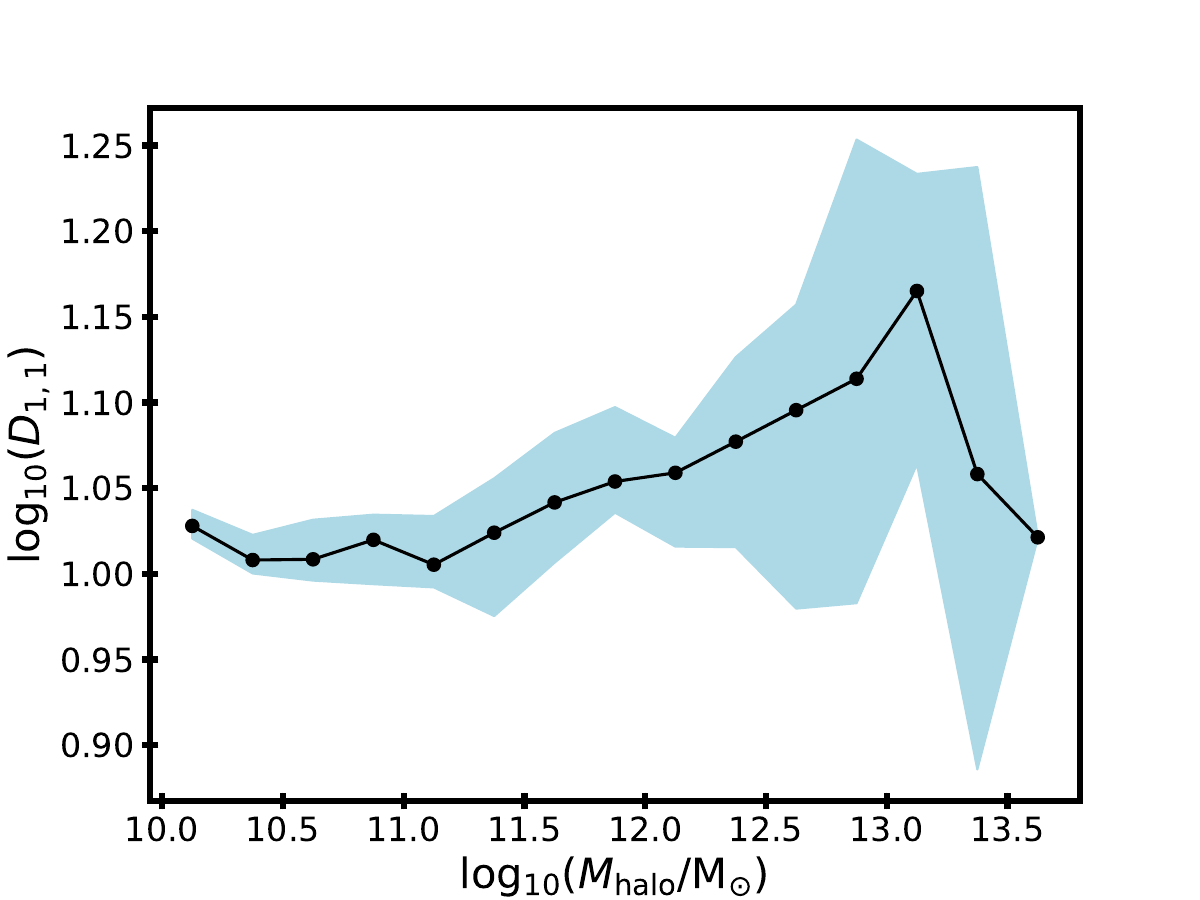}
    \caption{Local galaxy environmental density at $z=0$ as a function of halo mass for the $\delta_{10}$ (left) and $D_{1,1}$ (right) environment definitions (see Section 2.2 for details) in the SIMBA CV simulations. The black points represent the median overdensity in each mass bin, and the blue region represents the 25$^{\rm th}$ to 75$^{\rm th}$ percentile range quantifying cosmic variance across the CV set. There is a clear positive correlation between halo mass and $\delta_{10}$, with the most massive haloes usually forming in the densest regions. The reduced y-axis range in the right panel indicates that the definition of $D_{1,1}$ is far less sensitive to $M_{\rm halo}$ than $\delta_{10}$.}
    \label{fig:delta-DNF}
\end{figure*}

\begin{figure*}
    {\centerline{

    \includegraphics[width=0.495\textwidth]{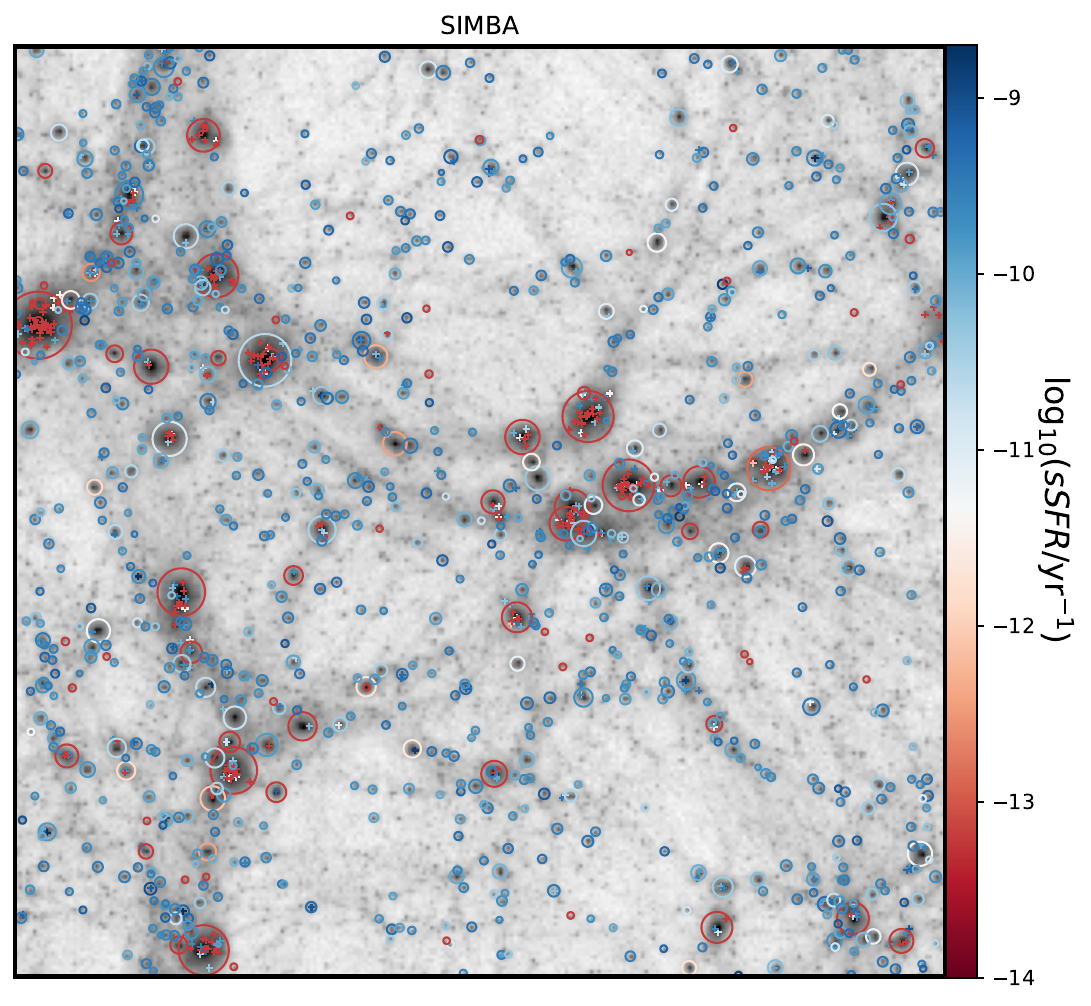}\hfill
    \includegraphics[width=0.495\textwidth]{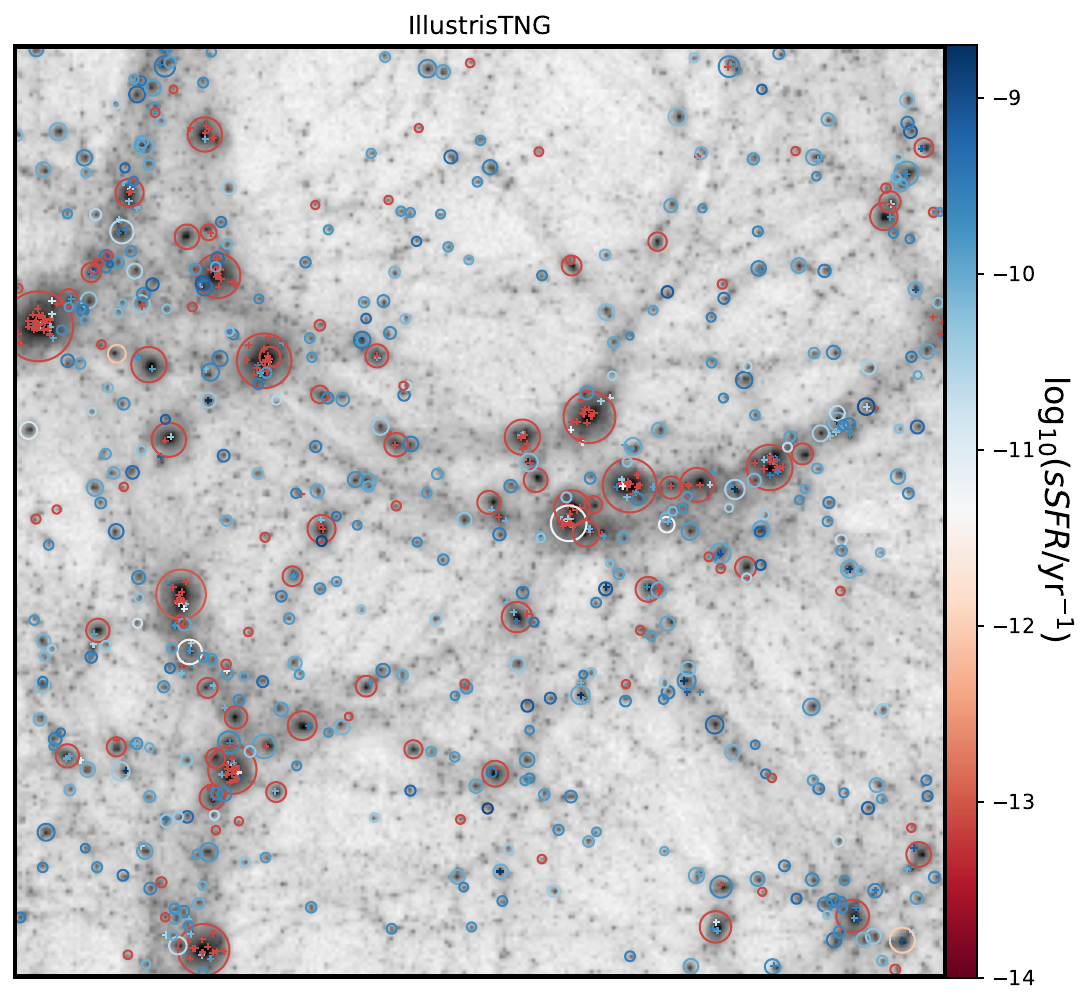}}}

    {\centerline{
    \includegraphics[width=0.495\textwidth]{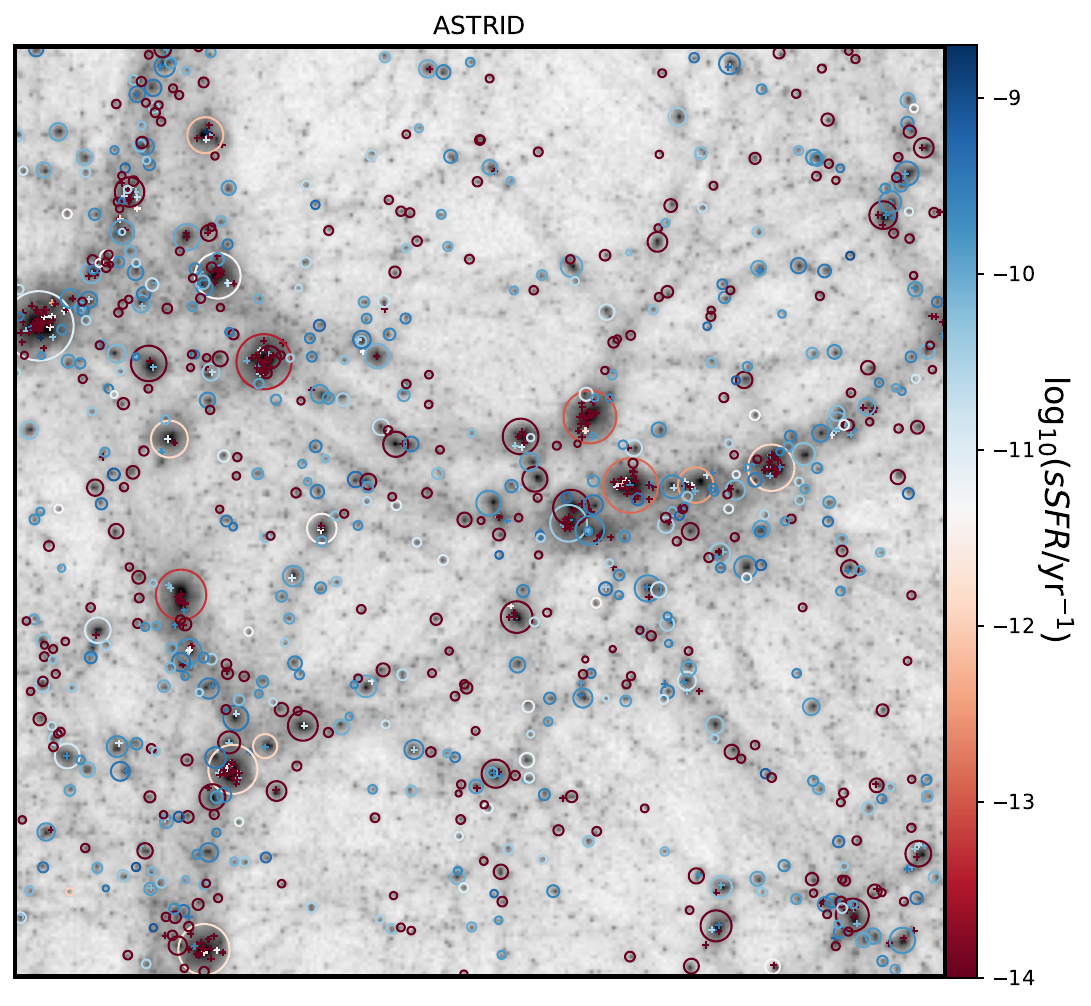}\hfill
    \includegraphics[width=0.495\textwidth]{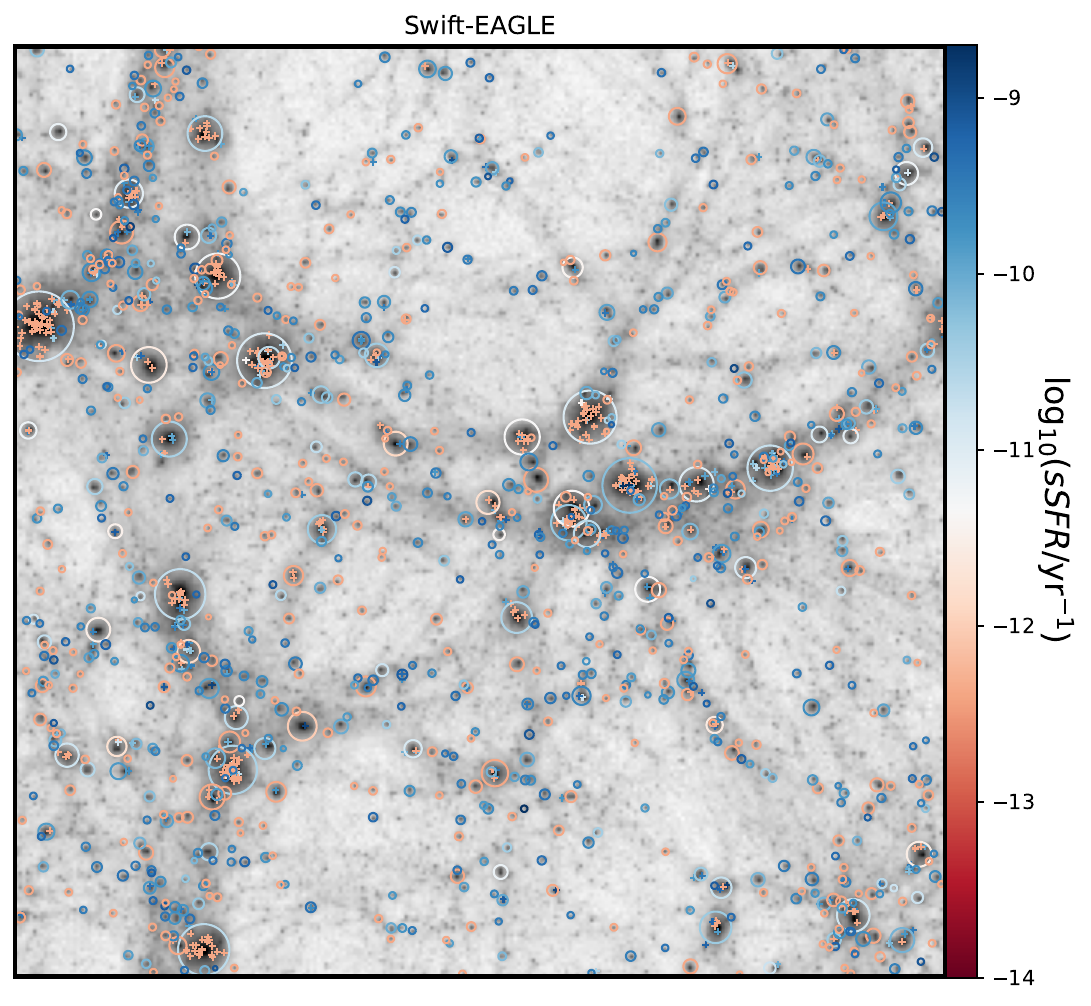}}}
    \caption{Distibution of galaxies for the CV\_0 simulation from the SIMBA (top left), IllustrisTNG (top right), ASTRID (bottom left), and Swift-EAGLE (bottom right) simulation suites in CAMELS. The background gray scale represents the projected dark-matter mass distribution in each $(25\,\hmpc)^3$ volume. Central galaxies are represented as circles with size equal to the virial radius of the corresponding host dark matter halo while satellite galaxies are represented by the plus signs, with their colors indicating each galaxy specific SFR. The most massive galaxies tend to be quenched and reside in the densest regions while the smaller and more actively star forming galaxies are more abundant in lower density regions. Satellite galaxies of massive haloes often display low sSFR. Despite these general trends, there are significant differences between models. SIMBA and IllustrisTNG similarly quench the most massive galaxies but SIMBA predicts a higher abundance of low-mass, star-forming galaxies. ASTRID and Swift-EAGLE are less efficient at suppressing star formation in massive galaxies while more efficiently quenching low-mass galaxies in under-dense environments.}
    \label{fig:simbaplot}
\end{figure*}

Multiple definitions of environment are explored in this work to test the sensitivity of the results to each measure. Primarily, we adopt the galaxy overdensity definition of \citet{2017MNRAS.466.3460V}, which is derived from the number of galaxies within a given volume. Number density is a common approach to defining the environment of galaxies, and galaxy overdensity has been shown to be a useful parameterization in cosmological hydrodynamic simulations. Here we define galaxy overdensity ($\delta_N$) as the number density of galaxies with stellar mass $M_\star > 10^8$\,M$_\odot$ in a given volume  normalized by the average galaxy density in the full simulation volume. For a given number $N$ of galaxy neighbours with $M_\star > 10^8$\,M$_\odot$, we first calculate the galaxy density, $\rho_{\rm N}$, as:

\begin{equation}
    \rho_{\rm N} = \frac{N}{\frac{4}{3}\pi R^3},
\end{equation}

\noindent
where R is the distance from the reference galaxy to its Nth nearest neighbour. This creates a spherical volume that is just large enough to encompass all $N$ of the reference galaxy's nearest neighbours. We then use $\rho_{\rm N}$ and $\rho_{\rm all}$, the average galaxy density for the entire simulation volume, to calculate $\delta_{\rm N}$:

\begin{equation}
    \delta_{\rm N} = \frac{\rho_{\rm N}}{\rho_{\rm all}}
\end{equation}

\noindent

As an alternative measure of environment, we use $D_{\rm N,f}$ from \citet{Haas2012}. This measure of environment is shown to be less sensitive to halo mass. It is defined as:

\begin{equation}
    D_{\rm N,f} = \frac{r_{\rm ngb}}{R_{\rm vir,ngb}}
\end{equation}

\noindent
where $r_{\rm ngb}$ is the distance from the reference halo to its Nth nearest neighbour of at least f times its mass and $R_{\rm vir,ngb}$ is defined as the spherical region with density 200 times the critical density.

For this work, we primarily use $N=10$ for the $\delta_{\rm N}$ defintion and we show in Appendix~\ref{sec: Appendix} that our main results are consistent regardless of the exact number of galaxy neighbours used. For the $D_{\rm N,f}$ definition of environment, we use $N=f=1$ to include every halo in the simulation. In Figure~\ref{fig:delta-DNF}, we show $\delta_{10}$ and $D_{1,1}$ as functions of halo mass in the SIMBA simulations, which we describe further in Section~\ref{subsec:local environment}.

\subsection{Halo and galaxy properties}

We quantify the impact of environment with the following halo and galaxy properties: halo baryon fraction ($f_{\rm B}$), CGM fraction ($f_{\rm CGM}$), galaxy gas fraction ($f_{\rm gas}$), and galaxy SFR. We compute halo properties ($f_{\rm B}$ and $f_{\rm CGM}$) based on the FOF Group halo definitions of Subfind. We compute halo baryon fraction as:

\begin{equation}\label{eq:f_B}
    f_{\rm B} = \frac{M_{\rm halo,gas} + M_{\mathrm {halo,stars}}}{M_{\mathrm {halo}}}
\end{equation}

\noindent
where $M_{\rm halo,gas}$ is the total mass of gas in the halo, $M_{\rm halo,stars}$ is the total mass of stars in the halo, and $M_{\rm halo}$ is the total mass of the entire halo. Additionally, we compute halo CGM fraction as: 

\begin{equation}
    f_{\rm CGM} = \frac{M_{\rm CGM,gas}}{M_{\rm halo}}
\end{equation}

\noindent
where $M_{\rm CGM,gas}$ is the mass of gas in the CGM. We approximate $M_{\rm CGM,gas}$ as the mass of gas bound to the halo minus the mass of gas in the half stellar mass radii of each subhalo bound to the main halo \citep[see also][]{Medlock2025}. Further, we compute galaxy gas fraction as: 

\begin{equation}
    f_{\rm gas} = \frac{M_{\rm gas}}{M_{\rm gas} + M_{\rm stars}}
\end{equation}

\noindent
where $M_{\rm gas}$ and $M_{\rm stars}$ refer to the mass of gas and stars within the half stellar mass radius of the galaxy. We limit the gas and stars to those within the half stellar mass radius to ensure we only analyze baryonic matter inside of the galaxy and our calculations do not include CGM gas or extended stellar haloes. Galaxy SFR refers to the sum of the individual star formation rates of all gas cells in the subhalo.

\subsection{Statistical analysis}
\label{subsec:statistics}
To isolate the intrinsic effect of environment from that of mass dependent effects, we directly compare galaxies/haloes with similar mass and different $\delta_{10}$ overdensity values in Section~\ref{sec: Impact of Environment}. We separate galaxies/haloes into logarithmically-spaced mass bins of 0.25 dex width, and we determine the median $\delta_{10}$ value as well as the $25^{\rm th}$ and $75^{\rm th}$ percentiles of the distribution of $\delta_{10}$ values within each mass bin. We then identify {\it overdense} galaxies/haloes with $\delta_{10}$ above the $75^{\rm th}$ percentile and {\it underdense} galaxies with $\delta_{10}$ below the $25^{\rm th}$ percentile. We can then identify trends related to the environment for galaxies of similar mass. These $\delta_{10}$ percentile thresholds defining the underdense and overdense classifications ensure that our analysis reflects the environments represented by half of the galaxies/haloes in each mass bin.
Additionally, we employ bootstrap error bars and Mann-Whitney-U tests to quantify the statistical significance of our results. The bootstrap error bars represent the 90\% confidence interval of the median values of a given galaxy/halo property in each mass bin when comparing the underdense and overdense populations. The Mann-Whitney-U test p-values ($p_{\rm MWU}$) quantify the probability that our underdense and overdense median values originate from the same distribution. We take $p_{\rm MWU} < 0.05$ to be the requirement to pass the significance test.

\begin{figure*}
    \includegraphics[width=0.3\textwidth]{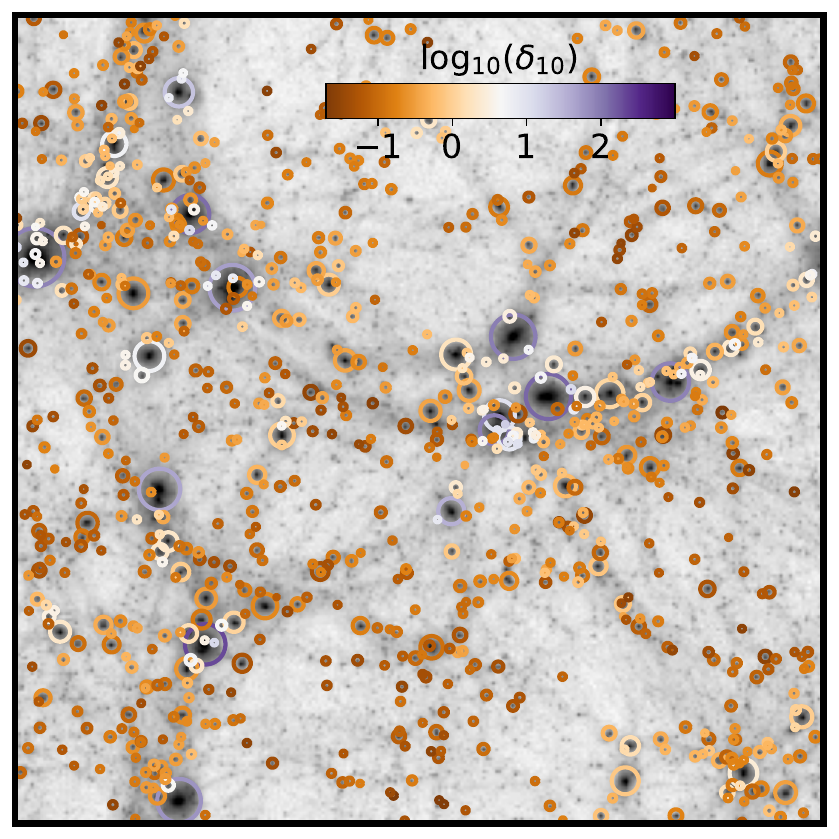}
    \includegraphics[width=0.3\textwidth]{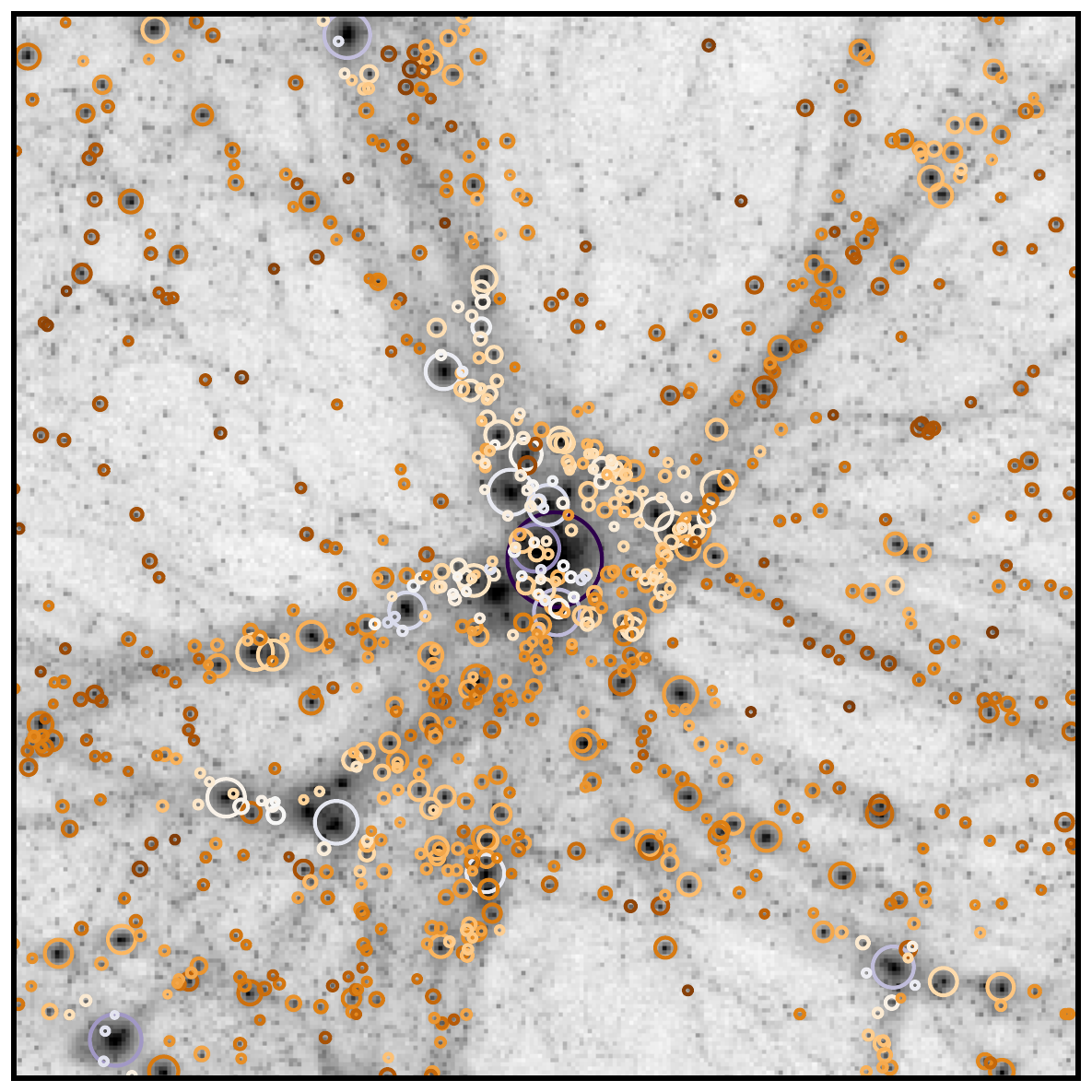}
    \includegraphics[width=0.3\textwidth]{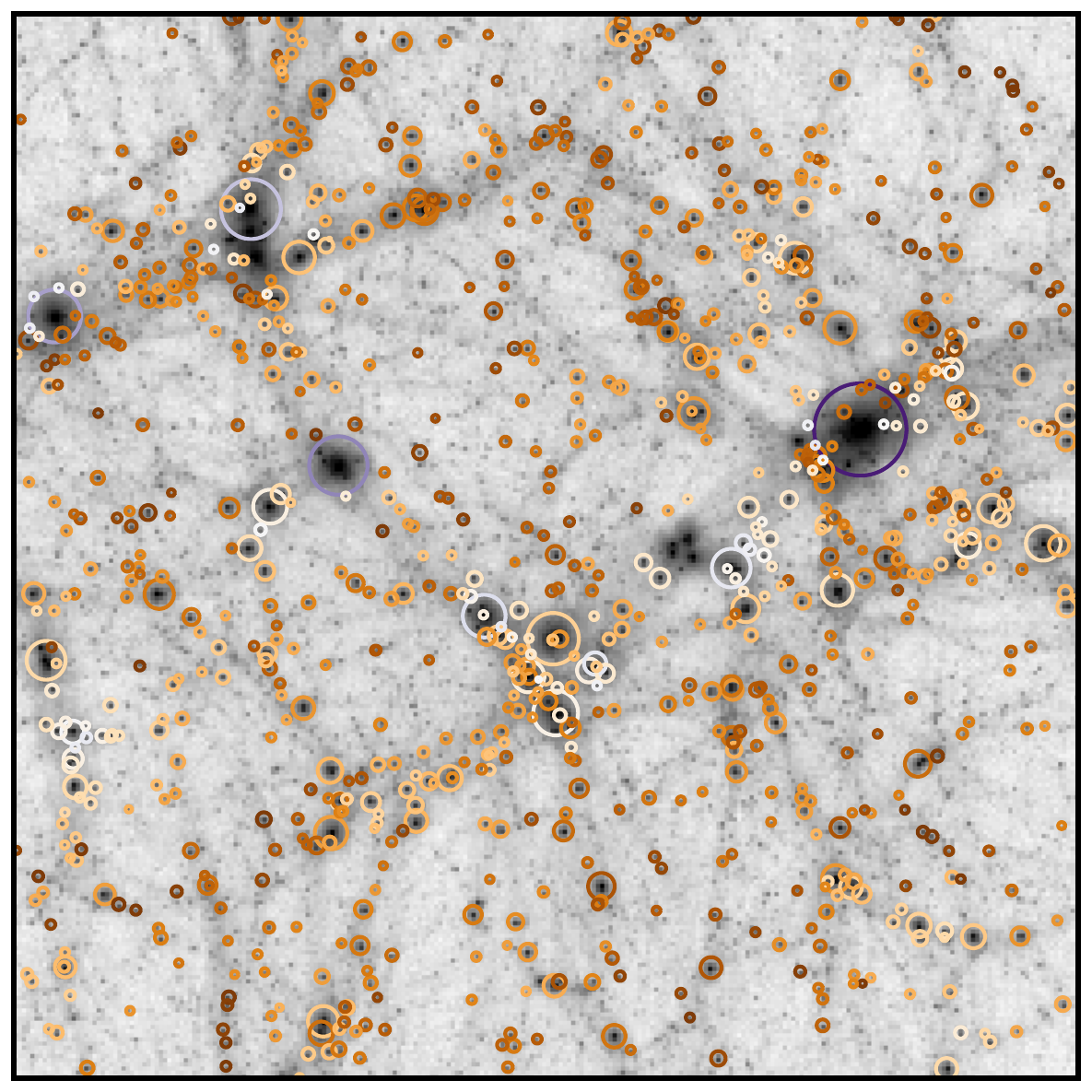}

    \includegraphics[width=0.3\textwidth]{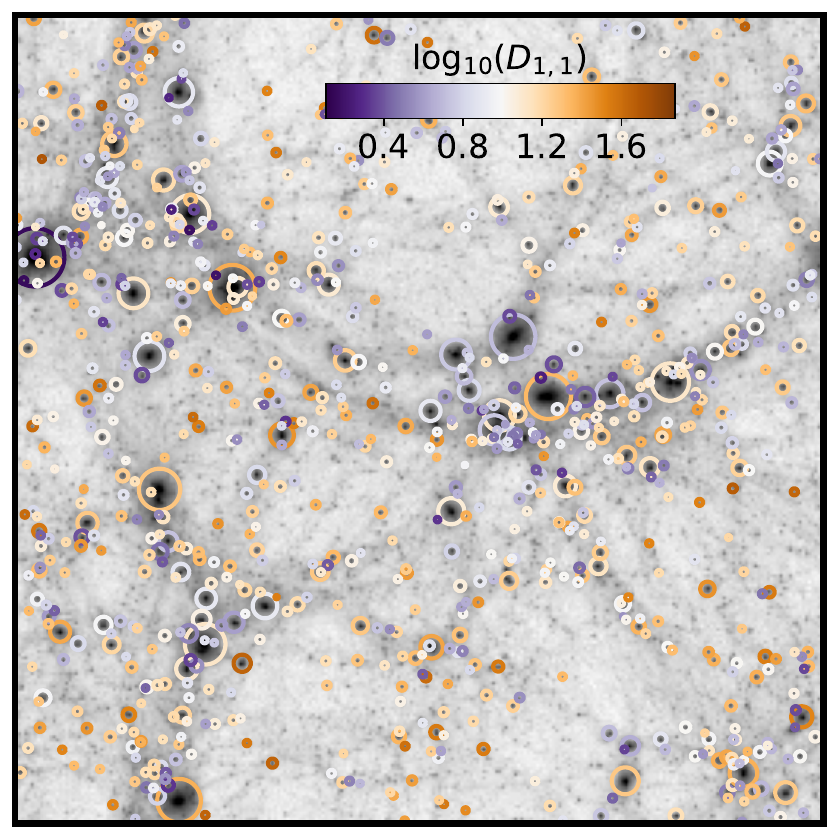}
    \includegraphics[width=0.3\textwidth]{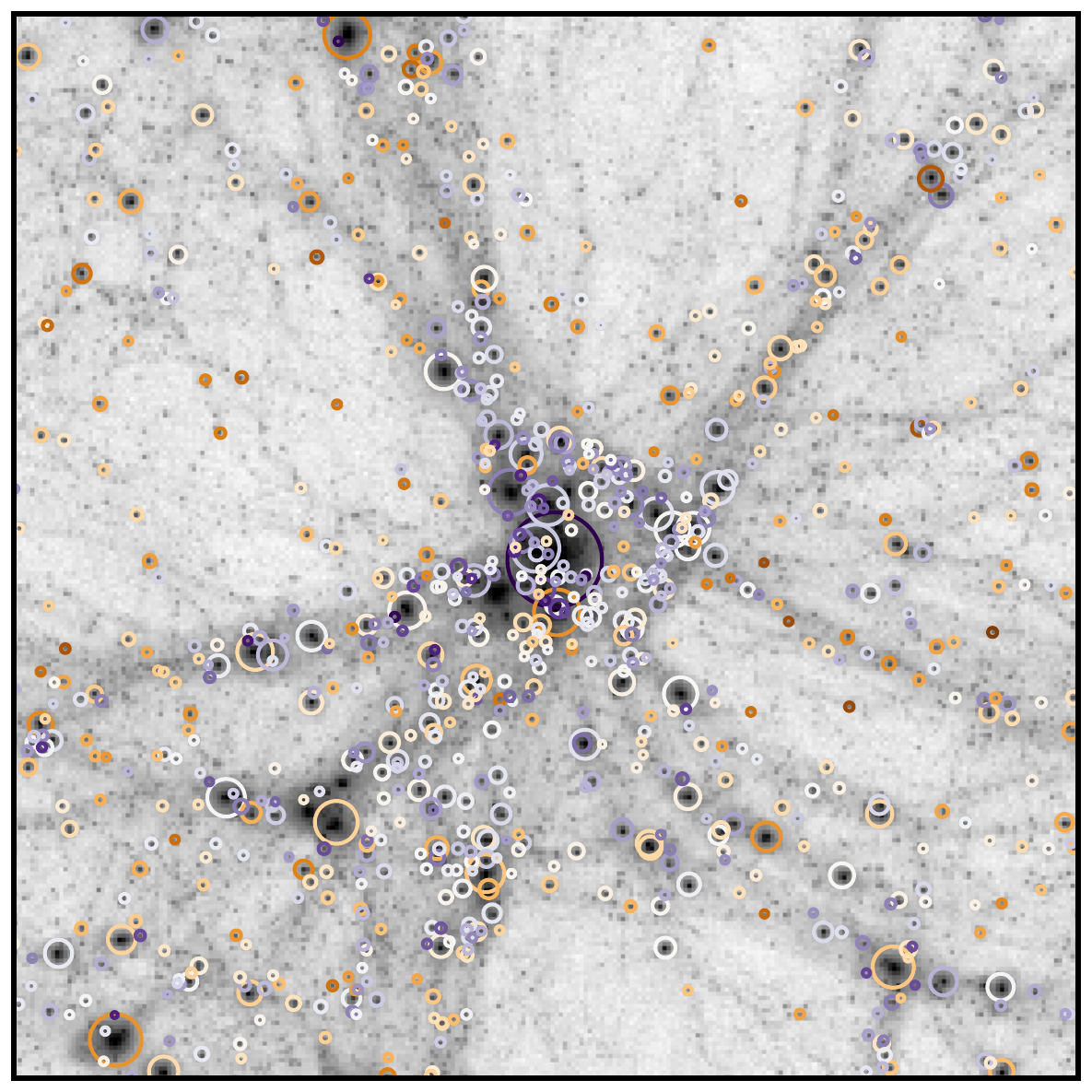}
    \includegraphics[width=0.3\textwidth]{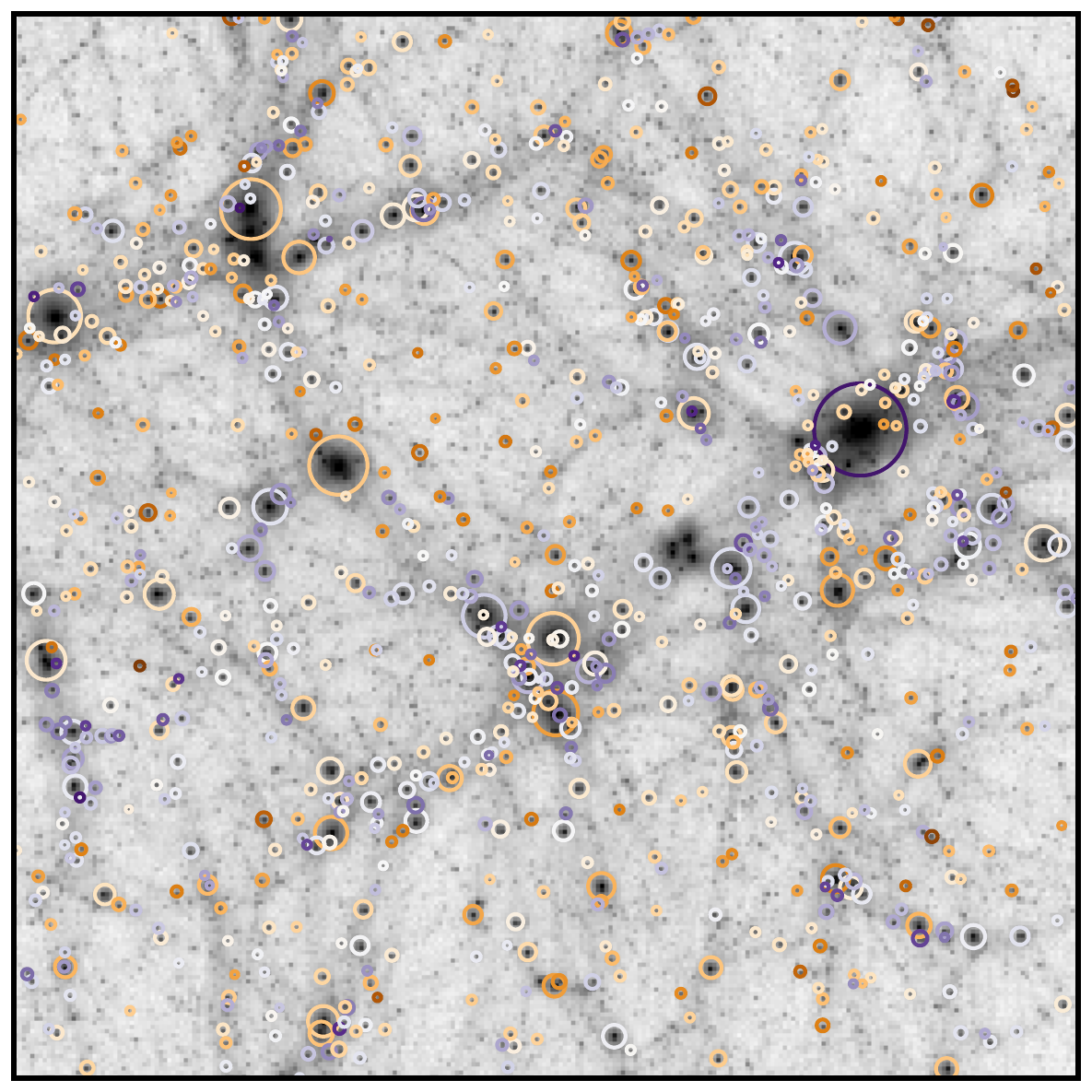}
    \caption{Visual illustration of central galaxy overdensities (top row) and $D_{1,1}$ (bottom row) for three representative simulations from the SIMBA suite (CV\_0, CV\_1, and CV\_2 from left to right). The background gray scale represents the projected dark matter distribution and the color of each halo indicates the $\delta_{10}$ or $D_{1,1}$ environment value of its central galaxy. The most massive haloes reside in the most overdense regions and have higher $\delta_{10}$ values/lower $D_{1,1}$ values.}
    \label{fig:envir_over}
\end{figure*}

\section{Overview of Simulations}
\label{sec: Overview of Simulations}

In this section, we illustrate the effects of environment on galaxy properties across different simulation models. Additionally, we show the variety of environments present across different simulations in the SIMBA CV set, quantifying the range of $\delta_{10}$ and $D_{1,1}$ values and their connection to halo mass.

\subsection{Galaxy formation models in CAMELS}

Figure~\ref{fig:simbaplot} shows a visual representation of the relationship between environment and specific SFR (sSFR) of galaxies in SIMBA, IllustrisTNG, ASTRID, and Swift-EAGLE at $z=0$. This figure shows central haloes (circles) and satellite galaxies (plus signs) color coded by their sSFR, and the color bars are identical across all four panels so that sSFR can be compared directly between galaxy formation models. The most massive, quenched haloes only reside in regions of the simulated volume that are densely populated with galaxies, while smaller, star-forming galaxies tend to reside in lower density environments. Despite this general trend, there are significant differences between models. SIMBA has the largest quantity of small, actively star forming galaxies when compared to the other three models. Particularly, low-mass galaxies in ASTRID and Swift-EAGLE tend to be quenched even in lower density regions while most small galaxies in IllustrisTNG tend to have active star formation with only a small minority of them being quenched. This implies that stellar feedback is far more effective at suppressing star formation in ASTRID and Swift-EAGLE than in SIMBA and IllutrisTNG, although the minimum sSFR that galaxies reach in Swift-EAGLE is higher compared to the other three models. There are also significant differences in the ability of models to quench massive galaxies, with TNG and SIMBA implementing more efficient AGN feedback compared to ASTRID and Swift-EAGLE. On the other hand, satellite galaxies in massive haloes tend to have low sSFR in all models. We next quantify the correlation between environment and halo mass for the specific case of the SIMBA simulations.

\subsection{Local environment in CAMELS}
\label{subsec:local environment}
The two definitions of environment, $\delta_{10}$ and $D_{1,1}$ are qualitatively consistent but vary in their connection to $M_{\rm halo}$. We quantify the dependence of $\delta_{10}$ and $D_{1,1}$ on halo mass in Figure~\ref{fig:delta-DNF} for the SIMBA CV simulations at $z=0$ (similar results apply to the other simulation models). The left panel shows median $\delta_{10}$ galaxy overdensities (black line) and $25^{\rm th}-75^{\rm th}$ percentile ranges (blue shaded region) within different host halo mass bins, with all galaxy environments in the range $\delta_{10} \sim 0.1$--$1000$. Low-mass haloes ($M_{\rm halo} \lesssim 10^{12}$\,M$_\odot$) preferentially populate lower density regions, with $\delta_{10} \sim 0.15$ roughly independent of $M_{\rm halo}$, but higher mass haloes show a strong correlation between $\delta_{10}$ and $M_{\rm halo}$, where the typical galaxy overdensity increases by three orders of magnitude in the range $M_{\rm halo} = 10^{12} \rightarrow 10^{14}$\,M$_\odot$. The right panel of Figure~\ref{fig:delta-DNF} shows $D_{1,1}$ as a function of $M_{\rm halo}$.
In this case, central galaxies tend to populate regions within the range $0.9 \lesssim D_{1,1} \lesssim 1.25$, with only a weak trend of increasing median $D_{1,1}$ values (by $<$20\%) in the range $M_{\rm halo} = 10^{10} \rightarrow 10^{13}$\,M$_\odot$. Comparing the two panels in Figure~\ref{fig:delta-DNF}, it is clear that $D_{1,1}$ is much less sensitive to $M_{\rm halo}$ than $\delta_{10}$. There is a fairly steep increase in $\delta_{10}$ after galaxies surpass $M_{\rm halo} \sim 10^{12}$\,M$_{\odot}$ while $D_{1,1}$ only slightly increases across the full range of $M_{\rm halo}$.
We show in Appendix~\ref{sec:Appendix_B} that $\delta_{10}$ and $D_{1,1}$ also correspond to different physical distances over which local environment is measured ($R_{\rm env}$) as a function of halo mass. Our measured values of $\delta_{10}$ correspond to scales decreasing from $R_{\rm env} \sim 3.5$\,Mpc for $M_{\rm halo} \lesssim 10^{11}$\,M$_{\odot}$ to $R_{\rm env} \lesssim 0.5$\,Mpc for $M_{\rm halo} > 10^{13}$\,M$_{\odot}$, while the corresponding environment distances for $D_{1,1}$ increase from $R_{\rm env} \sim 1 \rightarrow 10$\,Mpc in the same halo mass range.

Figure~\ref{fig:envir_over} visually illustrates the connection between $\delta_{10}$ and $D_{1,1}$ and the large scale environment of galaxies in three different SIMBA CV simulations (CV\_0, CV\_1, and CV\_2), where we show projected dark matter mass distributions (background gray scale) with central galaxies/haloes overlaid as circles of sizes representing their virial radii and color-coded by their corresponding $\delta_{10}$ (top row of Figure~\ref{fig:envir_over}) and $D_{1,1}$ (bottom row of Figure~\ref{fig:envir_over}) values. The different large-scale structures seen in the three CV simulations underscore the importance of cosmic variance in the CAMELS's (25\,$h^{-1}$Mpc)$^3$ simulated volumes. As expected from Figure~\ref{fig:delta-DNF}, larger haloes form in denser regions and have higher $\delta_{10}$ values. The $D_{1,1}$ environmental measure, instead, quantifies the proximity of haloes to others of similar or larger mass, which is roughly independent of mass. As such, low (high) mass galaxies with predominantly low (high) $\delta_{10}$ values can show a wide distribution of $D_{1,1}$ values.

\section{Impact of Environment}
\label{sec: Impact of Environment}

We investigate the effects of local environment on galaxy evolution by looking at various halo and galaxy global properties as a function of halo mass and stellar mass. Using the methodology described in Section~\ref{subsec:statistics}, we explore the impact of environment on $f_{\rm B}$, $f_{\rm CGM}$, $f_{\rm gas}$, and SFR.

\subsection{Halo baryon fraction}

The baryon fraction quantifies the ratio of baryonic mass to the total mass of a halo. The baryon fraction of a halo is highly dependent on halo mass and the efficiency of stellar and AGN feedback within the halo, providing a strong constraint for feedback models, but it can also depend on the environment. Therefore, it is useful to determine the impact of environment on baryon fraction for different feedback models, where the amount of baryons that haloes can retain may also correlate with the intrinsic feedback produced by nearby haloes.

\begin{figure*}
    \centering
    \includegraphics[width=\textwidth]{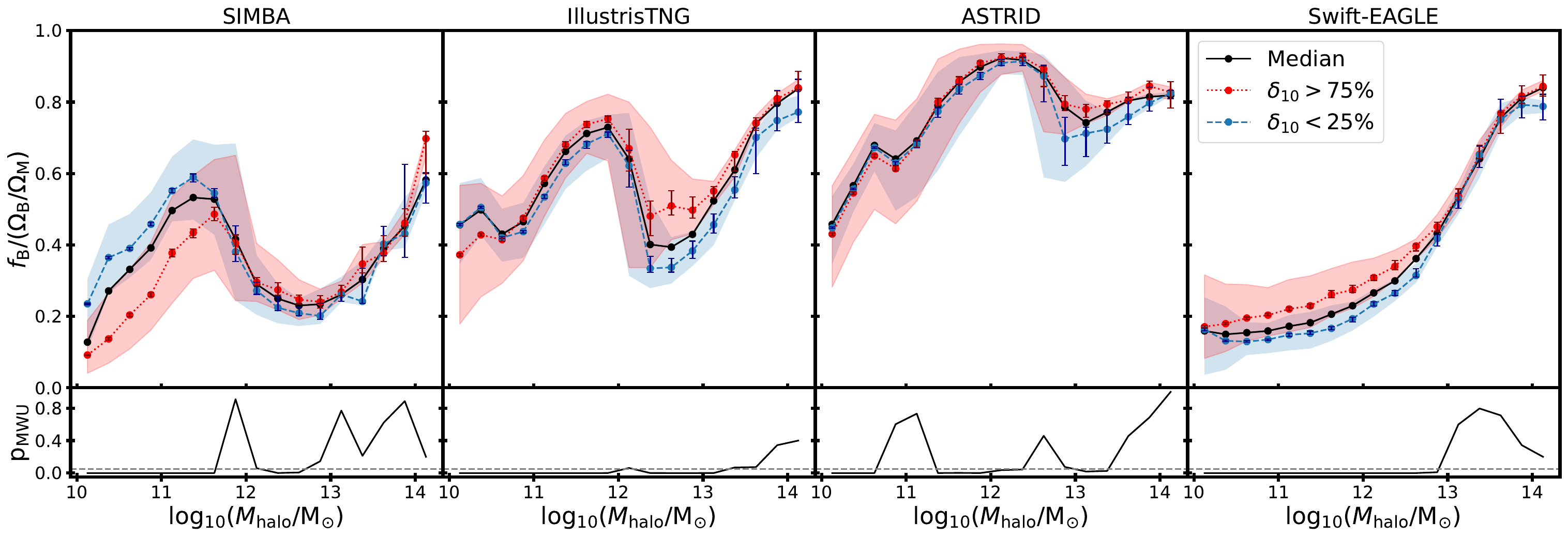}
    \vspace{-0.15in}
    \caption{Baryon fraction, $f_{\rm B}$, as a function of halo mass at $z=0$ in SIMBA, IllustrisTNG, ASTRID, and Swift-EAGLE (from left to right) for all simulations in each CAMELS CV set. We show the median $f_{\rm B}$ in each halo mass bin (black), the median $f_{\rm B}$ of underdense haloes with $\delta_{10}$ below the $25^{\rm th}$ percentile in each bin (blue), and the median $f_{\rm B}$ of overdense haloes with $\delta_{10}$ above the $75^{\rm th}$ percentile in each bin (red). The blue and red shaded regions indicate the $25^{\rm th}$ to $75^{\rm th}$ percentile ranges of underdense and overdense haloes, respectively. Error bars depict 90\% confidence intervals of bootstrap medians and the bottom panels show Mann-Whitney-U test p-values between underdense and overdense haloes as a function of $M_{\rm halo}$. In SIMBA, low-mass haloes ($M_{\rm halo} \lesssim 10^{12} {\rm M}_{\rm \odot}$) have higher $f_{\rm B}$ in lower density environments but this trend reverses for intermediate mass haloes ($10^{12} {\rm M}_{\rm \odot} \lesssim M_{\rm halo} \lesssim 10^{13} {\rm M}_{\rm \odot}$) and disappears for higher mass haloes. In IllustrisTNG and ASTRID, low-mass haloes are roughly independent of environment, showing instead a weak trend for higher $f_{\rm B}$ in overdense regions for massive haloes. In contrast, Swift-EAGLE haloes in overdense regions have systematically higher $f_{\rm B}$ across the full halo mass range with the trend weakening for haloes with $10^{13} {\rm M}_{\rm \odot} \lesssim M_{\rm halo} \lesssim 10^{13.5} {\rm M}_{\rm \odot}$.}
    \label{fig:f_b}
\end{figure*}

Figure~\ref{fig:f_b} shows baryon fraction as a function of halo mass for the fiducial models of SIMBA, IllustrisTNG, ASTRID, and Swift-EAGLE at $z=0$. We show the median $f_{\rm B}$--$M_{\rm halo}$ relation for all galaxies in the CV simulations of each model (black solid line) as well as the median relation for galaxies located in overdense regions (red dotted line) and underdense regions (blue dashed line). In all models, the baryon fraction is generally lowest in low-mass haloes and highest in high-mass haloes but the halo mass and environmental dependence of $f_{\rm B}$ is highly dependent on galaxy formation model.

SIMBA strongly suppresses the baryon fraction of low-mass haloes, with $f_{\rm B} \sim 0.2$ for $M_{\rm halo} \sim 10^{10}$\,M$_\odot$, owing to efficient star formation-driven winds evacuating gas from haloes. Higher mass haloes retain more baryons as the gravitational potential well deepens, reaching $f_{\rm B} \sim 0.5$ at $M_{\rm halo} \sim 10^{11.5}$\,M$_\odot$, but the onset of efficient AGN feedback reduces the baryon fraction at higher masses, with $f_{\rm B} \sim 0.2$--$0.3$ in the range $10^{12}$\,M$_\odot < M_{\rm halo} < 10^{13.5}$\,M$_\odot$. For the most massive haloes, AGN feedback becomes less effective at evacuating gas and the baryon fraction increases to $f_{\rm B} \sim 0.6$ at $M_{\rm halo} \sim 10^{14}$\,M$_\odot$. Comparing $f_{\rm B}$ for haloes in different environments at fixed $M_{\rm halo}$, we see that low-mass haloes ($M_{\rm halo}<10^{12}$\,M$_\odot$) tend to have higher baryon fractions when they are in less dense environments than their counterparts in denser environments. This trend reverses at $M_{\rm halo} \sim 10^{12}\,{\rm M}_{\odot}$--$10^{13}\,{\rm M}_{\odot}$. Above this mass, the trend is not significant as indicated by the $p_{\rm MWU} > 0.05$ values (bottom row).

\begin{figure*}
    \centering
    \includegraphics[width=\textwidth]{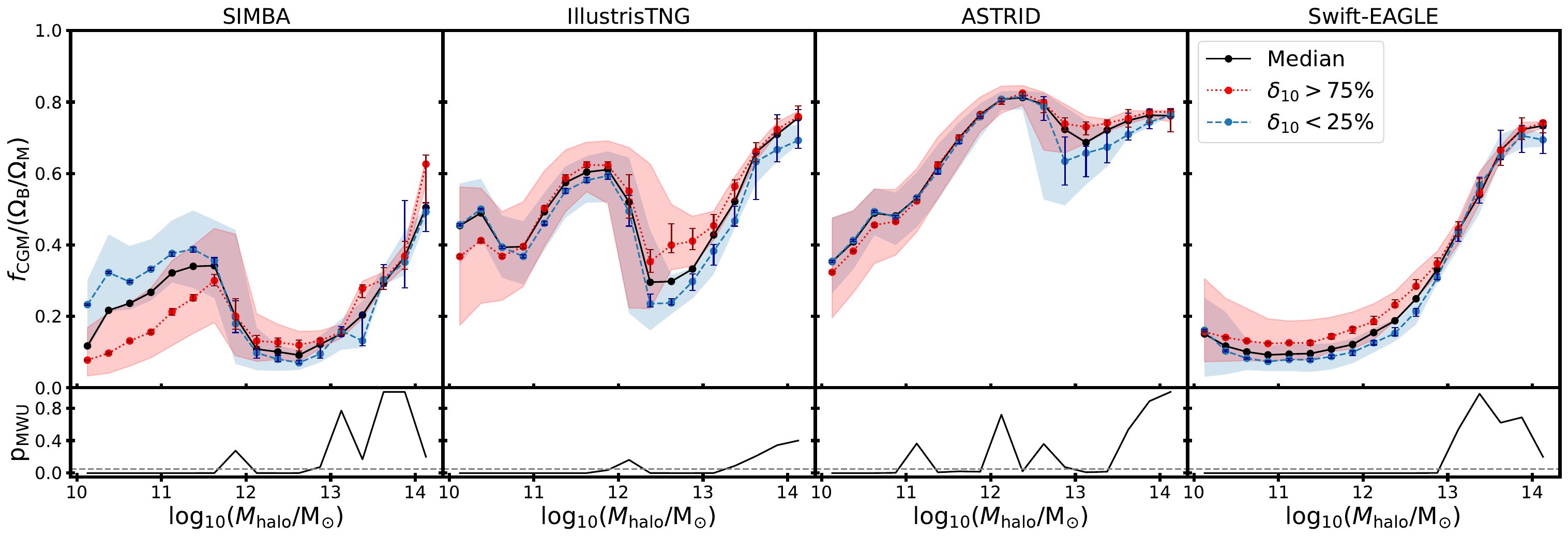}
    \vspace{-0.15in}
    \caption{Circumgalactic medium fraction, $f_{\rm CGM}$, as a function of halo mass at $z=0$ in SIMBA, IllustrisTNG, ASTRID, and Swift-EAGLE (from left to right) for all simulations in each CAMELS CV set. Black, blue, and red lines show median trends for all, underdense, and overdense haloes respectively, with 90\% bootstrap error bars and Mann-Whitney-U p-values shown as in Figure~\ref{fig:f_b}. In SIMBA, haloes with $M_{\rm halo} < 10^{11.5}\,{\rm M}_\odot$ have higher $f_{\rm CGM}$ in underdense environments compared to overdense environments. This trend reverses in the $10^{12}\,{\rm M}_\odot $--$10^{13}\,{\rm M}_\odot$ halo mass regime, and there is no trend with environment for higher masses. In IllustrisTNG and ASTRID, haloes with $M_{\rm halo} > 10^{12.5}\,{\rm M}_\odot$ have higher $f_{\rm CGM}$ in high density environments but there is only a weak or absent environmental impact at lower $M_{\rm halo}$. In Swift-EAGLE, haloes in overdense regions have systematically higher $f_{\rm CGM}$ compared to lower density regions until they reach a halo mass of $10^{13}\,{\rm M}_\odot$ with the trend reappearing at $M_{\rm halo} \sim 10^{14}\,{\rm M}_\odot$.}
    \label{fig:f_cgm}
\end{figure*}

IllustrisTNG predicts a qualitatively similar dependence of baryon fraction on halo mass as the SIMBA model, with the transition mass occurring at $M_{\rm halo} \sim 10^{12}\,{\rm M}_{\odot}$, but with higher overall baryon fractions across the full halo mass range. In this case, we find $p_{\rm MWU} < 0.05$ across the halo mass range, with the most significant difference between
the baryon fraction of underdense and overdense populations occurring at $M_{\rm halo} \gtrsim 10^{12.5}$\,M$_\odot$. 
Unlike SIMBA, $f_{\rm B}$ in the IllustrisTNG model is less sensitive to changes in environment at lower halo masses, though still statistically significant.

ASTRID shows again a qualitatively similar trend between baryon fraction and halo mass as IllustrisTNG and SIMBA, with the peak of $f_{\rm B}$ ocurring at $M_{\rm halo} \sim 10^{12}$\,M$_\odot$, but with significantly higher overall baryon fractions indicating weaker feedback efficiency across the full halo mass range. In terms of the impact of environment on halo baryon fractions, ASTRID predicts no significant differences in $f_{\rm B}$ between high $\delta_{10}$ and low $\delta_{10}$ haloes for $M_{\rm halo} \lesssim 10^{12}$\,M$_\odot$, but shows a systematic trend for higher $f_{\rm B}$ in overdense environments for higher mass haloes (qualitatively similar to IllustrisTNG).

The Swift-EAGLE model shows significantly different baryon fractions as a function of halo mass compared to the other three models, with $f_{\rm B} \sim 0.2$ weakly dependent on halo mass at $M_{\rm halo} \lesssim 10^{12}$\,M$_\odot$ and a steep increase at higher masses (rather than a turnover) reaching $f_{\rm B} \sim 0.8$ at $M_{\rm halo} \sim 10^{14}$\,M$_\odot$. The impact of environment is also qualitatively different in Swift-EAGLE compared to the other models, with systematically higher $f_{\rm B}$ in overdense regions at fixed $M_{\rm halo}$ across the full halo mass range with a weakening of the trend for haloes with $M_{\rm halo} \gtrsim 10^{13} {\rm M}_{\rm \odot}$.

\subsection{Halo CGM fraction}

CGM fraction is defined as the ratio of CGM gas mass to total halo mass \citep{Medlock2025}. Since the CGM is made up of non-star forming gas surrounding a galaxy, we estimate $f_{\rm CGM}$ by subtracting gas that is dense and most likely to form stars from the gas of the entire halo \citep{Tumlinson2017}. Like baryon fraction, $f_{\rm CGM}$ is a very informative measure of feedback efficiency \citep{davies2020, Oppenheimer2020, Medlock2025}, since the CGM gas mass that haloes can retain depends on the balance between gravitational potential and feedback strength, and it can also be a sensitive probe of environment.

Figure~\ref{fig:f_cgm} shows $f_{\rm CGM}$ as a function of halo mass at $z=0$ for the CV simulations in the SIMBA, IllustrisTNG, ASTRID, and Swift-EAGLE suites. As in Figure~\ref{fig:f_b}, we compare the median relations for all haloes in each CV set (black) to those for haloes in overdense (red) and underdense (blue) regions separately. As the gas mass in haloes dominates over the stellar component, $f_{\rm CGM}$ shows very similar trends with halo mass and environment as the halo baryon fraction (Figure~\ref{fig:f_b}), showing again very significant differences between models. In the SIMBA model (Figure~\ref{fig:f_cgm}; left panel), haloes in underdense regions (lower $\delta_{10}$) have higher $f_{\rm CGM}$ than haloes of similar mass in overdense regions (high $\delta_{10}$) at $M_{\rm halo} < 10^{11.5}$\,M$_\odot$, where $p_{\rm MWU} < 0.05$. This trend continues until haloes are massive enough to contain SMBHs and exert AGN feedback from within. Conversely, haloes in overdense regions have higher $f_{\rm CGM}$ than haloes of similar mass in underdense regions at $10^{12}$\,M$_\odot < M_{\rm halo} < 10^{13}$\,M$_\odot$. In contrast, IllustrisTNG shows little variation with environment for haloes with $M_{\rm halo} \lesssim 10^{12}\,{\rm M}_{\odot}$. At higher masses, haloes with high $\delta_{10}$ have higher $f_{\rm CGM}$ than haloes with low $\delta_{10}$. ASTRID shows a similar impact of environment on $f_{\rm CGM}$ as IllustrisTNG, but the variation of $f_{\rm CGM}$ with environment for large haloes is less drastic and occurs at higher masses. In ASTRID, the peak/turnaround in $f_{\rm CGM}$ occurs at $M_{\rm halo} \sim 10^{12.75}$, and the decrease of $f_{\rm CGM}$ at higher masses is less pronounced compared to SIMBA and IllustrisTNG, as seen for halo baryon fractions. 

Finally, as expected, Swift-EAGLE shows qualitatively different trends for $f_{\rm CGM}$ compared to the other models. Overdense haloes have higher $f_{\rm CGM}$ than underdense haloes until $M_{\rm halo} \sim 10^{13}$\,M$_\odot$. Above this mass, the $p_{\rm MWU}$ values indicate no trend with environment. While SIMBA, IllustrisTNG, and ASTRID show a peak in $f_{\rm CGM}$ at $M_{\rm halo} \sim 10^{12}$\,M$_\odot$, reflecting a transition from weakening stellar feedback efficiency to stronger AGN feedback relative to the increasing halo gravitational potential, Swift-EAGLE shows a monotonic increase in $f_{\rm CGM}$ with halo mass as a consequence of stronger stellar feedback at low $M_{\rm halo}$ and weaker AGN feedback at high $M_{\rm halo}$ \citep{Wright2024}. The environmental dependence of $f_{\rm CGM}$ in Swift-EAGLE is consistent with \citet{Shreeram25} who showed that galaxies in denser environments are X-ray brighter than galaxies in void environments. This result is shown to be partially due to higher hot CGM gas densities in dense environments.

\begin{figure*}
    \centering
    \includegraphics[width=\textwidth]{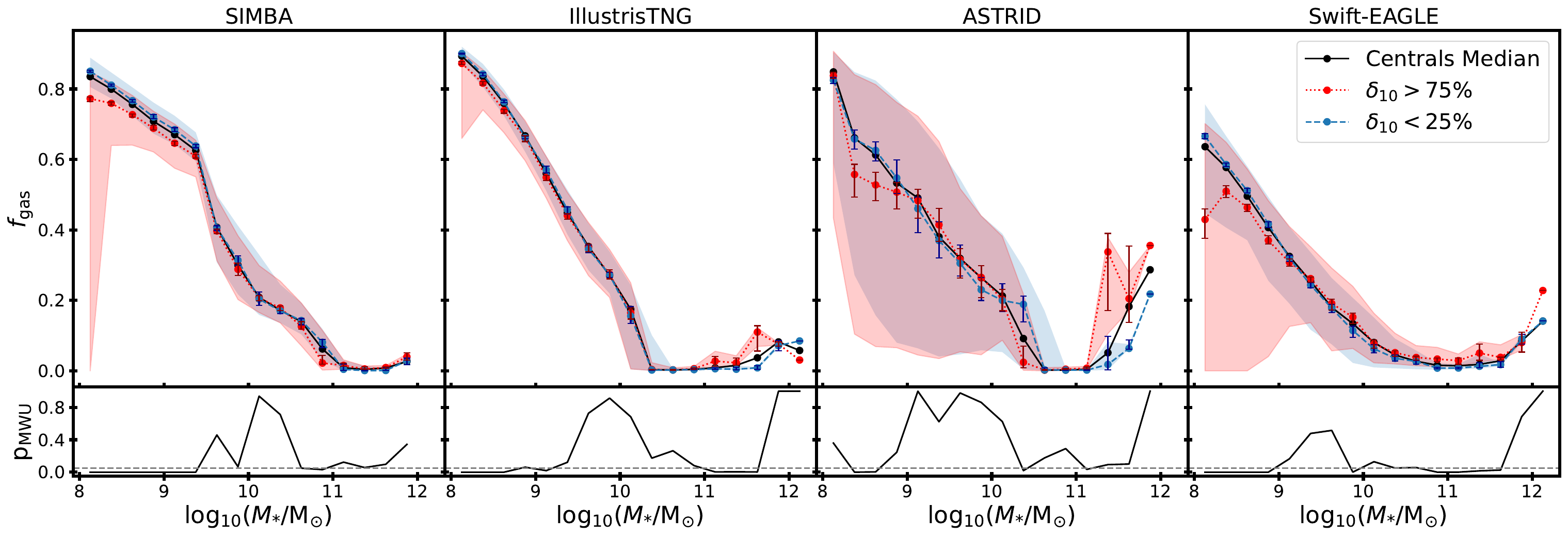}
    \vspace{0.2in}
    \includegraphics[width=\textwidth]{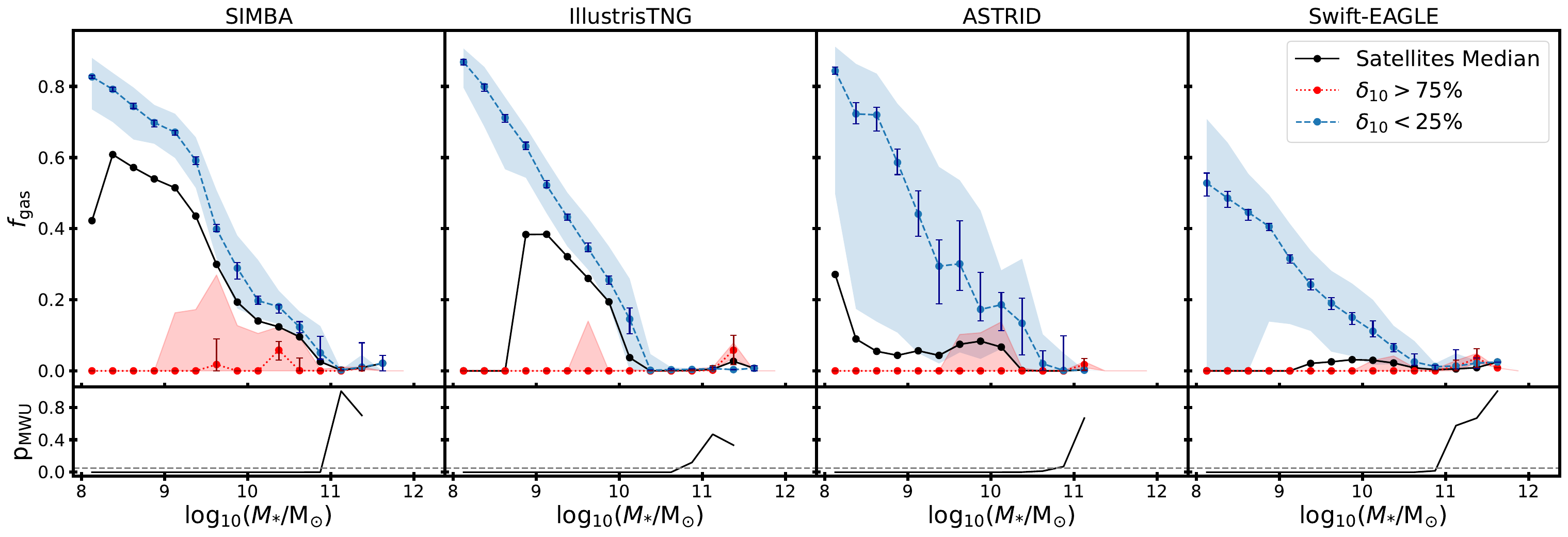}
    \vspace{-0.15in}
    \caption{Galaxy gas fraction, $f_{\rm gas}$, as a function of stellar mass at $z=0$ for central galaxies (top row) and satellite galaxies (bottom row) in SIMBA, IllustrisTNG, ASTRID, and Swift-EAGLE (from left to right) for all simulations in each CAMELS CV set. Black, blue, and red lines show median trends for all, underdense, and overdense galaxies respectively, with 90\% bootstrap error bars and Mann-Whitney-U p-values shown as in Figure~\ref{fig:f_b}. Central galaxy gas fractions are roughly insensitive to environment, with only a hint of lower $f_{\rm gas}$ in overdense environments for $M_\star \lesssim 10^{9.5}\, {\rm M}_{\odot}$ and the opposite trend at higher masses. In contrast, gas fraction in satellite galaxies is strongly dependent on environment, with significantly higher $f_{\rm gas}$ for satellites in underdense environments (until they reach $M_\star \gtrsim 10^{10.5} \,{\rm M}_{\odot}$, with $f_{\rm gas}\sim 0$ at higher masses regardless of environment). All simulation models (SIMBA, TNG, Astrid, Swift-EAGLE) predict qualitatively similar impact of environment on galaxy gas fractions.}
    \label{fig:f_g}
\end{figure*}
\subsection{Galaxy gas fraction}

Unlike $f_{\rm B}$ and $f_{\rm CGM}$, the gas fraction $f_{\rm gas}$ is defined as a galaxy property, and thus can be defined for satellite galaxies in addition to central galaxies. Therefore, we can use $f_{\rm gas}$ to investigate the sensitivity of both satellite galaxies and central galaxies to their local environments. 

In the top panels of Figure~\ref{fig:f_g}, we show the relationship between gas fraction and stellar mass at $z=0$ for all central galaxies in the CV simulations of SIMBA, IllustrisTNG, ASTRID, and Swift-EAGLE (black), comparing the median trends for central galaxies in overdense (red) and underdense (blue) regions. There is an overall decrease in $f_{\rm gas}$ as $M_\star$ increases in all galaxy formation models, with $f_{\rm gas} \sim 0.7$--$0.9$ at $M_\star \sim 10^8$\,M$_\odot$ decreasing down to $f_{\rm gas} \sim 0$ at $M_\star \gtrsim 10^{10}$--$10^{11}$\,M$_\odot$. The galaxy stellar mass at which its gas content vanishes depends on model implementation details, with IllustrisTNG reaching $f_{\rm gas}\sim 0$ as early as $M_\star \sim 10^{10.25}$\,M$_\odot$ compared to $M_\star \sim 10^{11}$\,M$_\odot$ in SIMBA. Interestingly, at the highest stellar masses there is a clear increase in the median $f_{\rm gas}$ values, more prominently in ASTRID and Swift-EAGLE, indicating that feedback is no longer able to evacuate gas from the central region of the most massive galaxies and correlating with the higher baryon fraction and $f_{\rm CGM}$ in the most massive haloes. In all models (to a different extent), we identify low-mass central galaxies in overdense regions having lower $f_{\rm gas}$ compared to underdense regions (with $p_{\rm MWU} < 0.05$ at $M_\star \lesssim 10^9$\,M$_\odot$), and a hint for the opposite trend (higher $f_{\rm gas}$ in overdense regions) for massive galaxies ($M_\star \sim 10^{11.5}$\,M$_\odot$).

\begin{figure*}
    \includegraphics[width=\textwidth]{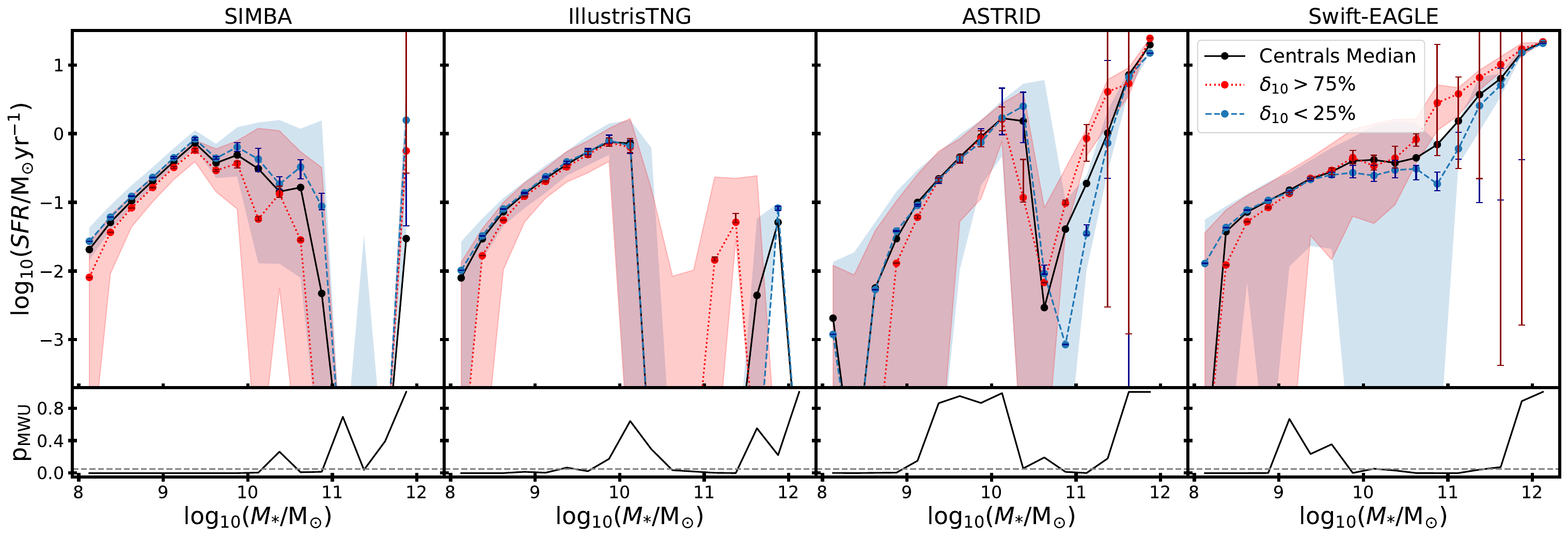}
    \vspace{0.2in}
    \includegraphics[width=\textwidth]{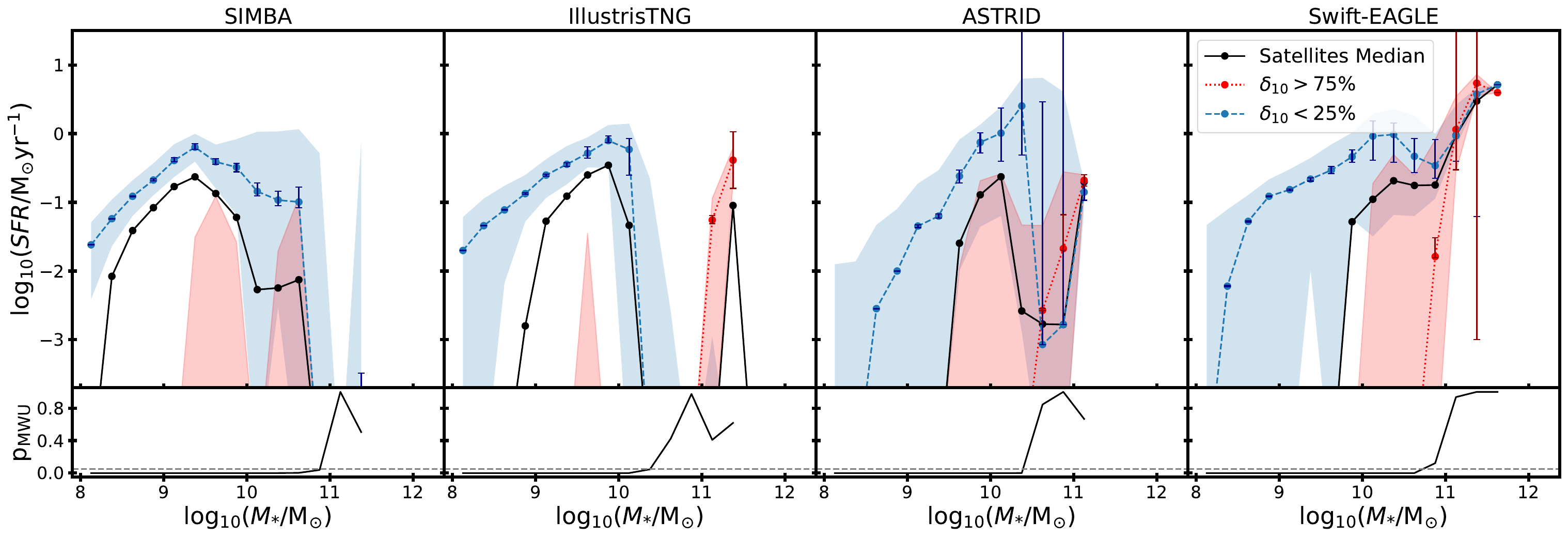}
    \vspace{-0.2in}
    \caption{Same as Figure~\ref{fig:f_g} for the SFR of central galaxies (top row) and satellite galaxies (bottom row) at $z=0$ in SIMBA, IllustrisTNG, ASTRID, and Swift-EAGLE (from left to right). Central galaxies show a weak correlation between SFR and environment, with lower SFR at low masses and a hint for higher SFR at high masses (depending on model) for galaxies in overdense regions. However, there is a very strong trend with the environment for satellite galaxies, displaying significantly higher SFR in low density environments at fixed $M_\star$ and a hint for the opposite trend at $M_\star > 10^{11} \,{\rm M}_{\odot}$ (depending on model).}
    \label{fig:sfrboth}
\end{figure*}

The shaded regions represent the $25^{\rm th}$ to $75^{\rm th}$ percentile range of variation for the gas fraction of overdense and underdense galaxies in a given stellar mass, showing interesting differences between models. ASTRID shows substantial stochasticity in galaxy gas fractions, with $f_{\rm gas}$ values greatly departing from median values for both underdense and overdense regions while SIMBA and IllustrisTNG yield significantly smaller variation in $f_{\rm gas}$ at fixed $M_\star$ across the entire mass range (except at $M_\star \sim 10^8$\,M$_\odot$ in SIMBA). As an intermediate case, Swift-EAGLE predicts relatively low scatter in $f_{\rm gas}$ values for underdense galaxies but as much stochasticity as ASTRID for overdense galaxies at $M_\star \lesssim 10^9$\,M$_\odot$. Interestingly, Swift-EAGLE seems to show higher $f_{\rm gas}$ in overdense galaxies relative to underdense galaxies for the most massive systems ($M_\star \gtrsim 10^{11}$\,M$_\odot$).

In the bottom panels of Figure~\ref{fig:f_g}, we show $f_{\rm gas}$ as a function of $M_\star$ for satellite galaxies in SIMBA, IllustrisTNG, ASTRID, and Swift-EAGLE, comparing the overall median trend for all satellites (black) to the selected populations of overdense (red) and underdense (blue) satellites. In all models, satellite galaxies have substantially lower gas fractions than centrals at fixed $M_\star$, with this difference becoming more obvious at lower masses. SIMBA predicts the highest satellite gas fractions, followed by IllustrisTNG, ASTRID, and finally Swift-EAGLE with the lowest median satellite gas fractions.

In addition, unlike centrals, the gas fraction of satellite galaxies is very sensitive to environment. Underdense satellite galaxies have much higher $f_{\rm gas}$ than overdense satellites with the same stellar mass. This trend passes the significance test ($p_{\rm MWU} < 0.05$) across the stellar mass range until $M_{\star} \sim 10^{11}$\,M$\odot$, where the difference in $f_{\rm gas}$ for high $\delta_{10}$ and low $\delta_{10}$ galaxies decreases as the overall gas fraction decreases at high stellar masses. Notably, in all models, the overdense satellite population is consistent with quenched galaxies with negligible gas content at all stellar masses.

\begin{figure*}
    \centering 
    \includegraphics[width=\textwidth]{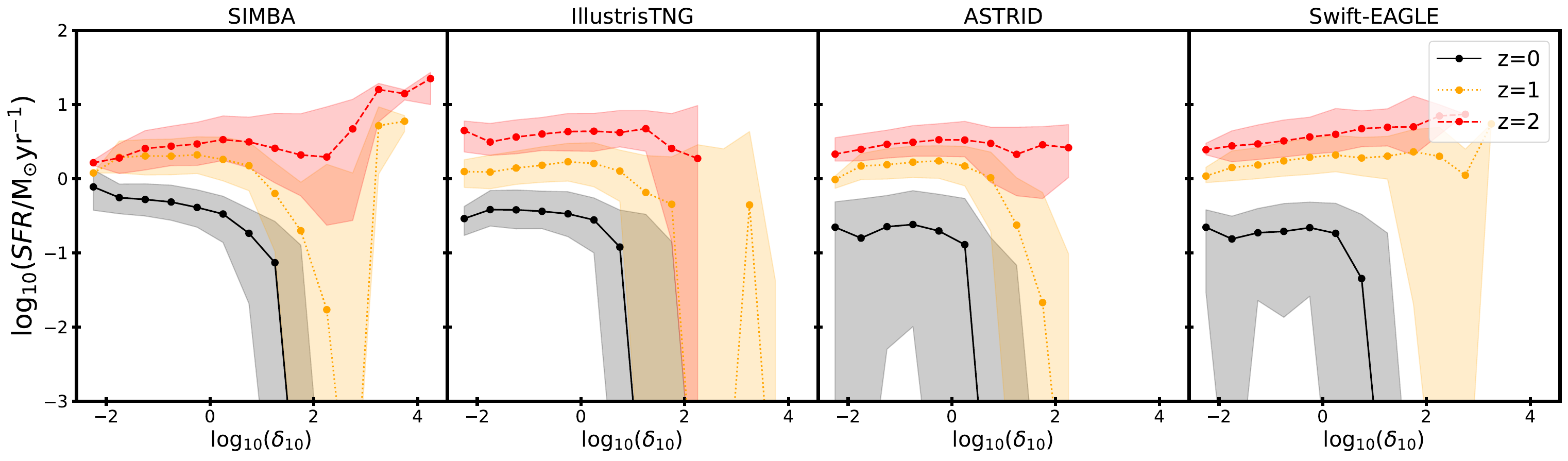}
    \vspace{-0.2in}
    \caption{Median SFR as a function of overdensity, $\delta_{10}$, for all galaxies in the mass range $M_\star=10^9$--$10^{10}\,{\rm M}_\odot$ at $z=0$\; (solid black), $z=1$\; (dotted orange), and $z=2$\; (dashed red). The shaded regions indicate the 25$^{\rm th}$ to 75$^{\rm th}$ percentile range of SFR for each overdensity bin. SFRs generally increase with redshift at fixed overdensity and galaxies quench at lower $\delta_{10}$ at lower redshift, but how overdensity impacts galaxy SFRs is redshift and model dependent.}
    \label{fig:z_sfr}
\end{figure*}

\subsection{Galaxy star formation rate}

For each galaxy, we define its SFR as the sum of the instantaneous SFRs of all gas particles bound to the corresponding subhalo \citep[e.g.,][]{Villaescusa-Navarro2023}. Like $f_{\rm gas}$, SFR can be computed for satellite galaxies as well as central galaxies. Therefore, we can use this quantity to compare the impact of environment on central galaxies and satellite galaxies.

The top panels of Figure~\ref{fig:sfrboth} show SFR as a function of $M_\star$ for central galaxies at $z=0$ in the SIMBA, IllustrisTNG, ASTRID, and Swift-EAGLE CV simulations (from left to right). The median SFR of central galaxies (black solid line) increases monotonically with stellar mass for star-forming galaxies in all simulation suites, reminiscent of the observed star formation main sequence \citep{Salim2016, Salim2018}, but there are significant differences between models. SIMBA shows a steady increase in SFR with stellar mass until reaching SFR$ \sim 1$\,M$_\odot/{\rm yr}$ at $M_\star \sim 10^{9.5}$\,M$_\odot$, followed by a slow decline in SFR at higher masses and the sudden quenching of star formation at $M_\star \gtrsim 10^{10.5}$\,M$_\odot$. IllustrisTNG shows qualitatively similar results to SIMBA, but reaching the peak of SFR at $M_\star \sim 10^{10}$\,M$_\odot$ and sharply quenching star formation at higher masses. In both cases, central galaxies reaching $M_\star \gtrsim 10^{11.5}$\,M$_\odot$ are able to recover some level of star formation owing to the difficulty of fully quenching the most massive systems given their AGN feedback model implementations. ASTRID shows qualitatively similar results but reaching significantly higher SFRs at $M_\star \sim 10^{10.5}$\,M$_\odot$, only partially quenching galaxies at higher masses and maintaining SFR $\sim 10$\,M$_\odot/{\rm yr}$ at $M_\star \gtrsim 10^{11}$\,M$_\odot$, owing to its weaker AGN feedback \citep{Ni2023}. Finally, Swift-EAGLE shows a qualitatively different trend, with a fairly consistent increase in SFR all the way to the most massive galaxies.

Comparing the median SFRs of central galaxies in overdense regions (red dotted lines) and underdense regions (blue dashed lines), we find that there is only a weak correlation between SFR and local galaxy environment at fixed $M_\star$. There is some indication of lower SFR in overdense regions in SIMBA (for $M_\star \lesssim 10^{11}$\,M$_\odot$) and IllustrisTNG (for $M_\star \lesssim 10^9$\,M$_\odot$), and the opposite trend with higher SFR for high mass galaxies ($M_\star \gtrsim 10^{10.5}$\,M$_\odot$) in overdense regions in the IllustrisTNG and ASTRID simulations. Swift-EAGLE shows substantial scatter in SFR at fixed $M_\star$ for galaxies in both underdense and overdense regions, and some indication for higher SFRs in overdense galaxies in the range $10^{9.5}$\,M$_{\odot} < M_\star < 10^{11.5}$\,M$_\odot$.

The bottom panels of Figure~\ref{fig:sfrboth} show SFR as a function of stellar mass selecting only satellite galaxies from the CV simulations of SIMBA, TNG, ASTRID, and Swift-EAGLE. As expected, the median SFRs of satellites (black solid lines) are significantly lower than that of central galaxies at fixed $M_\star$. While there are substantial differences between models, comparing the median SFRs of galaxies in overdense (red dotted lines) and underdense (blue dashed lines) regions shows that satellite galaxy SFRs are strongly impacted by the local environment. In SIMBA, underdense satellites have higher SFR than overdense satellites across the stellar mass range, passing the significance test ($p_{\rm MWU} < 0.05$) until $M_\star \gtrsim 10^{11}$\,M$_\odot$, where most satellite galaxies are quenched. In IllustrisTNG, underdense satellites have higher SFR than overdense satellites with $p_{\rm MWU} < 0.05$ at $M_\star \lesssim 10^{10.5}$\,M$_\odot$. However, unlike SIMBA, there is a hint for the opposite trend at higher masses, where satellites with $M_\star \gtrsim 10^{11}$\,M$_\odot$ show higher SFR in overdense regions compared to underdense regions, although the trend fails the significance test in this mass regime. While this trend is consistent with similar results for central galaxies, IllustrisTNG efficiently suppresses the growth of massive galaxies and there are very few satellites populating the highest stellar mass bins, cautioning against overinterpreting these results.

Similar to central galaxies, both ASTRID and Swift-EAGLE show significantly higher spread in the SFR of satellites at fixed $M_\star$ compared to SIMBA and IllustrisTNG. Low-mass satellites in underdense regions in ASTRID and Swift-EAGLE generally have higher SFR than satellites in overdense regions (which are mostly quenched) but they can nonetheless reach very low SFRs. As concluded in Figure~\ref{fig:f_g} based on galaxy gas fractions, the overdense satellite population is generally consistent (in all models) with primarily quenched galaxies with negligible SFR across the stellar mass range, except for rare massive satellite galaxies with ongoing star formation.

\section{Redshift Evolution}

\begin{figure*}
    \centering
    \includegraphics[width=\textwidth]{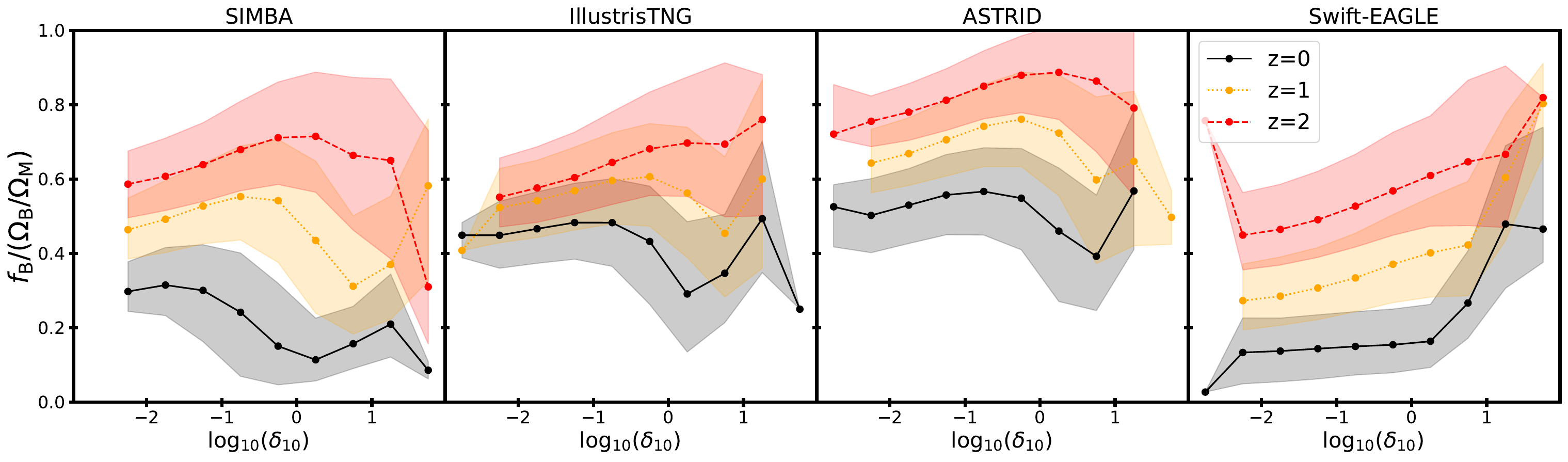}
    \vspace{-0.2in}
    \caption{Median baryon fraction, $f_{\rm B}$, as a function of overdensity, $\delta_{10}$, for haloes in the mass range $M_{\rm halo} = 10^{10}-10^{11}\,{\rm M}_{\odot}$ at $z=0$\; (solid black), $z=1$\; (dotted orange), and $z=2$\; (dashed red). The shaded regions indicate the 25$^{\rm th}$ to 75$^{\rm th}$ percentile range of $f_{\rm B}$ for each overdensity bin. In all models, $f_{\rm B}$ increases with redshift at fixed overdensity, but how $\delta_{10}$ impacts halo baryon fractions is redshift and model dependent.}
    \label{fig:z_fb}
\end{figure*}

\label{sec:Redshift}

In addition to identifying how global properties of galaxies and haloes are impacted by local environment, here we explore how these trends evolve over cosmic time. In this section we revisit previous results at $z=0,\; z=1,\; z=2$ to determine if local environment has a similar impact on galaxy properties at earlier times. As illustrative examples, we focus on galaxy SFRs and halo baryon fractions of SFR and $f_{\rm B}$ of similar mass galaxies and haloes as a function of overdensity at different redshifts.

\subsection{Galaxy SFRs at varying redshifts}
\label{subsec:Redshift SFR}
Figure~\ref{fig:z_sfr} shows the median SFR as a function of environment for all galaxies within the stellar mass range $M_\star=10^9$--$10^{10}$\,M$_\odot$ at $z=0$ (solid black line), $z=1$ (dotted orange line), and $z=2$ (dashed red line) for the SIMBA, IllustrisTNG, ASTRID, and Swift-EAGLE CV simulations (from left to right). We select this stellar mass range because low-mass galaxies tend to be the most sensitive to changes in local environment. In all four galaxy formation models, SFR increases with redshift, as expected given the increased normalization of the star formation main sequence and the cosmic SFR density at earlier times \citep{Madau&Dickinson2014, Salim2016}. The detailed redshift evolution of SFR is model dependent \citep{Ni2022}, with e.g. ASTRID and Swift-EAGLE showing stronger redshift evolution from $z = 1 \rightarrow 0$ compared to SIMBA and IllustrisTNG. Interestingly, the range of predicted galaxy overdensities at a given redshift is sensitive to galaxy formation model implementation, with e.g. SIMBA reaching ${\rm log}_{10}\delta_{10} > 4$ at $z=2$ compared to ${\rm log}_{10}\delta_{10} \lesssim 2$ in IllustrisTNG and ASTRID (likely related to their lower efficiency at forming satellite galaxies), and how the impact of environment depends on redshift is also model dependent.

In SIMBA, galaxies in the mass range $M_\star = 10^{9}$--$10^{10}$\,M$_{\odot}$ at $z=0$ show decreasing SFR with increasing local galaxy environmental density and rapidly quenching at overdensities ${\rm log}_{10}\delta_{10} \gtrsim 1.5$. At $z=1$, the SFR is roughly independent of environment at ${\rm log}_{10}\delta_{10} \lesssim 0.5$ but decreases rapidly at higher overdensity values, with most galaxies quenching at ${\rm log}_{10}\delta_{10} \sim 2$. Quenching is thus happening at higher overdensities at $z=1$ compared to $z=0$. Interestingly, galaxies at ${\rm log}_{10}\delta_{10} \gtrsim 2.5$ revert the trend and instead show increasing SFRs, reaching the highest SFRs at the highest overdensities. At $z=2$, the SFR--$\delta_{10}$ relation shows further differences relative to $z=0$, with SFRs mildly increasing with overdensity at ${\rm log}_{10}\delta_{10} \lesssim 0.5$, decreasing in the intermediate range $0.5 < {\rm log}_{10}\delta_{10} < 2$, with a smaller fraction of galaxies quenching, and rapidly increasing SFRs at the highest overdensities.

IllustrisTNG shows weaker trends in the SFR\,--$\delta_{10}$ relation and its redshift dependence compared to SIMBA, but with some similarities. At low environmental density compared to average (${\rm log}_{10}\delta_{10} \lesssim 0$), SFR is roughly independent of $\delta_{10}$ at $z=0$ but galaxies show mildly increasing SFR with overdensity at $z=1$--$2$. Similar to SIMBA, galaxies in the mass range $M_\star = 10^{9}$--$10^{10}$\,M$_{\odot}$ tend to quench at higher overdensities at higher redshifts, but IllustrisTNG does not produce rapidly star-forming galaxies at the highest overdensities.

ASTRID shows similar trends as IllustrisTNG, with the slope of the SFR--$\delta_{10}$ relation at ${\rm log}_{10}\delta_{10} \lesssim 0$ mildly increasing at higher redshifts, and galaxies quenching at higher overdensities when increasing redshift. Interestingly, galaxies in the mass range $M_\star = 10^{9}$--$10^{10}$\,M$_{\odot}$ at $z=0$--$1$ tend to quench at lower overdensities in ASTRID compared to SIMBA and TNG.

In this galaxy mass range, Swift-EAGLE predicts significant differences in the SFR--$\delta_{10}$ relation. In this case, SFR increases with overdensity at $z=1$--$2$. There are quenching thresholds of ${\rm log}_{10}\delta_{10} \gtrsim 1$ at $z=0$ and ${\rm log}_{10}\delta_{10} > 2$ at $z=1$. Swift-EAGLE is thus quenching low-mass galaxies at similar $\delta_{10}$ as ASTRID and lower overdensities than SIMBA and IllustrisTNG at $z=0$, but the quenching threshold is higher than the other three models at $z=1$.

\subsection{Halo Baryon fractions at varying redshifts}

Figure~\ref{fig:z_fb} shows the median baryon fraction for haloes in the mass range $M_{\rm halo}=10^{10}$--$10^{11}\,{\rm M}_\odot$ as a function of overdensity $\delta_{10}$ at $z=0$ (solid black), $z=1$ (dotted orange), and $z=2$ (dashed red) in the SIMBA, IllustrisTNG, ASTRID and Swift-EAGLE CV simulations. We follow the selection of haloes from Section~\ref{subsec:Redshift SFR}. In all four galaxy formation models, $f_{\rm B}$ increases with redshift at all overdensity values. However, the relationship between $f_{\rm B}$ and $\delta_{10}$ at a given redshift is dependent on galaxy formation model.

In SIMBA, the baryon fraction of haloes in the mass range $M_{\rm halo} = 10^{10}$--$10^{11}\,{\rm M}_{\odot}$ is strongly dependent on environment and redshift. At $z=0$, $f_{\rm B}$ decreases with increasing overdensity at ${\rm log}_{10}\delta_{10} \lesssim 0$ with a hint for the reverse trend at higher $\delta_{10}$. In contrast, at earlier times $f_{\rm B}$ steadily increases with overdensity reaching peak values at ${\rm log}_{10}\delta_{10} \sim -0.5$ at $z=1$ and ${\rm log}_{10}\delta_{10} \sim 0.5$ at $z=2$, subsequently decreasing at higher overdensities. 

In IllustrisTNG, there is significantly more overlap in $f_{\rm B}$ values at different redshifts given the scatter at fixed $\delta_{10}$ (as indicated by the shaded regions), particularly at low overdensities. This indicates that there is weaker redshift evolution of halo baryon fractions at this mass range in IllustrisTNG compared to SIMBA (and the other models). In IllustrisTNG, $f_{\rm B}$ increases with $\delta_{10}$ at all redshifts for low to intermediate overdensities, with a redshift-dependent transition point at ${\rm log}_{10}\delta_{10} \sim -0.5 \rightarrow 0.5$ for $z=0 \rightarrow 2$ at which baryon fractions begin to decrease at higher overdensities, and a subsequent upturn of increasing $f_{\rm B}$ at the highest overdensities, with these variations strongest at $z=0$ and weakest at $z=2$.

ASTRID shows qualitatively similar trends to IllustrisTNG but with stronger redshift evolution of median baryon fractions relative to the scatter at fixed overdensity. As in IllustrisTNG, the baryon fraction of low-mass haloes increases with $\delta_{10}$ until reaching a peak at a redhift-dependent overdensity, and decreasing at higher $\delta_{10}$. ASTRID also shows indication for a reversed upward trend at the highest overdensites but weaker than IllustrisTNG.

In Swift-EAGLE, $f_{\rm B}$ generally increases with $\delta_{10}$ at all redshifts with a minor decrease at ${\rm log}_{10}\delta_{10} \gtrsim 1$ at $z=0$ and a sharp decrease at ${\rm log}_{10}\delta_{10} \lesssim -2.5$ at $z=2$. In this case, there is significantly higher baryon fractions at higher redshifts for fixed $\delta_{10}$, except at the highest overdensities where $f_{\rm B}$ at $z=1$ can become as high as at $z=2$. The slope of the $f_{\rm B}$--$\delta_{10}$ relation increases with redshift at low to intermediate overdensities, with a transition to stronger dependence of $f_{\rm B}$ on $\delta_{10}$ (i.e. steeper slope) occurring at higher overdensity at early times.

\section{Definition of Environment}
\label{sec:Definition of Environment}

\begin{figure*}
    \centering
    \includegraphics[width=\textwidth]{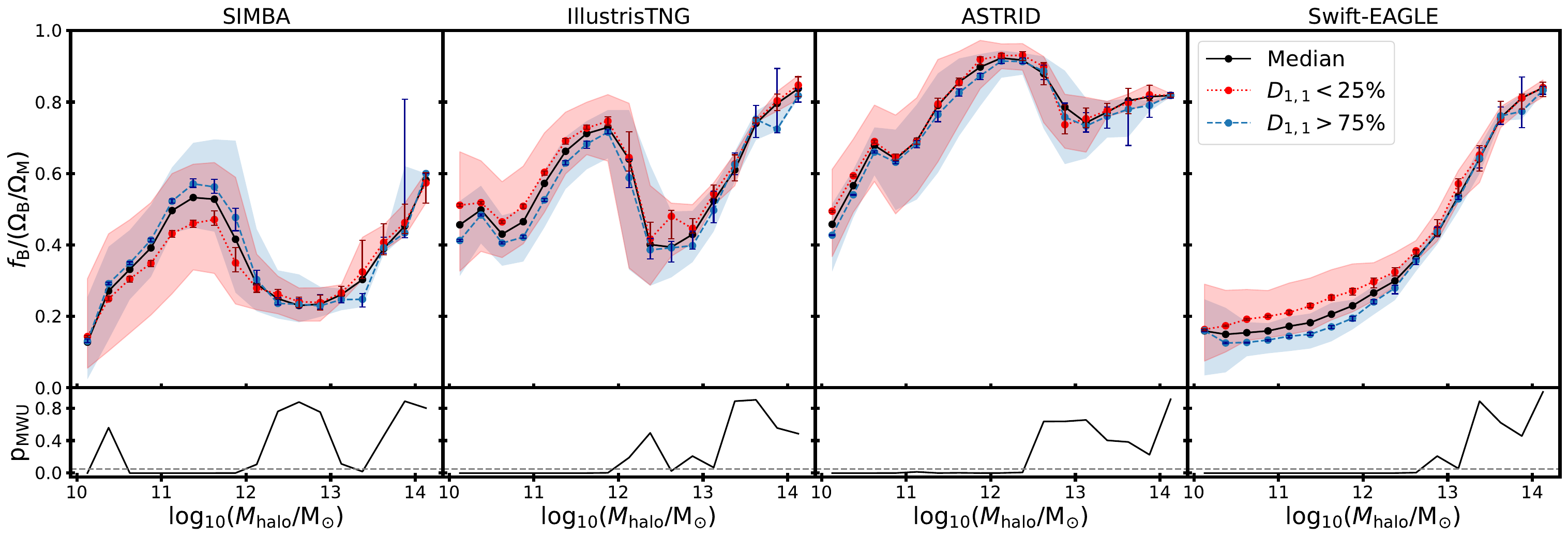}
    \vspace{-0.2in}
    \caption{Baryon fraction as a function of halo mass and environment at $z=0$, similar to Figure~\ref{fig:f_b} but using $D_{1,1}$ as the measure of the environment. The impact of environment on $f_{\rm B}$ as measured by $D_{1,1}$ is qualitatively consistent with the results based on $\delta_{10}$ in Figure~\ref{fig:f_b}, with e.g., overdense environments ($D_{1,1} < 25\%$; dashed red line) suppressing the baryon fraction for $M_{\rm halo} \lesssim 10^{12}\,{\rm M}_{\odot}$ in SIMBA while increasing $f_{\rm B}$ in Swift-EAGLE, but the amplitude of the effect decreases with $D_{1,1}$ as the measure of environment compared to $\delta_{10}$.}
    \label{fig:dnfbary}
\end{figure*}

\begin{figure*}
    \centering
    \includegraphics[width=\textwidth]{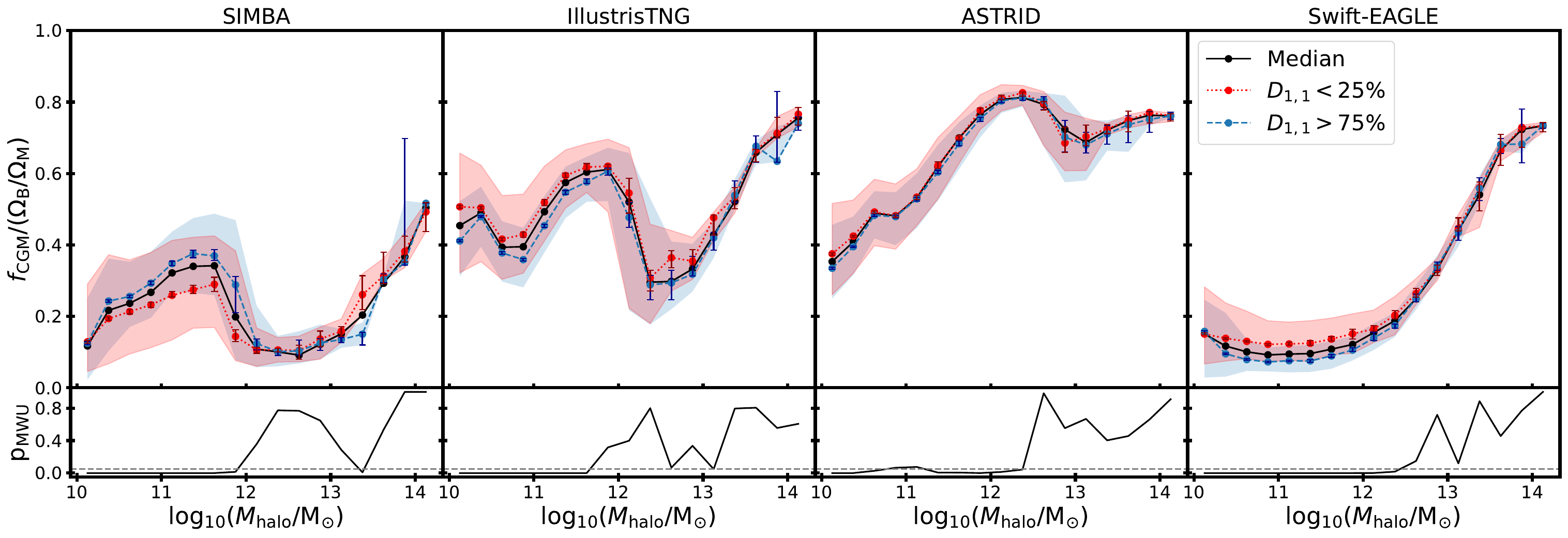}
    \vspace{-0.2in}
    \caption{CGM fraction as a function of $M_{\rm halo}$ and environment at $z=0$ for central galaxies in SIMBA, IllustrisTNG, ASTRID, and Swift-EAGLE (from left to right), similar to Figure~\ref{fig:f_cgm} but using $D_{1,1}$ as the measure of environment. The impact of environment on $f_{\rm CGM}$ as measured by $D_{1,1}$ is qualitatively consistent with the results based on $\delta_{10}$ in Figure~\ref{fig:f_cgm}. In SIMBA, CGM fraction is suppressed in overdense environments ($D_{1,1}<25\%$; dashed red line) for $M_{\rm halo} \lesssim 10^{12}\,{\rm M}_{\odot}$, but increased in Swift-EAGLE. The impact of environment on $f_{\rm CGM}$ as measured by $D_{1,1}$ is lesser than that of $\delta_{10}$.}
    \label{fig:dnfcgm}
\end{figure*}

In Sections~\ref{sec: Impact of Environment} and~\ref{sec:Redshift}, we have investigated the impact of environment on global galaxy/halo properties using $\delta_{10}$ as a measure of environment, corresponding to the overdensity of galaxies within a sphere containing the 10 nearest galaxy neighbours. In this section, we explore the dependence of results on the definition of galaxy environment employing $D_{1,1}$ as the measure of environment, which quantifies the proximity of haloes to others with similar or larger mass. The $D_{1,1}$ definition of environment is less sensitive to halo mass than $\delta_{10}$, as shown in Figure~\ref{fig:delta-DNF}.
Therefore, we can use $D_{1,1}$ to test the dependency of our previous results (Section~\ref{sec: Impact of Environment}) on halo mass.

Figure~\ref{fig:dnfbary} shows baryon fraction as a function of halo mass at $z=0$ in the SIMBA, IllustrisTNG, ASTRID and Swift-EAGLE simulations (from left to right). As in Figure~\ref{fig:f_b}, we show the median $f_{\rm B}$($M_{\rm halo}$) trend for all haloes (solid black line) compared to the results for haloes in overdense ($D_{1,1}$ < 25\%; dashed red line) and underdense ($D_{1,1}$ > 75\%; dotted blue line) regions, where we are now using $D_{1,1}$ rather than $\delta_{10}$ as the measure of environment. In this case, higher $D_{1,1}$ values ($D_{1,1}$ > 75\%) correspond to lower density environments, as $D_{1,1}$ is proportional to the distance between similar mass haloes. 

SIMBA predicts a qualitatively similar but weaker dependence of halo baryon fraction on $D_{1,1}$ compared to $\delta_{10}$. While there is substantial overlap between haloes in underdense and overdense regions, there is a trend for lower $f_{\rm B}$ in overdense regions compared at $10^{10.5}$\,M$_\odot \lesssim M_{\rm halo} \lesssim 10^{12}$\,M$_\odot$, most notable near the peak of $f_{\rm B}$ at $M_{\rm halo} \sim 10^{11.5}$\,M$_\odot$, and a hint for the opposite trend for massive haloes with $M_{\rm halo} \sim 10^{13.5}$\,M$_\odot$.

In IllustrisTNG, low-mass haloes tend to have higher median baryon fraction in overdense regions (low $D_{1,1}$) compared to underdense regions (high $D_{1,1}$), with this trend weakening at $M_{\rm halo} > 10^{12}$\,M$_\odot$. While there is significant overlap between the $25^{\rm th}$ to $75^{\rm th}$ percentiles of these two populations, the bootstrap error bars do not overlap. These trends are qualitatively similar to the results presented in Figure~\ref{fig:f_b} using $\delta_{10}$ to quantify the local environment of haloes. At $M_{\rm halo} < 10^{12}$\,M$_\odot$, low $D_{1,1}$ haloes tend to have higher median $f_{\rm B}$ than high $D_{1,1}$ haloes, but the 25$^{\rm th}$ to 75$^{\rm th}$ percentiles as well as bootstrap error bars overlap between the two populations. 

Unlike the other models, ASTRID does not show any clear dependence of $f_{\rm B}$ with $D_{1,1}$ across the full halo mass range. This is in contrast to the $f_{\rm B}$--$\delta_{10}$ results in Figure~\ref{fig:f_b}, where we show that massive haloes ($M_{\rm halo} > 10^{12.5}$\,M$_\odot$) tend to have higher baryon fraction in overdense regions compared to underdense regions according to the $\delta_{10}$ definition.

Finally, Swift-EAGLE shows a systematic trend of higher baryon fraction for haloes in overdense regions (lower $D_{1,1}$ values) at $M_{\rm halo} < 10^{12.5}$\,M$_\odot$, with increasing overlap between haloes with low and high $D_{1,1}$ values at higher masses. This trend is qualitatively consistent with the dependence of $f_{\rm B}$ on $\delta_{10}$ at $M_{\rm halo} < 10^{12.5}$\,M$_\odot$, which showed a more systematic increase in baryon fraction for overdense regions throughout the entire halo mass range.

Figure~\ref{fig:dnfcgm} shows CGM fraction as a function of halo mass at $z=0$ in the SIMBA, IllustrisTNG, ASTRID, and Swift-EAGLE simulations (from left to right). As in Figure~\ref{fig:dnfbary}, we compare the median relations for all haloes in each CV set (black) to those for haloes in low $D_{1,1}$ (red) and high $D_{1,1}$ (blue) regions separately. In SIMBA, low-mass haloes ($M_{\rm halo} \lesssim 10^{12}$\,M$_\odot$) tend to have higher $f_{\rm CGM}$ in underdense environments than in overdense environments as measured by $D_{1,1}$, although there is substantial overlap between the two regimes. There is a hint for the opposite trend for haloes with $M_{\rm halo} \sim 10^{13.5}$\,M$_\odot$. IllustrisTNG shows a qualitatively similar trend to Figure~\ref{fig:dnfbary}, where haloes have higher $f_{\rm CGM}$ in overdense regions than in underdense regions. This trend weakens for haloes with halo mass $M_{\rm halo} \gtrsim 10^{12}$\,M$_\odot$. Similarly to Figure~\ref{fig:dnfbary}, ASTRID does not predict a dependence of $f_{\rm CGM}$ with $D_{1,1}$ across the full halo mass range. Finally, Swift-EAGLE predicts a systematic trend of higher $f_{\rm CGM}$ in overdense regions for haloes with $M_{\rm halo} \lesssim 10^{12.5}$\,M$_\odot$, while the significance test fails for populations of overdense and underdense haloes at $M_{\rm halo} \gtrsim 10^{12.5}$\,M$_\odot$.

\section{Discussion}
\label{sec:Discussion}

\subsection{Comparison with observations} 
The results presented here, based on the analysis of >100 CAMELS simulations using the fiducial SIMBA, IllustrisTNG, ASTRID, and Swift-EAGLE galaxy formation models, generally follow expectations from observations relating the properties of galaxies and their environments. For example, it was shown in \citet{Kauffman2004} that galaxies with a lower number of local neighbours will contain a higher fraction of the total stellar mass in the local universe than galaxies with more local neighbours at $M_\star < 10^{11}$\,M$_\odot$. However, this trend tends to flip at $M_\star > 10^{11}$\,M$_\odot$. The variation in galaxy abundance as a function of environment is qualitatively consistent with our results for the $\delta_{10}$--$M_{\rm halo}$ correlation. We show in Figure~\ref{fig:delta-DNF} that low-mass haloes tend to have low $\delta_{10}$ and high-mass haloes have high $\delta_{10}$ in SIMBA. This same qualitative trend is also shown for IllustrisTNG, ASTRID, and Swift-EAGLE in Figure~\ref{fig:simbaplot}. Furthermore, observations indicate that more massive galaxies tend to form in denser environments and show lower gas fractions and specific SFRs, while lower mass
central galaxies typically form in lower density environments and have comparatively higher SFRs \citep{Dressler1980, Blanton2009, Etherington17, Wells22}, in qualitative agreement with our results in Figures~\ref{fig:delta-DNF},~\ref{fig:f_g}, and~\ref{fig:sfrboth}.

For a given stellar mass, observations have shown that satellite galaxies are very strongly impacted by environmental effects, with significantly lower gas fractions and SFRs in overdense regions compared to similar-mass satellite galaxies in underdense regions \citep{Peng2012, Brown2017, Davies_LJM2019}. Satellites are more severely impacted by environment because they are affected by certain astrophysical processes, such as tidal stripping and ram pressure stripping, that would not affect central galaxies \citep{Putman2021}. Satellites lose gas while moving against the CGM of the host halo strongly enough to overcome the satellite's potential well \citep{Gunn&Gott1972, Broderick25, Xu25}. This is generally consistent with our findings in Figures~\ref{fig:f_g} and~\ref{fig:sfrboth}, where satellite galaxies tend to be quenched in overdense regions and have higher gas fraction and SFR in underdense regions. In comparison, $f_{\rm gas}$ and SFR show significantly less variation with $\delta_{10}$ in central galaxies across all four galaxy formation models.

\subsection{Comparison with previous simulations}

Previous studies have used cosmological hydrodynamic simulations to investigate the impact of large scale environment on galaxy evolution, generally focusing on a single galaxy formation model.

Using the original EAGLE simulation, \citet{2017MNRAS.466.3460V} investigated how galaxy overdensity impacts gas accretion onto galaxies, explicitly showing that at fixed stellar mass, gas accretion in central galaxies is higher in high $\delta_{10}$ regions.
We do not directly investigate gas accretion in this work, but the results of \citet{2017MNRAS.466.3460V} are in general agreement with Figures~\ref{fig:f_b} and~\ref{fig:f_cgm}, where we show that $f_{\rm B}$ and $f_{\rm CGM}$ (which are affected by how much gas haloes can accrete) are higher in overdense regions at fixed halo mass in the CAMELS Swift-EAGLE simulations. However, the different dependence of $f_{\rm B}$ and $f_{\rm CGM}$ with $\delta_{10}$ seen for SIMBA, IllustrisTNG, and ASTRID  implies that the impact of environment on galaxy and halo gas accretion rates depends on the details of the galaxy formation model.
Other simulation-based studies have used distance to neighbouring cosmic web filaments as a measure of environment. Using the SIMBA simulations, \citet{bulichi24} found that galaxies that are close to cosmic web filaments or nodes have suppressed specific SFR and cold gas fraction at low redshift, qualitatively consistent with our results. Likewise, using the IllustrisTNG simulations, \citet{Hasan2023} showed that low $M_\star$ galaxies have suppressed specific SFR when they are located closer to cosmic web nodes at low redshift, with the trend driven by satellites.

The differences in environmental impact between central and satellite galaxies have also been explored through the use of cosmological hydrodynamic simulations. Using the TNG100 simulation, \citet{Stevens21} found that AGN feedback from more massive galaxies can displace gas from nearby neighbours and that satellites are particularly inefficient at re-accreting their displaced gas. Similarly, using the EAGLE simulations, \citet{Wright22} reported the suppression of gas accretion rates for satellite galaxies within higher-mass host haloes. Other studies using the original IllustrisTNG simulations \citep{Rohr2023} and the high-resolution AGORA simulations \citep{Rodriguez2025} have shown that ram pressure stripping is a significant driver in the removal of gas from satellite galaxies (see also \citealt{Tonnesen2009,Kulier23,Roy2024,Pathak2025,Lizhi25}). This negative impact of environment on satellites is consistent with the higher gas fraction and SFR that we find for  satellite galaxies in underdense regions across all four galaxy formation models (see Section~\ref{sec: Impact of Environment}). We show in Figures~\ref{fig:f_g} and~\ref{fig:sfrboth} that $f_{\rm gas}$ and SFR for satellites are far more sensitive to environment than that of central galaxies (where $f_{\rm gas}$ and SFR are generally weakly correlated with environment in all models). Satellite galaxies have systematically higher SFR in lower density environments across the full mass range except for the most massive galaxies ($M_{\star} \gtrsim 10^{11}$\,$M_\odot$) in IllustrisTNG, ASTRID, and Swift-EAGLE, where the trend reverses. A plausible explanation for the reversal of this trend is that massive satellites are affected by their own AGN feedback and may thus benefit from overdense environments where the larger effective gravitational potential makes it easier to retain gas and form stars, as seen in IllustrisTNG, ASTRID, and Swift-EAGLE for massive central galaxies. However, this conclusion may be affected by the lack of galaxy cluster environments in the CAMELS simulations analyzed here. 

\subsection{Model dependencies}

Low-mass central galaxies in high-density environments can be more vulnerable than higher-mass galaxies to the suppression of growth owing to competition for gas accretion with neighbouring galaxies \citep{Aragon-calvo2019}, ram pressure stripping in the surroundings of galaxy groups and clusters \citep{Boselli2022}, and AGN feedback from nearby galaxies \citep{Martin2019, Martin2021}. Recent high-resolution cosmological hydrodynamic simulations have shown that low-mass central galaxies in underdense environments have higher gas fractions and SFRs compared to similar-mass galaxies in denser environments \citep{Christensen24}, suggesting that differences in halo growth histories correlated with environment can also play a role in the emerging galaxy properties.

Here we have used CAMELS to test these effects for multiple galaxy formation models based on several halo and galaxy properties. We show in Figure~\ref{fig:f_b} that low-mass haloes in SIMBA have systematically higher baryon fraction in underdense regions compared to similar mass halos in overdense regions. 
However, other simulation models show either no impact of environment on low-mass haloes (IllustrisTNG and ASTRID) or the opposite effect (Swift-EAGLE). Given their similarity in gravity solver and identical initial conditions (and thus large-scale structure and overall halo growth histories) for the CV simulations analyzed here, a plausible explanation for these conflicting results is their different implementations of feedback. 

SIMBA is characterized by strong kinetic AGN feedback with long-range preventative effects \citep{Angles-Alcazar2017a, SIMBA.PAPER, Christiansen2020, Tillman23b, Tillman2023, Sorini24}, where powerful jets can push gas up to few Mpc away from massive haloes \citep{Borrow20, Gebhardt2024}. It is thus plausible that AGN feedback exerted by massive galaxies, preferentially located in overdense regions, can also directly impact nearby lower-mass haloes, contributing to the suppressed baryon fraction (and $f_{\rm CGM}$) of low-mass halos in overdense regions that we find in SIMBA. In contrast, AGN feedback in IllustrisTNG, ASTRID, and Swift-EAGLE is weaker than that of SIMBA \citep{Ni2023, Gebhardt2024, Wright2024, Medlock2025}, which would make low-mass haloes less affected by external AGN feedback in high density regions.  IllustrisTNG and ASTRID indeed show little difference in the baryon fraction of low-mass haloes in overdense compared to underdense regions, but Swift-EAGLE yields the opposite trend with higher baryon fraction for low-mass haloes in overdense regions.   
In this case, the contradictory Swift-EAGLE results could be attributed to the implementation of stellar feedback, which drives outflows to several virial radii stronger than other models \citep{Wright2024}. The gravitational potential of low-mass haloes is not deep enough to overcome stellar feedback, resulting in halo baryon fractions significanty suppressed in Swift-EAGLE compared to other models. However, the stronger effective gravitational potential of low-mass haloes in overdense regions may enable them to retain more of their baryons, potentially explaining the higher baryon fraction of low-mass haloes in overdense environments. In contrast, the weaker stellar feedback in SIMBA, IllustrisTNG, and ASTRID does not appear to affect the impact of environment on low-mass haloes.

Interestingly, while higher-mass haloes are not expected to be impacted by external feedback, we have shown that their environment can still significantly impact the evolution of massive haloes. Figures~\ref{fig:f_b} and~\ref{fig:f_cgm} show that the baryon and CGM mass fractions of massive haloes ($M_{\rm halo} \gtrsim 10^{12.5}\,{\rm M}_\odot$) increase in overdense regions across all four galaxy formation models, with the trend being strongest in IllustrisTNG and ASTRID and weakest in Swift-EAGLE and SIMBA. This positive environmental dependence appears even as the deepening gravitational potentials of massive haloes make internal feedback less efficient at ejecting gas, causing their halo and baryon fractions to rise rapidly with increasing mass in all models. As in the case of massive satellite galaxies (with higher SFR in overdense regions), a plausible explanation for the higher baryon fraction of massive haloes in higher-density environments is the larger effective gravitational potential in overdense regions enabling the more efficient retention of baryons. 

\subsection{Redshift evolution}

Using the original SIMBA simulation, \citet{Ghodsi2024} showed that the star formation efficiency (SFE) decreases with increasing galaxy overdensity at $z=0$. This is consistent with our results in Figure~\ref{fig:sfrboth}, where we show that overdense galaxies in CAMELS-SIMBA (both centrals and satellites) have lower SFR than their underdense counterparts. In addition, \citet{Ghodsi2024} showed that SFR increases with redshift for galaxies in similar environments for the original SIMBA simulation, which is consistent with the trend shown in Figure~\ref{fig:z_sfr} across all four simulation models. This result is broadly consistent with the redshift evolution of the observed star-forming main sequence \citep{Popesso23}, in which sSFRs increase toward earlier epochs at fixed stellar mass, a trend also reproduced in previous simulations (e.g., \citealt{Cochrane2018} for the EAGLE simulation and \citealt{SIMBA.PAPER} for SIMBA).

Interestingly, we find that, at fixed stellar mass, SFR in CAMELS-SIMBA decreases with increasing $\delta_{10}$ at low redshift but increases with increasing $\delta_{10}$ at higher redshift, indicating that the impact of environment depends on cosmic epoch. This trend is also seen for IllustrisTNG and ASTRID but is absent in EAGLE, where SFR increases with $\delta_{10}$ at all redshifts (for star-forming galaxies).  As expected, our CAMELS-SIMBA results are in qualitative agreement with \citet{Ghodsi2024} for the original SIMBA simulation, although the decrease in SFR with $\delta_{10}$ at $z=0$ is more drastic in our analysis of CAMELS-SIMBA simulations. In addition to our different measures of environment, these differences can be attributed to the range of galaxy stellar masses considered. In their analysis, \citet{Ghodsi2024} considered galaxies with stellar mass $M_{\star} > 10^{9}$\,M$_\odot$, while we focus on a narrower stellar mass range $M_{\star} = 10^9$--$10^{10}$\,M$_\odot$ to investigate the redshift evolution of environmental effects at fixed stellar mass.  
Effects that could increase the SFR in overdense regions at higher redshift include higher merger rates \citep{Rodriguez-Gomez2015} and higher specific gas inflow rates onto haloes \citep{dekel2009, vandeVoort11, Correa15}. Galaxy mergers are known to boost SFR \citep{Ellison08, Moreno2019} and faster inflows of gas in overdense regions will lead to higher SFR. 
Given the comparatively weaker impact of feedback at earlier times for a given halo mass, these results suggest that overdense environments at high redshift provide an overabundance of cool gas that can enhance (rather than suppress) SFRs. 

These results are also supported by the redshift evolution of the $f_{\rm B}$--$\delta_{10}$ relation (Figure~\ref{fig:z_fb}), where the baryon fraction of low-mass haloes increases with overdensity at $z=2$ but decreases with overdensity at $z=0$ in all galaxy formation models, with the exception of SIMBA (and ASTRID to a lesser extent) at the highest overdensities.  Indeed, we find that galaxies in the mass range $M_{\star} = 10^9$--$10^{10}$\,M$_\odot$ tend to quench at sufficiently high overdensity value, with the $\delta_{10}$ threshold for quenching increasing with redshift across all four galaxy formation models. Overall, these results suggest a transition from overdense environments enhancing the gas supply and star formation rate of galaxies at early times to suppressing star formation at late times. Despite the widely different scales, this is qualitatively consistent with the observed enhanced star formation rate and AGN activity of galaxies in early protocluster environments ($z>2$) compared to the field, while galaxy clusters are predominantly populated by quenched/inactive galaxies in the low-redshift universe \citep{Kauffman2004, Overzier2016, Vito2024, Traina2025}.

\subsection{Halo mass--environment correlation}

As has been pointed out in previous work, there is a clear correlation between halo mass and many common definitions of environment \citep{Haas2012}. In Figure~\ref{fig:delta-DNF}, we show that the correlation between $\delta_{10}$ and halo mass is particularly clear for high-mass haloes, where $\delta_{10}$ steeply increases with halo mass at $M_{\rm halo} \gtrsim 10^{12.5}$\,M$_\odot$. In order to understand the intrinsic impact of environment in galaxy evolution, it is imperative to disentangle this correlation and separate truly environmental effects from the overall halo mass dependence of galaxy properties. By directly comparing galaxies and haloes of similar mass, we have identified the intrinsic impact of $\delta_{10}$ on galaxy and halo properties. In addition, we have also used an alternative definition of environment, $D_{1,1}$, which is far less sensitive to $M_{\rm halo}$ than $\delta_{10}$ \citep{Haas2012}. We find that the same qualitative trends with environment are present for the halo-based properties analyzed here ($f_{\rm B}$ and $f_{\rm CGM}$) using both $\delta_{10}$ and $D_{1,1}$, indicating that our results are robust with respect to the definitions of environment (see Figures~\ref{fig:f_b},~\ref{fig:f_cgm},~\ref{fig:dnfbary}, and~\ref{fig:dnfcgm}).
However, $D_{1,1}$ can only be defined for central galaxies/haloes, which are less sensitive to environment than satellites, and our results emphasize the importance of disentangling intrinsic
environmental effects from otherwise halo-mass effects.

\subsection{Galaxy--halo connection}
We have shown that both halo and galaxy properties can be heavily impacted by local environment. It follows that environment may play a role in the connection between galaxies and their host haloes. Indeed, some studies have attempted to account for the impact of environment when investigating the galaxy--halo connection and its potential implications for galaxy assembly bias \citep[e.g.,][]{Artale2018,Dragomir2018, Jo2019}. Using the original EAGLE and Illustris simulations, \citet{Artale2018} found that low-mass haloes in high-density environments are more likely to host central galaxies (and they are more massive) than their underdense halo counterparts. Future work could investigate the galaxy-halo connection as a function of cosmological and astrophysical parameters in CAMELS while incorporating measures of local environment such as those explored in this work. 

\section{Conclusions}
\label{sec:Conclusions}

We have performed a systematic comparison of the dependence of various halo and galaxy properties on local environment across four different galaxy formation models in the CAMELS simulations. Our main results are summarized as follows:

\begin{itemize}
    \item Halo baryon fraction $f_{\rm B}$ and CGM mass fraction $f_{\rm CGM}$ at $z=0$ are very sensitive to environmental effects. Low-mass haloes ($M_{\rm halo} \lesssim 10^{12}$\,M$_\odot$) in SIMBA exhibit higher $f_{\rm B}$ and $f_{\rm CGM}$ in underdense environments, whereas IllustrisTNG and ASTRID show higher $f_{\rm B}$ and $f_{\rm CGM}$ for massive haloes in overdense regions. Swift-EAGLE yields systematically higher $f_{\rm B}$ and $f_{\rm CGM}$ in overdense regions in low-mass haloes.

    \item The gas fraction $f_{\rm gas}$ of central galaxies at $z=0$ is less sensitive to local environment than $f_{\rm B}$ and $f_{\rm CGM}$, but still displays clear systematic trends with overdensity. Centrals in overdense regions tend to have lower $f_{\rm gas}$ at low masses (in all models) and higher $f_{\rm gas}$ at high masses (in IllustrisTNG, ASTRID, and Swift-EAGLE) compared to their underdense counterparts.

    \item Central galaxy SFRs are generally suppressed in overdense environments in SIMBA (and for low-mass centrals in other models), while IllustrisTNG and ASTRID show enhanced SFRs in overdense regions for the most massive central galaxies.

    \item Satellite galaxies are strongly impacted by environmental effects, with the gas fraction and SFR of satellites in overdense regions significantly reduced compared to underdense environments in all galaxy formation models.

    \item While satellite galaxies in overdense environments are predominantly quenched, the most massive satellites in TNG, ASTRID, and Swift-EAGLE show higher $f_{\rm gas}$ and SFR in overdense regions compared to lower-density environments.

    \item The impact of environmental effects on global properties of haloes and galaxies depends non-trivially on redshift for all galaxy formation models. 

    \item The baryon fraction of low-mass haloes in SIMBA increases with overdensity at $z=2$ but decreases with overdensity at $z=0$, showing a turnover at $\delta_{10} \gtrsim 0.5$ where $f_{\rm B}$ continues to show opposite trends at both redshifts. In other models, $f_{\rm B}$ also often responds differently to the environment at different redshifts.

    \item Low-mass galaxies quench rapidly after reaching a critical overdensity that depends on galaxy formation model (trend dominated by satellites), where this overdensity threshold for quenching generally increases at higher redshifts.

    \item Our main results are robust relative to the definition of environment and emphasize the importance of disentangling intrinsic environmental effects from otherwise primarily halo-mass effects.
\end{itemize}

Overall, our results demonstrate that the impact of environmental effects on galaxy evolution depends significantly on the details of the galaxy formation sub-grid model implementation. In follow-up work, we leverage the thousands of simulations with varying cosmological and astrophysical parameters available in CAMELS to explore their indirect impact on galaxy evolution by modifying local environments. Future work should also revisit our results using larger-volume simulations. Due to the small (25\,$h^{-1}$Mpc)$^3$ volumes of the current CAMELS simulations, our analysis is limited in the range of environments sampled, particularly in the absence of rich clusters and large voids. Second-generation CAMELS simulations with (50\,$h^{-1}$Mpc)$^3$ volumes will provide a unique opportunity to extend this work, allowing more comprehensive tests of how environment shapes galaxy evolution across cosmological parameters and feedback models.
  
\section*{Acknowledgements}
\label{sec: Acknowledgements}

We thank Matt Gebhardt, Joop Schaye, Matthieu Schaller, Romeel Dav\'e, and Shy Genel for comments that helped improve the paper.
We thank the Flatiron Institute for hosting the CAMELS Public Data Repository, which is supported by the Simons Foundation. DAA acknowledges support from NSF grant AST-2108944 and CAREER award AST-2442788, NASA grant ATP23-0156, STScI JWST grants AR-04357.001-A, and AR-05366.005-A, an Alfred P. Sloan Research Fellowship, and Cottrell Scholar Award CS-CSA-2023-028 by the Research Corporation for Science Advancement. D.N. and I.M. acknowledge support by the NSF grant AST-2511137.

%%%%%%%%%%%%%%%%%%%%%%%%%%%%%%%%%%%%%%%%%%%%%%%%%%
\section*{Data Availability}

The CAMELS data is publicly available and can be accessed through different methods \citep{Ni2023,Villaescusa-Navarro2023}. Information on how to access and use the data can be found at https://camels.readthedocs.io.

%%%%%%%%%%%%%%%%%%%% REFERENCES %%%%%%%%%%%%%%%%%%

\bibliographystyle{mnras}
\bibliography{refs} 

@ARTICLE{Christensen24,
       author = {{Christensen}, Charlotte R. and {Brooks}, Alyson M. and {Munshi}, Ferah and {Riggs}, Claire and {Van Nest}, Jordan and {Akins}, Hollis and {Quinn}, Thomas R. and {Chamberland}, Lucas},
        title = "{Environment Matters: Predicted Differences in the Stellar Mass{\textendash}Halo Mass Relation and History of Star Formation for Dwarf Galaxies}",
      journal = {\apj},
         year = 2024,
        month = feb,
       volume = {961},
       number = {2},
          eid = {236},
        pages = {236},
          doi = {10.3847/1538-4357/ad0c5a},
       adsurl = {https://ui.adsabs.harvard.edu/abs/2024ApJ...961..236C}
}

@ARTICLE{Madau&Dickinson2014,
       author = {{Madau}, Piero and {Dickinson}, Mark},
        title = "{Cosmic Star-Formation History}",
      journal = {\araa},
         year = 2014,
        month = aug,
       volume = {52},
        pages = {415-486},
          doi = {10.1146/annurev-astro-081811-125615},
archivePrefix = {arXiv},
       eprint = {1403.0007},
 primaryClass = {astro-ph.CO},
       adsurl = {https://ui.adsabs.harvard.edu/abs/2014ARA&A..52..415M}
}

@ARTICLE{Putman2021,
       author = {{Putman}, Mary E. and {Zheng}, Yong and {Price-Whelan}, Adrian M. and {Grcevich}, Jana and {Johnson}, Amalya C. and {Tollerud}, Erik and {Peek}, Joshua E.~G.},
        title = "{The Gas Content and Stripping of Local Group Dwarf Galaxies}",
      journal = {\apj},
         year = 2021,
        month = may,
       volume = {913},
       number = {1},
          eid = {53},
        pages = {53},
          doi = {10.3847/1538-4357/abe391},
archivePrefix = {arXiv},
       eprint = {2101.07809},
 primaryClass = {astro-ph.GA},
       adsurl = {https://ui.adsabs.harvard.edu/abs/2021ApJ...913...53P}
}

@ARTICLE{Hassan2022,
       author = {{Hassan}, Sultan and {Villaescusa-Navarro}, Francisco and {Wandelt}, Benjamin and {Spergel}, David N. and {Angl{\'e}s-Alc{\'a}zar}, Daniel and {Genel}, Shy and {Cranmer}, Miles and {Bryan}, Greg L. and {Dav{\'e}}, Romeel and {Somerville}, Rachel S. and {Eickenberg}, Michael and {Narayanan}, Desika and {Ho}, Shirley and {Andrianomena}, Sambatra},
        title = "{HIFLOW: Generating Diverse HI Maps and Inferring Cosmology while Marginalizing over Astrophysics Using Normalizing Flows}",
      journal = {\apj},
         year = 2022,
        month = oct,
       volume = {937},
       number = {2},
          eid = {83},
        pages = {83},
          doi = {10.3847/1538-4357/ac8b09},
archivePrefix = {arXiv},
       eprint = {2110.02983},
 primaryClass = {astro-ph.CO},
       adsurl = {https://ui.adsabs.harvard.edu/abs/2022ApJ...937...83H}
}

@ARTICLE{Jo2019,
       author = {{Jo}, Yongseok and {Kim}, Ji-hoon},
        title = "{Machine-assisted semi-simulation model (MSSM): estimating galactic baryonic properties from their dark matter using a machine trained on hydrodynamic simulations}",
      journal = {\mnras},
         year = 2019,
        month = nov,
       volume = {489},
       number = {3},
        pages = {3565-3581},
          doi = {10.1093/mnras/stz2304},
archivePrefix = {arXiv},
       eprint = {1908.09844},
 primaryClass = {astro-ph.GA},
       adsurl = {https://ui.adsabs.harvard.edu/abs/2019MNRAS.489.3565J}
}

@ARTICLE{Medlock2025b,
       author = {{Medlock}, Isabel and {Nagai}, Daisuke and {Angl{\'e}s-Alc{\'a}zar}, Daniel and {Gebhardt}, Matthew},
        title = "{Constraining Baryonic Feedback Effects on the Matter Power Spectrum with Fast Radio Bursts}",
      journal = {\apj},
         year = 2025,
        month = apr,
       volume = {983},
       number = {1},
          eid = {46},
        pages = {46},
          doi = {10.3847/1538-4357/adbc9c},
archivePrefix = {arXiv},
       eprint = {2501.17922},
 primaryClass = {astro-ph.CO},
       adsurl = {https://ui.adsabs.harvard.edu/abs/2025ApJ...983...46M}
}

@article{Medlock2025,
   title={Quantifying Baryonic Feedback on the Warm–Hot Circumgalactic Medium in CAMELS Simulations},
   volume={980},
   ISSN={1538-4357},
   url={http://dx.doi.org/10.3847/1538-4357/ada442},
   DOI={10.3847/1538-4357/ada442},
   number={1},
   journal={The Astrophysical Journal},
   publisher={American Astronomical Society},
   author={Medlock, Isabel and Neufeld, Chloe and Nagai, Daisuke and Anglés-Alcázar, Daniel and Genel, Shy and Oppenheimer, Benjamin D. and Sims, Xavier and Singh, Priyanka and Villaescusa-Navarro, Francisco},
   year={2025},
   month=feb, pages={61} }

@ARTICLE{Crain&voort2023,
       author = {{Crain}, Robert A. and {van de Voort}, Freeke},
        title = "{Hydrodynamical Simulations of the Galaxy Population: Enduring Successes and Outstanding Challenges}",
      journal = {\araa},
         year = 2023,
        month = aug,
       volume = {61},
        pages = {473-515},
          doi = {10.1146/annurev-astro-041923-043618},
archivePrefix = {arXiv},
       eprint = {2309.17075},
 primaryClass = {astro-ph.GA},
       adsurl = {https://ui.adsabs.harvard.edu/abs/2023ARA&A..61..473C}
}

@ARTICLE{Boselli2022,
       author = {{Boselli}, Alessandro and {Fossati}, Matteo and {Sun}, Ming},
        title = "{Ram pressure stripping in high-density environments}",
      journal = {\aapr},
         year = 2022,
        month = dec,
       volume = {30},
       number = {1},
          eid = {3},
        pages = {3},
          doi = {10.1007/s00159-022-00140-3},
archivePrefix = {arXiv},
       eprint = {2109.13614},
 primaryClass = {astro-ph.GA},
       adsurl = {https://ui.adsabs.harvard.edu/abs/2022A&ARv..30....3B}
}

@ARTICLE{Rodriguez2025,
       author = {{Rodr{\'\i}guez-Cardoso}, Ram{\'o}n and {Roca-F{\`a}brega}, Santi and {Jung}, Minyong and {Nguyễn}, Thịnh H. and {Kim}, Ji-Hoon and {Primack}, Joel and {Agertz}, Oscar and {Barrow}, Kirk S.~S. and {Gallego}, Jesus and {Nagamine}, Kentaro and {Powell}, Johnny W. and {Revaz}, Yves and {Vel{\'a}zquez}, Hector and {Genina}, Anna and {Kim}, Hyeonyong and {Lupi}, Alessandro and {Abel}, Tom and {Cen}, Renyue and {Ceverino}, Daniel and {Dekel}, Avishai and {Oh}, Boon Kiat and {Quinn}, Thomas R. and {The Agora Collaboration}},
        title = "{The AGORA High-Resolution Galaxy Simulations Comparison Project: VII. Satellite quenching in zoom-in simulation of a Milky Way-mass halo}",
      journal = {\aap},
         year = 2025,
        month = jun,
       volume = {698},
          eid = {A303},
        pages = {A303},
          doi = {10.1051/0004-6361/202453639},
archivePrefix = {arXiv},
       eprint = {2505.05844},
 primaryClass = {astro-ph.GA},
       adsurl = {https://ui.adsabs.harvard.edu/abs/2025A&A...698A.303R}
}

@ARTICLE{Rodriguez-Gomez2015,
       author = {{Rodriguez-Gomez}, Vicente and {Genel}, Shy and {Vogelsberger}, Mark and {Sijacki}, Debora and {Pillepich}, Annalisa and {Sales}, Laura V. and {Torrey}, Paul and {Snyder}, Greg and {Nelson}, Dylan and {Springel}, Volker and {Ma}, Chung-Pei and {Hernquist}, Lars},
        title = "{The merger rate of galaxies in the Illustris simulation: a comparison with observations and semi-empirical models}",
      journal = {\mnras},
         year = 2015,
        month = may,
       volume = {449},
       number = {1},
        pages = {49-64},
          doi = {10.1093/mnras/stv264},
archivePrefix = {arXiv},
       eprint = {1502.01339},
 primaryClass = {astro-ph.GA},
       adsurl = {https://ui.adsabs.harvard.edu/abs/2015MNRAS.449...49R}
}

@ARTICLE{Oppenheimer2020,
       author = {{Oppenheimer}, Benjamin D. and {Davies}, Jonathan J. and {Crain}, Robert A. and {Wijers}, Nastasha A. and {Schaye}, Joop and {Werk}, Jessica K. and {Burchett}, Joseph N. and {Trayford}, James W. and {Horton}, Ryan},
        title = "{Feedback from supermassive black holes transforms centrals into passive galaxies by ejecting circumgalactic gas}",
      journal = {\mnras},
         year = 2020,
        month = jan,
       volume = {491},
       number = {2},
        pages = {2939-2952},
          doi = {10.1093/mnras/stz3124},
archivePrefix = {arXiv},
       eprint = {1904.05904},
 primaryClass = {astro-ph.GA},
       adsurl = {https://ui.adsabs.harvard.edu/abs/2020MNRAS.491.2939O}
}

@ARTICLE{Rohr2023,
       author = {{Rohr}, Eric and {Pillepich}, Annalisa and {Nelson}, Dylan and {Zinger}, Elad and {Joshi}, Gandhali D. and {Ayromlou}, Mohammadreza},
        title = "{Jellyfish galaxies with the IllustrisTNG simulations - when, where, and for how long does ram pressure stripping of cold gas occur?}",
      journal = {\mnras},
         year = 2023,
        month = sep,
       volume = {524},
       number = {3},
        pages = {3502-3525},
          doi = {10.1093/mnras/stad2101},
archivePrefix = {arXiv},
       eprint = {2304.09196},
 primaryClass = {astro-ph.GA},
       adsurl = {https://ui.adsabs.harvard.edu/abs/2023MNRAS.524.3502R}
}

@ARTICLE{Dragomir2018,
       author = {{Dragomir}, Radu and {Rodr{\'\i}guez-Puebla}, Aldo and {Primack}, Joel R. and {Lee}, Christoph T.},
        title = "{Does the galaxy-halo connection vary with environment?}",
      journal = {\mnras},
         year = 2018,
        month = may,
       volume = {476},
       number = {1},
        pages = {741-758},
          doi = {10.1093/mnras/sty283},
archivePrefix = {arXiv},
       eprint = {1710.09392},
 primaryClass = {astro-ph.GA},
       adsurl = {https://ui.adsabs.harvard.edu/abs/2018MNRAS.476..741D}
}

@ARTICLE{Roy2024,
       author = {{Roy}, Manami and {Su}, Kung-Yi and {Tonnesen}, Stephanie and {Fielding}, Drummond B. and {Faucher-Gigu{\`e}re}, Claude-Andr{\'e}},
        title = "{Seeding the CGM: how satellites populate the cold phase of milky way haloes}",
      journal = {\mnras},
         year = 2024,
        month = jan,
       volume = {527},
       number = {1},
        pages = {265-280},
          doi = {10.1093/mnras/stad3142},
archivePrefix = {arXiv},
       eprint = {2310.04404},
 primaryClass = {astro-ph.GA},
       adsurl = {https://ui.adsabs.harvard.edu/abs/2024MNRAS.527..265R}
}

@ARTICLE{Steinwandel2024,
       author = {{Steinwandel}, Ulrich P. and {Kim}, Chang-Goo and {Bryan}, Greg L. and {Ostriker}, Eve C. and {Somerville}, Rachel S. and {Fielding}, Drummond B.},
        title = "{The Structure and Composition of Multiphase Galactic Winds in a Large Magellanic Cloud Mass Simulated Galaxy}",
      journal = {\apj},
         year = 2024,
        month = jan,
       volume = {960},
       number = {2},
          eid = {100},
        pages = {100},
          doi = {10.3847/1538-4357/ad09e1},
archivePrefix = {arXiv},
       eprint = {2212.03898},
 primaryClass = {astro-ph.GA},
       adsurl = {https://ui.adsabs.harvard.edu/abs/2024ApJ...960..100S}
}

@ARTICLE{Wright2024,
       author = {{Wright}, Ruby J. and {Somerville}, Rachel S. and {Lagos}, Claudia del P. and {Schaller}, Matthieu and {Dav{\'e}}, Romeel and {Angl{\'e}s-Alc{\'a}zar}, Daniel and {Genel}, Shy},
        title = "{The baryon cycle in modern cosmological hydrodynamical simulations}",
      journal = {\mnras},
         year = 2024,
        month = aug,
       volume = {532},
       number = {3},
        pages = {3417-3440},
          doi = {10.1093/mnras/stae1688},
archivePrefix = {arXiv},
       eprint = {2402.08408},
 primaryClass = {astro-ph.GA},
       adsurl = {https://ui.adsabs.harvard.edu/abs/2024MNRAS.532.3417W}
}

@ARTICLE{Ni2023,
       author = {{Ni}, Yueying and {Genel}, Shy and {Angl{\'e}s-Alc{\'a}zar}, Daniel and {Villaescusa-Navarro}, Francisco and {Jo}, Yongseok and {Bird}, Simeon and {Di Matteo}, Tiziana and {Croft}, Rupert and {Chen}, Nianyi and {de Santi}, Natal{\'\i} S.~M. and {Gebhardt}, Matthew and {Shao}, Helen and {Pandey}, Shivam and {Hernquist}, Lars and {Dave}, Romeel},
        title = "{The CAMELS Project: Expanding the Galaxy Formation Model Space with New ASTRID and 28-parameter TNG and SIMBA Suites}",
      journal = {\apj},
         year = 2023,
        month = dec,
       volume = {959},
       number = {2},
          eid = {136},
        pages = {136},
          doi = {10.3847/1538-4357/ad022a},
archivePrefix = {arXiv},
       eprint = {2304.02096},
 primaryClass = {astro-ph.CO},
       adsurl = {https://ui.adsabs.harvard.edu/abs/2023ApJ...959..136N}
}

@inbook{Hlavacek_Larrondo_2022,
   title={AGN Feedback in Groups and Clusters of Galaxies},
   ISBN={9789811645440},
   url={http://dx.doi.org/10.1007/978-981-16-4544-0_122-1},
   DOI={10.1007/978-981-16-4544-0_122-1},
   booktitle={Handbook of X-ray and Gamma-ray Astrophysics},
   publisher={Springer Nature Singapore},
   author={Hlavacek-Larrondo, Julie and Li, Yuan and Churazov, Eugene},
   year={2022},
   pages={1–66} }

@ARTICLE{Martin2021,
       author = {{Mart{\'\i}n-Navarro}, Ignacio and {Pillepich}, Annalisa and {Nelson}, Dylan and {Rodriguez-Gomez}, Vicente and {Donnari}, Martina and {Hernquist}, Lars and {Springel}, Volker},
        title = "{Anisotropic satellite galaxy quenching modulated by black hole activity}",
      journal = {\nat},
         year = 2021,
        month = jun,
       volume = {594},
       number = {7862},
        pages = {187-190},
          doi = {10.1038/s41586-021-03545-9},
archivePrefix = {arXiv},
       eprint = {2106.04587},
 primaryClass = {astro-ph.GA},
       adsurl = {https://ui.adsabs.harvard.edu/abs/2021Natur.594..187M}
}

@ARTICLE{Gilli2019,
       author = {{Gilli}, R. and {Mignoli}, M. and {Peca}, A. and {Nanni}, R. and {Prandoni}, I. and {Liuzzo}, E. and {D'Amato}, Q. and {Brusa}, M. and {Calura}, F. and {Caminha}, G.~B. and {Chiaberge}, M. and {Comastri}, A. and {Cucciati}, O. and {Cusano}, F. and {Grandi}, P. and {Decarli}, R. and {Lanzuisi}, G. and {Mannucci}, F. and {Pinna}, E. and {Tozzi}, P. and {Vanzella}, E. and {Vignali}, C. and {Vito}, F. and {Balmaverde}, B. and {Citro}, A. and {Cappelluti}, N. and {Zamorani}, G. and {Norman}, C.},
        title = "{Discovery of a galaxy overdensity around a powerful, heavily obscured FRII radio galaxy at z = 1.7: star formation promoted by large-scale AGN feedback?}",
      journal = {\aap},
         year = 2019,
        month = dec,
       volume = {632},
          eid = {A26},
        pages = {A26},
          doi = {10.1051/0004-6361/201936121},
archivePrefix = {arXiv},
       eprint = {1909.00814},
 primaryClass = {astro-ph.GA},
       adsurl = {https://ui.adsabs.harvard.edu/abs/2019A&A...632A..26G}
}

@ARTICLE{Martin2019,
       author = {{Mart{\'\i}n-Navarro}, I. and {Burchett}, Joseph N. and {Mezcua}, Mar},
        title = "{Quantifying the Effect of Black Hole Feedback from the Central Galaxy on the Satellite Populations of Groups and Clusters}",
      journal = {\apjl},
         year = 2019,
        month = oct,
       volume = {884},
       number = {2},
          eid = {L45},
        pages = {L45},
          doi = {10.3847/2041-8213/ab4885},
archivePrefix = {arXiv},
       eprint = {1909.12841},
 primaryClass = {astro-ph.GA},
       adsurl = {https://ui.adsabs.harvard.edu/abs/2019ApJ...884L..45M}
}

@ARTICLE{Sorini24,
       author = {{Sorini}, Daniele and {Bose}, Sownak and {Dav{\'e}}, Romeel and {Angl{\'e}s-Alc{\'a}zar}, Daniel},
        title = "{The impact of feedback on the evolution of gas density profiles from galaxies to clusters: a universal fitting formula from the Simba suite of simulations}",
      journal = {The Open Journal of Astrophysics},
         year = 2024,
        month = dec,
       volume = {7},
          eid = {115},
        pages = {115},
          doi = {10.33232/001c.126621},
archivePrefix = {arXiv},
       eprint = {2409.05815},
 primaryClass = {astro-ph.GA},
       adsurl = {https://ui.adsabs.harvard.edu/abs/2024OJAp....7E.115S}
}

@ARTICLE{Christiansen2020,
       author = {{Christiansen}, Jacob F. and {Dav{\'e}}, Romeel and {Sorini}, Daniele and {Angl{\'e}s-Alc{\'a}zar}, Daniel},
        title = "{Jet feedback and the photon underproduction crisis in SIMBA}",
      journal = {\mnras},
         year = 2020,
        month = dec,
       volume = {499},
       number = {2},
        pages = {2617-2635},
          doi = {10.1093/mnras/staa3007},
archivePrefix = {arXiv},
       eprint = {1911.01343},
 primaryClass = {astro-ph.GA},
       adsurl = {https://ui.adsabs.harvard.edu/abs/2020MNRAS.499.2617C}
}

@ARTICLE{Kulier23,
       author = {{Kulier}, Andrea and {Poggianti}, Bianca and {Tonnesen}, Stephanie and {Smith}, Rory and {Ignesti}, Alessandro and {Akerman}, Nina and {Marasco}, Antonino and {Vulcani}, Benedetta and {Moretti}, Alessia and {Wolter}, Anna},
        title = "{Ram Pressure Stripping in the EAGLE Simulation}",
      journal = {\apj},
         year = 2023,
        month = sep,
       volume = {954},
       number = {2},
          eid = {177},
        pages = {177},
          doi = {10.3847/1538-4357/aceda3},
archivePrefix = {arXiv},
       eprint = {2305.03758},
 primaryClass = {astro-ph.GA},
       adsurl = {https://ui.adsabs.harvard.edu/abs/2023ApJ...954..177K}
}

@ARTICLE{Vayner2024,
       author = {{Vayner}, Andrey and {D{\'\i}az-Santos}, Tanio and {Eisenhardt}, Peter R.~M. and {Stern}, Daniel and {Armus}, Lee and {Angl{\'e}s-Alc{\'a}zar}, Daniel and {Assef}, Roberto J. and {Fern{\'a}ndez Aranda}, Rom{\'a}n and {Blain}, Andrew W. and {Jun}, Hyunsung D. and {Tsai}, Chao-Wei and {Roy}, Niranjan Chandra and {Brisbin}, Drew and {Ferkinhoff}, Carl D. and {Aravena}, Manuel and {Gonz{\'a}lez-L{\'o}pez}, Jorge and {Li}, Guodong and {Liao}, Mai and {Shobhana}, Devika and {Wu}, Jingwen and {Zewdie}, Dejene},
        title = "{Powerful Nuclear Outflows and Circumgalactic Medium Shocks Driven by the Most Luminous Known Obscured Quasar in the Universe}",
      journal = {\apj},
         year = 2025,
        month = aug,
       volume = {989},
       number = {2},
          eid = {230},
        pages = {230},
          doi = {10.3847/1538-4357/addbdd},
archivePrefix = {arXiv},
       eprint = {2412.02862},
 primaryClass = {astro-ph.GA},
       adsurl = {https://ui.adsabs.harvard.edu/abs/2025ApJ...989..230V}
}

@ARTICLE{kraljic2020,
       author = {{Kraljic}, Katarina and {Dav{\'e}}, Romeel and {Pichon}, Christophe},
        title = "{And yet it flips: connecting galactic spin and the cosmic web}",
      journal = {\mnras},
         year = 2020,
        month = mar,
       volume = {493},
       number = {1},
        pages = {362-381},
          doi = {10.1093/mnras/staa250},
archivePrefix = {arXiv},
       eprint = {1906.01623},
 primaryClass = {astro-ph.GA},
       adsurl = {https://ui.adsabs.harvard.edu/abs/2020MNRAS.493..362K}
}

@ARTICLE{SWIFT,
       author = {{Schaller}, Matthieu and {Borrow}, Josh and {Draper}, Peter W. and {Ivkovic}, Mladen and {McAlpine}, Stuart and {Vandenbroucke}, Bert and {Bah{\'e}}, Yannick and {Chaikin}, Evgenii and {Chalk}, Aidan B.~G. and {Chan}, Tsang Keung and {Correa}, Camila and {van Daalen}, Marcel and {Elbers}, Willem and {Gonnet}, Pedro and {Hausammann}, Lo{\"\i}c and {Helly}, John and {Hu{\v{s}}ko}, Filip and {Kegerreis}, Jacob A. and {Nobels}, Folkert S.~J. and {Ploeckinger}, Sylvia and {Revaz}, Yves and {Roper}, William J. and {Ruiz-Bonilla}, Sergio and {Sandnes}, Thomas D. and {Uyttenhove}, Yolan and {Willis}, James S. and {Xiang}, Zhen},
        title = "{SWIFT: A modern highly-parallel gravity and smoothed particle hydrodynamics solver for astrophysical and cosmological applications}",
      journal = {\mnras},
         year = 2024,
        month = may,
       volume = {530},
       number = {2},
        pages = {2378-2419},
          doi = {10.1093/mnras/stae922},
archivePrefix = {arXiv},
       eprint = {2305.13380},
 primaryClass = {astro-ph.IM},
       adsurl = {https://ui.adsabs.harvard.edu/abs/2024MNRAS.530.2378S}
}

@ARTICLE{vandeVoort11,
       author = {{van de Voort}, Freeke and {Schaye}, Joop and {Booth}, C.~M. and {Haas}, Marcel R. and {Dalla Vecchia}, Claudio},
        title = "{The rates and modes of gas accretion on to galaxies and their gaseous haloes}",
      journal = {\mnras},
         year = 2011,
        month = jul,
       volume = {414},
       number = {3},
        pages = {2458-2478},
          doi = {10.1111/j.1365-2966.2011.18565.x},
archivePrefix = {arXiv},
       eprint = {1011.2491},
 primaryClass = {astro-ph.CO},
       adsurl = {https://ui.adsabs.harvard.edu/abs/2011MNRAS.414.2458V}
}

@ARTICLE{Popesso23,
       author = {{Popesso}, P. and {Concas}, A. and {Cresci}, G. and {Belli}, S. and {Rodighiero}, G. and {Inami}, H. and {Dickinson}, M. and {Ilbert}, O. and {Pannella}, M. and {Elbaz}, D.},
        title = "{The main sequence of star-forming galaxies across cosmic times}",
      journal = {\mnras},
         year = 2023,
        month = feb,
       volume = {519},
       number = {1},
        pages = {1526-1544},
          doi = {10.1093/mnras/stac3214},
archivePrefix = {arXiv},
       eprint = {2203.10487},
 primaryClass = {astro-ph.GA},
       adsurl = {https://ui.adsabs.harvard.edu/abs/2023MNRAS.519.1526P}
}

@ARTICLE{Haas2012,
       author = {{Haas}, Marcel R. and {Schaye}, Joop and {Jeeson-Daniel}, Akila},
        title = "{Disentangling galaxy environment and host halo mass}",
      journal = {\mnras},
         year = 2012,
        month = jan,
       volume = {419},
       number = {3},
        pages = {2133-2146},
          doi = {10.1111/j.1365-2966.2011.19863.x},
archivePrefix = {arXiv},
       eprint = {1103.0547},
 primaryClass = {astro-ph.CO},
       adsurl = {https://ui.adsabs.harvard.edu/abs/2012MNRAS.419.2133H}
}

@ARTICLE{Springel2001,
       author = {{Springel}, Volker and {White}, Simon D.~M. and {Tormen}, Giuseppe and {Kauffmann}, Guinevere},
        title = "{Populating a cluster of galaxies - I. Results at z=0}",
      journal = {\mnras},
         year = 2001,
        month = dec,
       volume = {328},
       number = {3},
        pages = {726-750},
          doi = {10.1046/j.1365-8711.2001.04912.x},
archivePrefix = {arXiv},
       eprint = {astro-ph/0012055},
 primaryClass = {astro-ph},
       adsurl = {https://ui.adsabs.harvard.edu/abs/2001MNRAS.328..726S}
}

@ARTICLE{Springel2018,
       author = {{Springel}, Volker and {Pakmor}, R{\"u}diger and {Pillepich}, Annalisa and {Weinberger}, Rainer and {Nelson}, Dylan and {Hernquist}, Lars and {Vogelsberger}, Mark and {Genel}, Shy and {Torrey}, Paul and {Marinacci}, Federico and {Naiman}, Jill},
        title = "{First results from the IllustrisTNG simulations: matter and galaxy clustering}",
      journal = {\mnras},
         year = 2018,
        month = mar,
       volume = {475},
       number = {1},
        pages = {676-698},
          doi = {10.1093/mnras/stx3304},
archivePrefix = {arXiv},
       eprint = {1707.03397},
 primaryClass = {astro-ph.GA},
       adsurl = {https://ui.adsabs.harvard.edu/abs/2018MNRAS.475..676S}
}

@ARTICLE{Hasan2023,
       author = {{Hasan}, Farhanul and {Burchett}, Joseph N. and {Abeyta}, Alyssa and {Hellinger}, Douglas and {Mandelker}, Nir and {Primack}, Joel R. and {Faber}, S.~M. and {Koo}, David C. and {Elek}, Oskar and {Nagai}, Daisuke},
        title = "{The Evolving Effect of Cosmic Web Environment on Galaxy Quenching}",
      journal = {\apj},
         year = 2023,
        month = jun,
       volume = {950},
       number = {2},
          eid = {114},
        pages = {114},
          doi = {10.3847/1538-4357/acd11c},
archivePrefix = {arXiv},
       eprint = {2303.08088},
 primaryClass = {astro-ph.GA},
       adsurl = {https://ui.adsabs.harvard.edu/abs/2023ApJ...950..114H}
}

@ARTICLE{Crain2023,
       author = {{Crain}, Robert A. and {van de Voort}, Freeke},
        title = "{Hydrodynamical Simulations of the Galaxy Population: Enduring Successes and Outstanding Challenges}",
      journal = {\araa},
         year = 2023,
        month = aug,
       volume = {61},
        pages = {473-515},
          doi = {10.1146/annurev-astro-041923-043618},
archivePrefix = {arXiv},
       eprint = {2309.17075},
 primaryClass = {astro-ph.GA},
       adsurl = {https://ui.adsabs.harvard.edu/abs/2023ARA&A..61..473C}
}

@ARTICLE{Moreno2022,
       author = {{Moreno}, Jorge and {Danieli}, Shany and {Bullock}, James S. and {Feldmann}, Robert and {Hopkins}, Philip F. and {{\c{c}}atmabacak}, Onur and {Gurvich}, Alexander and {Lazar}, Alexandres and {Klein}, Courtney and {Hummels}, Cameron B. and {Hafen}, Zachary and {Mercado}, Francisco J. and {Yu}, Sijie and {Jiang}, Fangzhou and {Wheeler}, Coral and {Wetzel}, Andrew and {Angl{\'e}s-Alc{\'a}zar}, Daniel and {Boylan-Kolchin}, Michael and {Quataert}, Eliot and {Faucher-Gigu{\`e}re}, Claude-Andr{\'e} and {Kere{\v{s}}}, Du{\v{s}}an},
        title = "{Galaxies lacking dark matter produced by close encounters in a cosmological simulation}",
      journal = {Nature Astronomy},
         year = 2022,
        month = apr,
       volume = {6},
        pages = {496-502},
          doi = {10.1038/s41550-021-01598-4},
archivePrefix = {arXiv},
       eprint = {2202.05836},
 primaryClass = {astro-ph.GA},
       adsurl = {https://ui.adsabs.harvard.edu/abs/2022NatAs...6..496M}
}

@ARTICLE{Habouzit2022,
       author = {{Habouzit}, M{\'e}lanie and {Somerville}, Rachel S. and {Li}, Yuan and {Genel}, Shy and {Aird}, James and {Angl{\'e}s-Alc{\'a}zar}, Daniel and {Dav{\'e}}, Romeel and {Georgiev}, Iskren Y. and {McAlpine}, Stuart and {Rosas-Guevara}, Yetli and {Dubois}, Yohan and {Nelson}, Dylan and {Banados}, Eduardo and {Hernquist}, Lars and {Peirani}, S{\'e}bastien and {Vogelsberger}, Mark},
        title = "{Supermassive black holes in cosmological simulations - II: the AGN population and predictions for upcoming X-ray missions}",
      journal = {\mnras},
         year = 2022,
        month = jan,
       volume = {509},
       number = {2},
        pages = {3015-3042},
          doi = {10.1093/mnras/stab3147},
archivePrefix = {arXiv},
       eprint = {2111.01802},
 primaryClass = {astro-ph.GA},
       adsurl = {https://ui.adsabs.harvard.edu/abs/2022MNRAS.509.3015H}
}

@ARTICLE{Habouzit2021,
       author = {{Habouzit}, M{\'e}lanie and {Li}, Yuan and {Somerville}, Rachel S. and {Genel}, Shy and {Pillepich}, Annalisa and {Volonteri}, Marta and {Dav{\'e}}, Romeel and {Rosas-Guevara}, Yetli and {McAlpine}, Stuart and {Peirani}, S{\'e}bastien and {Hernquist}, Lars and {Angl{\'e}s-Alc{\'a}zar}, Daniel and {Reines}, Amy and {Bower}, Richard and {Dubois}, Yohan and {Nelson}, Dylan and {Pichon}, Christophe and {Vogelsberger}, Mark},
        title = "{Supermassive black holes in cosmological simulations I: M$_{BH}$ - M$_{{\ensuremath{\star}}}$ relation and black hole mass function}",
      journal = {\mnras},
         year = 2021,
        month = may,
       volume = {503},
       number = {2},
        pages = {1940-1975},
          doi = {10.1093/mnras/stab496},
archivePrefix = {arXiv},
       eprint = {2006.10094},
 primaryClass = {astro-ph.GA},
       adsurl = {https://ui.adsabs.harvard.edu/abs/2021MNRAS.503.1940H}
}

@ARTICLE{Mercedes-feliz2023,
       author = {{Mercedes-Feliz}, Jonathan and {Angl{\'e}s-Alc{\'a}zar}, Daniel and {Hayward}, Christopher C. and {Cochrane}, Rachel K. and {Terrazas}, Bryan A. and {Wellons}, Sarah and {Richings}, Alexander J. and {Faucher-Gigu{\`e}re}, Claude-Andr{\'e} and {Moreno}, Jorge and {Su}, Kung Yi and {Hopkins}, Philip F. and {Quataert}, Eliot and {Kere{\v{s}}}, Du{\v{s}}an},
        title = "{Local positive feedback in the overall negative: the impact of quasar winds on star formation in the FIRE cosmological simulations}",
      journal = {\mnras},
         year = 2023,
        month = sep,
       volume = {524},
       number = {3},
        pages = {3446-3463},
          doi = {10.1093/mnras/stad2079},
archivePrefix = {arXiv},
       eprint = {2301.01784},
 primaryClass = {astro-ph.GA},
       adsurl = {https://ui.adsabs.harvard.edu/abs/2023MNRAS.524.3446M}
}

@ARTICLE{Medlock2024,
       author = {{Medlock}, Isabel and {Nagai}, Daisuke and {Singh}, Priyanka and {Oppenheimer}, Benjamin and {Angl{\'e}s-Alc{\'a}zar}, Daniel and {Villaescusa-Navarro}, Francisco},
        title = "{Probing the Circumgalactic Medium with Fast Radio Bursts: Insights from CAMELS}",
      journal = {\apj},
         year = 2024,
        month = may,
       volume = {967},
       number = {1},
          eid = {32},
        pages = {32},
          doi = {10.3847/1538-4357/ad3070},
archivePrefix = {arXiv},
       eprint = {2403.02313},
 primaryClass = {astro-ph.GA},
       adsurl = {https://ui.adsabs.harvard.edu/abs/2024ApJ...967...32M}
}

@ARTICLE{Gebhardt2024,
       author = {{Gebhardt}, Matthew and {Angl{\'e}s-Alc{\'a}zar}, Daniel and {Borrow}, Josh and {Genel}, Shy and {Villaescusa-Navarro}, Francisco and {Ni}, Yueying and {Lovell}, Christopher C. and {Nagai}, Daisuke and {Dav{\'e}}, Romeel and {Marinacci}, Federico and {Vogelsberger}, Mark and {Hernquist}, Lars},
        title = "{Cosmological baryon spread and impact on matter clustering in CAMELS}",
      journal = {\mnras},
         year = 2024,
        month = apr,
       volume = {529},
       number = {4},
        pages = {4896-4913},
          doi = {10.1093/mnras/stae817},
archivePrefix = {arXiv},
       eprint = {2307.11832},
 primaryClass = {astro-ph.GA},
       adsurl = {https://ui.adsabs.harvard.edu/abs/2024MNRAS.529.4896G}
}

@ARTICLE{Lovell2025,
       author = {{Lovell}, Christopher C. and {Starkenburg}, Tjitske and {Ho}, Matthew and {Angl{\'e}s-Alc{\'a}zar}, Daniel and {Dav{\'e}}, Romeel and {Gabrielpillai}, Austen and {Iyer}, Kartheik G. and {Matthews}, Alice E. and {Roper}, William J. and {Somerville}, Rachel S. and {Sommovigo}, Laura and {Villaescusa-Navarro}, Francisco},
        title = "{Learning the Universe: cosmological and astrophysical parameter inference with galaxy luminosity functions and colours}",
      journal = {\mnras},
         year = 2025,
        month = dec,
       volume = {544},
       number = {4},
        pages = {3949-3979},
          doi = {10.1093/mnras/staf1888},
archivePrefix = {arXiv},
       eprint = {2411.13960},
 primaryClass = {astro-ph.GA},
       adsurl = {https://ui.adsabs.harvard.edu/abs/2025MNRAS.544.3949L}
}

@ARTICLE{DeSanti2023,
       author = {{de Santi}, Natal{\'\i} S.~M. and {Shao}, Helen and {Villaescusa-Navarro}, Francisco and {Abramo}, L. Raul and {Teyssier}, Romain and {Villanueva-Domingo}, Pablo and {Ni}, Yueying and {Angl{\'e}s-Alc{\'a}zar}, Daniel and {Genel}, Shy and {Hern{\'a}ndez-Mart{\'\i}nez}, Elena and {Steinwandel}, Ulrich P. and {Lovell}, Christopher C. and {Dolag}, Klaus and {Castro}, Tiago and {Vogelsberger}, Mark},
        title = "{Robust Field-level Likelihood-free Inference with Galaxies}",
      journal = {\apj},
         year = 2023,
        month = jul,
       volume = {952},
       number = {1},
          eid = {69},
        pages = {69},
          doi = {10.3847/1538-4357/acd1e2},
archivePrefix = {arXiv},
       eprint = {2302.14101},
 primaryClass = {astro-ph.CO},
       adsurl = {https://ui.adsabs.harvard.edu/abs/2023ApJ...952...69D}
}

@ARTICLE{Villaescusa-Navarro2022,
       author = {{Villaescusa-Navarro}, Francisco and {Ding}, Jupiter and {Genel}, Shy and {Tonnesen}, Stephanie and {La Torre}, Valentina and {Spergel}, David N. and {Teyssier}, Romain and {Li}, Yin and {Heneka}, Caroline and {Lemos}, Pablo and {Angl{\'e}s-Alc{\'a}zar}, Daniel and {Nagai}, Daisuke and {Vogelsberger}, Mark},
        title = "{Cosmology with One Galaxy?}",
      journal = {\apj},
         year = 2022,
        month = apr,
       volume = {929},
       number = {2},
          eid = {132},
        pages = {132},
          doi = {10.3847/1538-4357/ac5d3f},
archivePrefix = {arXiv},
       eprint = {2201.02202},
 primaryClass = {astro-ph.CO},
       adsurl = {https://ui.adsabs.harvard.edu/abs/2022ApJ...929..132V}
}

@ARTICLE{Nicola2022,
       author = {{Nicola}, Andrina and {Villaescusa-Navarro}, Francisco and {Spergel}, David N. and {Dunkley}, Jo and {Angl{\'e}s-Alc{\'a}zar}, Daniel and {Dav{\'e}}, Romeel and {Genel}, Shy and {Hernquist}, Lars and {Nagai}, Daisuke and {Somerville}, Rachel S. and {Wandelt}, Benjamin D.},
        title = "{Breaking baryon-cosmology degeneracy with the electron density power spectrum}",
      journal = {\jcap},
         year = 2022,
        month = apr,
       volume = {2022},
       number = {4},
          eid = {046},
        pages = {046},
          doi = {10.1088/1475-7516/2022/04/046},
archivePrefix = {arXiv},
       eprint = {2201.04142},
 primaryClass = {astro-ph.CO},
       adsurl = {https://ui.adsabs.harvard.edu/abs/2022JCAP...04..046N}
}

@ARTICLE{Villaescusa-Navarro2021arx,
       author = {{Villaescusa-Navarro}, Francisco and {Angl{\'e}s-Alc{\'a}zar}, Daniel and {Genel}, Shy and {Spergel}, David N. and {Li}, Yin and {Wandelt}, Benjamin and {Nicola}, Andrina and {Thiele}, Leander and {Hassan}, Sultan and {Zorrilla Matilla}, Jose Manuel and {Narayanan}, Desika and {Dave}, Romeel and {Vogelsberger}, Mark},
        title = "{Multifield Cosmology with Artificial Intelligence}",
      journal = {arXiv e-prints},
         year = 2021,
        month = sep,
          eid = {arXiv:2109.09747},
        pages = {arXiv:2109.09747},
          doi = {10.48550/arXiv.2109.09747},
archivePrefix = {arXiv},
       eprint = {2109.09747},
 primaryClass = {astro-ph.CO},
       adsurl = {https://ui.adsabs.harvard.edu/abs/2021arXiv210909747V}
}

@ARTICLE{Dalla_Vecchia&Schaye,
       author = {{Dalla Vecchia}, Claudio and {Schaye}, Joop},
        title = "{Simulating galactic outflows with thermal supernova feedback}",
      journal = {\mnras},
         year = 2012,
        month = oct,
       volume = {426},
       number = {1},
        pages = {140-158},
          doi = {10.1111/j.1365-2966.2012.21704.x},
archivePrefix = {arXiv},
       eprint = {1203.5667},
 primaryClass = {astro-ph.GA},
       adsurl = {https://ui.adsabs.harvard.edu/abs/2012MNRAS.426..140D}
}

@ARTICLE{gonzalez2025,
       author = {{Gonz{\'a}lez Lobos}, Jay and {Arrigoni Battaia}, Fabrizio and {Obreja}, Aura and {Kauffmann}, Guinevere and {Farina}, Emanuele Paolo and {Costa}, Tiago},
        title = "{QSO MUSEUM III: the circumgalactic medium in Ly$α$ emission around 120 $z\sim3$ quasars covering the SDSS parameter space. Witnessing the instantaneous AGN feedback on halo scales}",
      journal = {arXiv e-prints},
         year = 2025,
        month = jul,
          eid = {arXiv:2507.16898},
        pages = {arXiv:2507.16898},
          doi = {10.48550/arXiv.2507.16898},
archivePrefix = {arXiv},
       eprint = {2507.16898},
 primaryClass = {astro-ph.GA},
       adsurl = {https://ui.adsabs.harvard.edu/abs/2025arXiv250716898G}
}

@ARTICLE{Harrison2017,
       author = {{Harrison}, C.~M.},
        title = "{Impact of supermassive black hole growth on star formation}",
      journal = {Nature Astronomy},
         year = 2017,
        month = jul,
       volume = {1},
          eid = {0165},
        pages = {0165},
          doi = {10.1038/s41550-017-0165},
archivePrefix = {arXiv},
       eprint = {1703.06889},
 primaryClass = {astro-ph.GA},
       adsurl = {https://ui.adsabs.harvard.edu/abs/2017NatAs...1E.165H}
}

@ARTICLE{Lee2024,
       author = {{Lee}, Max E. and {Genel}, Shy and {Wandelt}, Benjamin D. and {Zhang}, Benjamin and {Delgado}, Ana Maria and {Pandey}, Shivam and {Lau}, Erwin T. and {Carr}, Christopher and {Cook}, Harrison and {Nagai}, Daisuke and {Angles-Alcazar}, Daniel and {Villaescusa-Navarro}, Francisco and {Bryan}, Greg L.},
        title = "{Zooming by in the CARPoolGP Lane: New CAMELS-TNG Simulations of Zoomed-in Massive Halos}",
      journal = {\apj},
         year = 2024,
        month = jun,
       volume = {968},
       number = {1},
          eid = {11},
        pages = {11},
          doi = {10.3847/1538-4357/ad3d4a},
archivePrefix = {arXiv},
       eprint = {2403.10609},
 primaryClass = {astro-ph.GA},
       adsurl = {https://ui.adsabs.harvard.edu/abs/2024ApJ...968...11L}
}

@ARTICLE{Villaescusa-Navarro2023,
       author = {{Villaescusa-Navarro}, Francisco and {Genel}, Shy and {Angl{\'e}s-Alc{\'a}zar}, Daniel and {Perez}, Lucia A. and {Villanueva-Domingo}, Pablo and {Wadekar}, Digvijay and {Shao}, Helen and {Mohammad}, Faizan G. and {Hassan}, Sultan and {Moser}, Emily and {Lau}, Erwin T. and {Machado Poletti Valle}, Luis Fernando and {Nicola}, Andrina and {Thiele}, Leander and {Jo}, Yongseok and {Philcox}, Oliver H.~E. and {Oppenheimer}, Benjamin D. and {Tillman}, Megan and {Hahn}, ChangHoon and {Kaushal}, Neerav and {Pisani}, Alice and {Gebhardt}, Matthew and {Delgado}, Ana Maria and {Caliendo}, Joyce and {Kreisch}, Christina and {Wong}, Kaze W.~K. and {Coulton}, William R. and {Eickenberg}, Michael and {Parimbelli}, Gabriele and {Ni}, Yueying and {Steinwandel}, Ulrich P. and {La Torre}, Valentina and {Dave}, Romeel and {Battaglia}, Nicholas and {Nagai}, Daisuke and {Spergel}, David N. and {Hernquist}, Lars and {Burkhart}, Blakesley and {Narayanan}, Desika and {Wandelt}, Benjamin and {Somerville}, Rachel S. and {Bryan}, Greg L. and {Viel}, Matteo and {Li}, Yin and {Irsic}, Vid and {Kraljic}, Katarina and {Marinacci}, Federico and {Vogelsberger}, Mark},
        title = "{The CAMELS Project: Public Data Release}",
      journal = {\apjs},
         year = 2023,
        month = apr,
       volume = {265},
       number = {2},
          eid = {54},
        pages = {54},
          doi = {10.3847/1538-4365/acbf47},
archivePrefix = {arXiv},
       eprint = {2201.01300},
 primaryClass = {astro-ph.CO},
       adsurl = {https://ui.adsabs.harvard.edu/abs/2023ApJS..265...54V}
}

@ARTICLE{Delgado2023,
       author = {{Delgado}, Ana Maria and {Angl{\'e}s-Alc{\'a}zar}, Daniel and {Thiele}, Leander and {Pandey}, Shivam and {Lehman}, Kai and {Somerville}, Rachel S. and {Ntampaka}, Michelle and {Genel}, Shy and {Villaescusa-Navarro}, Francisco and {Hernquist}, Lars},
        title = "{Predicting the impact of feedback on matter clustering with machine learning in CAMELS}",
      journal = {\mnras},
         year = 2023,
        month = dec,
       volume = {526},
       number = {4},
        pages = {5306-5325},
          doi = {10.1093/mnras/stad2992},
archivePrefix = {arXiv},
       eprint = {2301.02231},
 primaryClass = {astro-ph.GA},
       adsurl = {https://ui.adsabs.harvard.edu/abs/2023MNRAS.526.5306D}
}

@ARTICLE{CAMELS_PAPER,
       author = {{Villaescusa-Navarro}, Francisco and {Angl{\'e}s-Alc{\'a}zar}, Daniel and {Genel}, Shy and {Spergel}, David N. and {Somerville}, Rachel S. and {Dave}, Romeel and {Pillepich}, Annalisa and {Hernquist}, Lars and {Nelson}, Dylan and {Torrey}, Paul and {Narayanan}, Desika and {Li}, Yin and {Philcox}, Oliver and {La Torre}, Valentina and {Maria Delgado}, Ana and {Ho}, Shirley and {Hassan}, Sultan and {Burkhart}, Blakesley and {Wadekar}, Digvijay and {Battaglia}, Nicholas and {Contardo}, Gabriella and {Bryan}, Greg L.},
        title = "{The CAMELS Project: Cosmology and Astrophysics with Machine-learning Simulations}",
      journal = {\apj},
         year = 2021,
        month = jul,
       volume = {915},
       number = {1},
          eid = {71},
        pages = {71},
          doi = {10.3847/1538-4357/abf7ba},
archivePrefix = {arXiv},
       eprint = {2010.00619},
 primaryClass = {astro-ph.CO},
       adsurl = {https://ui.adsabs.harvard.edu/abs/2021ApJ...915...71V}
}

@ARTICLE{SIMBA.PAPER,
       author = {{Dav{\'e}}, Romeel and {Angl{\'e}s-Alc{\'a}zar}, Daniel and {Narayanan}, Desika and {Li}, Qi and {Rafieferantsoa}, Mika H. and {Appleby}, Sarah},
        title = "{SIMBA: Cosmological simulations with black hole growth and feedback}",
      journal = {\mnras},
         year = 2019,
        month = jun,
       volume = {486},
       number = {2},
        pages = {2827-2849},
          doi = {10.1093/mnras/stz937},
archivePrefix = {arXiv},
       eprint = {1901.10203},
 primaryClass = {astro-ph.GA},
       adsurl = {https://ui.adsabs.harvard.edu/abs/2019MNRAS.486.2827D}
}

@ARTICLE{Salim2018,
       author = {{Salim}, Samir and {Boquien}, M{\'e}d{\'e}ric and {Lee}, Janice C.},
        title = "{Dust Attenuation Curves in the Local Universe: Demographics and New Laws for Star-forming Galaxies and High-redshift Analogs}",
      journal = {\apj},
         year = 2018,
        month = may,
       volume = {859},
       number = {1},
          eid = {11},
        pages = {11},
          doi = {10.3847/1538-4357/aabf3c},
archivePrefix = {arXiv},
       eprint = {1804.05850},
 primaryClass = {astro-ph.GA},
       adsurl = {https://ui.adsabs.harvard.edu/abs/2018ApJ...859...11S}
}

@ARTICLE{Salim2016,
       author = {{Salim}, Samir and {Lee}, Janice C. and {Janowiecki}, Steven and {da Cunha}, Elisabete and {Dickinson}, Mark and {Boquien}, M{\'e}d{\'e}ric and {Burgarella}, Denis and {Salzer}, John J. and {Charlot}, St{\'e}phane},
        title = "{GALEX-SDSS-WISE Legacy Catalog (GSWLC): Star Formation Rates, Stellar Masses, and Dust Attenuations of 700,000 Low-redshift Galaxies}",
      journal = {\apjs},
         year = 2016,
        month = nov,
       volume = {227},
       number = {1},
          eid = {2},
        pages = {2},
          doi = {10.3847/0067-0049/227/1/2},
archivePrefix = {arXiv},
       eprint = {1610.00712},
 primaryClass = {astro-ph.GA},
       adsurl = {https://ui.adsabs.harvard.edu/abs/2016ApJS..227....2S}
}

@ARTICLE{Hopkins2008,
       author = {{Hopkins}, Philip F. and {Hernquist}, Lars and {Cox}, Thomas J. and {Kere{\v{s}}}, Du{\v{s}}an},
        title = "{A Cosmological Framework for the Co-Evolution of Quasars, Supermassive Black Holes, and Elliptical Galaxies. I. Galaxy Mergers and Quasar Activity}",
      journal = {\apjs},
         year = 2008,
        month = apr,
       volume = {175},
       number = {2},
        pages = {356-389},
          doi = {10.1086/524362},
archivePrefix = {arXiv},
       eprint = {0706.1243},
 primaryClass = {astro-ph},
       adsurl = {https://ui.adsabs.harvard.edu/abs/2008ApJS..175..356H}
}

@ARTICLE{Ellison2019,
       author = {{Ellison}, Sara L. and {Viswanathan}, Akshara and {Patton}, David R. and {Bottrell}, Connor and {McConnachie}, Alan W. and {Gwyn}, Stephen and {Cuillandre}, Jean-Charles},
        title = "{A definitive merger-AGN connection at z {\ensuremath{\sim}} 0 with CFIS: mergers have an excess of AGN and AGN hosts are more frequently disturbed}",
      journal = {\mnras},
         year = 2019,
        month = aug,
       volume = {487},
       number = {2},
        pages = {2491-2504},
          doi = {10.1093/mnras/stz1431},
archivePrefix = {arXiv},
       eprint = {1905.08830},
 primaryClass = {astro-ph.GA},
       adsurl = {https://ui.adsabs.harvard.edu/abs/2019MNRAS.487.2491E}
}

@ARTICLE{Moreno2019,
       author = {{Moreno}, Jorge and {Torrey}, Paul and {Ellison}, Sara L. and {Patton}, David R. and {Hopkins}, Philip F. and {Bueno}, Michael and {Hayward}, Christopher C. and {Narayanan}, Desika and {Kere{\v{s}}}, Du{\v{s}}an and {Bluck}, Asa F.~L. and {Hernquist}, Lars},
        title = "{Interacting galaxies on FIRE-2: the connection between enhanced star formation and interstellar gas content}",
      journal = {\mnras},
         year = 2019,
        month = may,
       volume = {485},
       number = {1},
        pages = {1320-1338},
          doi = {10.1093/mnras/stz417},
archivePrefix = {arXiv},
       eprint = {1902.02305},
 primaryClass = {astro-ph.GA},
       adsurl = {https://ui.adsabs.harvard.edu/abs/2019MNRAS.485.1320M}
}

@ARTICLE{Bulichi24,
       author = {{Bulichi}, Teodora-Elena and {Dav{\'e}}, Romeel and {Kraljic}, Katarina},
        title = "{How galaxy properties vary with filament proximity in the SIMBA simulations}",
      journal = {\mnras},
         year = 2024,
        month = apr,
       volume = {529},
       number = {3},
        pages = {2595-2610},
          doi = {10.1093/mnras/stae667},
archivePrefix = {arXiv},
       eprint = {2309.03282},
 primaryClass = {astro-ph.GA},
       adsurl = {https://ui.adsabs.harvard.edu/abs/2024MNRAS.529.2595B}
}

@ARTICLE{Bird2022,
       author = {{Bird}, Simeon and {Ni}, Yueying and {Di Matteo}, Tiziana and {Croft}, Rupert and {Feng}, Yu and {Chen}, Nianyi},
        title = "{The ASTRID simulation: galaxy formation and reionization}",
      journal = {\mnras},
         year = 2022,
        month = may,
       volume = {512},
       number = {3},
        pages = {3703-3716},
          doi = {10.1093/mnras/stac648},
archivePrefix = {arXiv},
       eprint = {2111.01160},
 primaryClass = {astro-ph.GA},
       adsurl = {https://ui.adsabs.harvard.edu/abs/2022MNRAS.512.3703B}
}

@ARTICLE{Ni2022,
       author = {{Ni}, Yueying and {Di Matteo}, Tiziana and {Bird}, Simeon and {Croft}, Rupert and {Feng}, Yu and {Chen}, Nianyi and {Tremmel}, Michael and {DeGraf}, Colin and {Li}, Yin},
        title = "{The ASTRID simulation: the evolution of supermassive black holes}",
      journal = {\mnras},
         year = 2022,
        month = jun,
       volume = {513},
       number = {1},
        pages = {670-692},
          doi = {10.1093/mnras/stac351},
archivePrefix = {arXiv},
       eprint = {2110.14154},
 primaryClass = {astro-ph.GA},
       adsurl = {https://ui.adsabs.harvard.edu/abs/2022MNRAS.513..670N}
}

@ARTICLE{Davies_LJM2019,
       author = {{Davies}, L.~J.~M. and {Robotham}, A.~S.~G. and {Lagos}, C. del P. and {Driver}, S.~P. and {Stevens}, A.~R.~H. and {Bah{\'e}}, Y.~M. and {Alpaslan}, M. and {Bremer}, M.~N. and {Brown}, M.~J.~I. and {Brough}, S. and {Bland-Hawthorn}, J. and {Cortese}, L. and {Elahi}, P. and {Grootes}, M.~W. and {Holwerda}, B.~W. and {Ludlow}, A.~D. and {McGee}, S. and {Owers}, M. and {Phillipps}, S.},
        title = "{Galaxy and Mass Assembly (GAMA): environmental quenching of centrals and satellites in groups}",
      journal = {\mnras},
         year = 2019,
        month = mar,
       volume = {483},
       number = {4},
        pages = {5444-5458},
          doi = {10.1093/mnras/sty3393},
archivePrefix = {arXiv},
       eprint = {1901.01640},
 primaryClass = {astro-ph.GA},
       adsurl = {https://ui.adsabs.harvard.edu/abs/2019MNRAS.483.5444D}
}

@ARTICLE{Borrow2022,
       author = {{Borrow}, Josh and {Schaller}, Matthieu and {Bower}, Richard G. and {Schaye}, Joop},
        title = "{SPHENIX: smoothed particle hydrodynamics for the next generation of galaxy formation simulations}",
      journal = {\mnras},
         year = 2022,
        month = apr,
       volume = {511},
       number = {2},
        pages = {2367-2389},
          doi = {10.1093/mnras/stab3166},
archivePrefix = {arXiv},
       eprint = {2012.03974},
 primaryClass = {astro-ph.GA},
       adsurl = {https://ui.adsabs.harvard.edu/abs/2022MNRAS.511.2367B}
}

@ARTICLE{Brown2017,
       author = {{Brown}, Toby and {Catinella}, Barbara and {Cortese}, Luca and {Lagos}, Claudia del P. and {Dav{\'e}}, Romeel and {Kilborn}, Virginia and {Haynes}, Martha P. and {Giovanelli}, Riccardo and {Rafieferantsoa}, Mika},
        title = "{Cold gas stripping in satellite galaxies: from pairs to clusters}",
      journal = {\mnras},
         year = 2017,
        month = apr,
       volume = {466},
       number = {2},
        pages = {1275-1289},
          doi = {10.1093/mnras/stw2991},
archivePrefix = {arXiv},
       eprint = {1611.00896},
 primaryClass = {astro-ph.GA},
       adsurl = {https://ui.adsabs.harvard.edu/abs/2017MNRAS.466.1275B}
}

@ARTICLE{Ghodsi2024,
       author = {{Ghodsi}, Laya and {Man}, Allison W.~S. and {Donevski}, Darko and {Dav{\'e}}, Romeel and {Lim}, Seunghwan and {Lovell}, Christopher C. and {Narayanan}, Desika},
        title = "{Star formation efficiency across large-scale galactic environments}",
      journal = {\mnras},
         year = 2024,
        month = mar,
       volume = {528},
       number = {3},
        pages = {4393-4408},
          doi = {10.1093/mnras/stae279},
archivePrefix = {arXiv},
       eprint = {2309.01277},
 primaryClass = {astro-ph.GA},
       adsurl = {https://ui.adsabs.harvard.edu/abs/2024MNRAS.528.4393G}
}

@ARTICLE{Schneider2020,
       author = {{Schneider}, Evan E. and {Ostriker}, Eve C. and {Robertson}, Brant E. and {Thompson}, Todd A.},
        title = "{The Physical Nature of Starburst-driven Galactic Outflows}",
      journal = {\apj},
         year = 2020,
        month = may,
       volume = {895},
       number = {1},
          eid = {43},
        pages = {43},
          doi = {10.3847/1538-4357/ab8ae8},
archivePrefix = {arXiv},
       eprint = {2002.10468},
 primaryClass = {astro-ph.GA},
       adsurl = {https://ui.adsabs.harvard.edu/abs/2020ApJ...895...43S}
}

@ARTICLE{Cochrane2018,
       author = {{Cochrane}, R.~K. and {Best}, P.~N.},
        title = "{Dissecting the roles of mass and environment quenching in galaxy evolution with EAGLE}",
      journal = {\mnras},
         year = 2018,
        month = oct,
       volume = {480},
       number = {1},
        pages = {864-878},
          doi = {10.1093/mnras/sty1708},
archivePrefix = {arXiv},
       eprint = {1806.11120},
 primaryClass = {astro-ph.GA},
       adsurl = {https://ui.adsabs.harvard.edu/abs/2018MNRAS.480..864C}
}

@ARTICLE{2017MNRAS.466.3460V,
       author = {{van de Voort}, Freeke and {Bah{\'e}}, Yannick M. and {Bower}, Richard G. and {Correa}, Camila A. and {Crain}, Robert A. and {Schaye}, Joop and {Theuns}, Tom},
        title = "{The environmental dependence of gas accretion on to galaxies: quenching satellites through starvation}",
      journal = {\mnras},
         year = 2017,
        month = apr,
       volume = {466},
       number = {3},
        pages = {3460-3471},
          doi = {10.1093/mnras/stw3356},
archivePrefix = {arXiv},
       eprint = {1611.03870},
 primaryClass = {astro-ph.GA},
       adsurl = {https://ui.adsabs.harvard.edu/abs/2017MNRAS.466.3460V}
}

@ARTICLE{Aragon-calvo2019,
       author = {{Aragon Calvo}, Miguel A. and {Neyrinck}, Mark C. and {Silk}, Joseph},
        title = "{Galaxy Quenching from Cosmic Web Detachment}",
      journal = {The Open Journal of Astrophysics},
         year = 2019,
        month = jul,
       volume = {2},
       number = {1},
          eid = {7},
        pages = {7},
          doi = {10.21105/astro.1697.07881},
archivePrefix = {arXiv},
       eprint = {1607.07881},
 primaryClass = {astro-ph.GA},
       adsurl = {https://ui.adsabs.harvard.edu/abs/2019OJAp....2E...7A}
}

@ARTICLE{Correa15,
       author = {{Correa}, Camila A. and {Wyithe}, J. Stuart B. and {Schaye}, Joop and {Duffy}, Alan R.},
        title = "{The accretion history of dark matter haloes - I. The physical origin of the universal function}",
      journal = {\mnras},
         year = 2015,
        month = jun,
       volume = {450},
       number = {2},
        pages = {1514-1520},
          doi = {10.1093/mnras/stv689},
archivePrefix = {arXiv},
       eprint = {1409.5228},
 primaryClass = {astro-ph.GA},
       adsurl = {https://ui.adsabs.harvard.edu/abs/2015MNRAS.450.1514C}
}

@ARTICLE{Ellison08,
       author = {{Ellison}, Sara L. and {Patton}, David R. and {Simard}, Luc and {McConnachie}, Alan W.},
        title = "{Galaxy Pairs in the Sloan Digital Sky Survey. I. Star Formation, Active Galactic Nucleus Fraction, and the Mass-Metallicity Relation}",
      journal = {\aj},
         year = 2008,
        month = may,
       volume = {135},
       number = {5},
        pages = {1877-1899},
          doi = {10.1088/0004-6256/135/5/1877},
archivePrefix = {arXiv},
       eprint = {0803.0161},
 primaryClass = {astro-ph},
       adsurl = {https://ui.adsabs.harvard.edu/abs/2008AJ....135.1877E}
}

@ARTICLE{Pillepich2018,
       author = {{Pillepich}, Annalisa and {Springel}, Volker and {Nelson}, Dylan and {Genel}, Shy and {Naiman}, Jill and {Pakmor}, R{\"u}diger and {Hernquist}, Lars and {Torrey}, Paul and {Vogelsberger}, Mark and {Weinberger}, Rainer and {Marinacci}, Federico},
        title = "{Simulating galaxy formation with the IllustrisTNG model}",
      journal = {\mnras},
         year = 2018,
        month = jan,
       volume = {473},
       number = {3},
        pages = {4077-4106},
          doi = {10.1093/mnras/stx2656},
archivePrefix = {arXiv},
       eprint = {1703.02970},
 primaryClass = {astro-ph.GA},
       adsurl = {https://ui.adsabs.harvard.edu/abs/2018MNRAS.473.4077P}
}

@ARTICLE{Angles-Alcazar2017a,
       author = {{Angl{\'e}s-Alc{\'a}zar}, Daniel and {Dav{\'e}}, Romeel and {Faucher-Gigu{\`e}re}, Claude-Andr{\'e} and {{\"O}zel}, Feryal and {Hopkins}, Philip F.},
        title = "{Gravitational torque-driven black hole growth and feedback in cosmological simulations}",
      journal = {\mnras},
         year = 2017,
        month = jan,
       volume = {464},
       number = {3},
        pages = {2840-2853},
          doi = {10.1093/mnras/stw2565},
archivePrefix = {arXiv},
       eprint = {1603.08007},
 primaryClass = {astro-ph.GA},
       adsurl = {https://ui.adsabs.harvard.edu/abs/2017MNRAS.464.2840A}
}

@ARTICLE{Angles-Alcazar2017b,
       author = {{Angl{\'e}s-Alc{\'a}zar}, Daniel and {Faucher-Gigu{\`e}re}, Claude-Andr{\'e} and {Kere{\v{s}}}, Du{\v{s}}an and {Hopkins}, Philip F. and {Quataert}, Eliot and {Murray}, Norman},
        title = "{The cosmic baryon cycle and galaxy mass assembly in the FIRE simulations}",
      journal = {\mnras},
         year = 2017,
        month = oct,
       volume = {470},
       number = {4},
        pages = {4698-4719},
          doi = {10.1093/mnras/stx1517},
archivePrefix = {arXiv},
       eprint = {1610.08523},
 primaryClass = {astro-ph.GA},
       adsurl = {https://ui.adsabs.harvard.edu/abs/2017MNRAS.470.4698A}
}

@ARTICLE{Kuutma2017,
       author = {{Kuutma}, Teet and {Tamm}, Antti and {Tempel}, Elmo},
        title = "{From voids to filaments: environmental transformations of galaxies in the SDSS}",
      journal = {\aap},
         year = 2017,
        month = apr,
       volume = {600},
          eid = {L6},
        pages = {L6},
          doi = {10.1051/0004-6361/201730526},
archivePrefix = {arXiv},
       eprint = {1703.04338},
 primaryClass = {astro-ph.GA},
       adsurl = {https://ui.adsabs.harvard.edu/abs/2017A&A...600L...6K}
}

@ARTICLE{Tumlinson2017,
       author = {{Tumlinson}, Jason and {Peeples}, Molly S. and {Werk}, Jessica K.},
        title = "{The Circumgalactic Medium}",
      journal = {\araa},
         year = 2017,
        month = aug,
       volume = {55},
       number = {1},
        pages = {389-432},
          doi = {10.1146/annurev-astro-091916-055240},
archivePrefix = {arXiv},
       eprint = {1709.09180},
 primaryClass = {astro-ph.GA},
       adsurl = {https://ui.adsabs.harvard.edu/abs/2017ARA&A..55..389T}
}

@ARTICLE{Weinberger2017,
       author = {{Weinberger}, Rainer and {Springel}, Volker and {Hernquist}, Lars and {Pillepich}, Annalisa and {Marinacci}, Federico and {Pakmor}, R{\"u}diger and {Nelson}, Dylan and {Genel}, Shy and {Vogelsberger}, Mark and {Naiman}, Jill and {Torrey}, Paul},
        title = "{Simulating galaxy formation with black hole driven thermal and kinetic feedback}",
      journal = {\mnras},
         year = 2017,
        month = mar,
       volume = {465},
       number = {3},
        pages = {3291-3308},
          doi = {10.1093/mnras/stw2944},
archivePrefix = {arXiv},
       eprint = {1607.03486},
 primaryClass = {astro-ph.GA},
       adsurl = {https://ui.adsabs.harvard.edu/abs/2017MNRAS.465.3291W}
}

@ARTICLE{Dave2016,
       author = {{Dav{\'e}}, Romeel and {Thompson}, Robert and {Hopkins}, Philip F.},
        title = "{MUFASA: galaxy formation simulations with meshless hydrodynamics}",
      journal = {\mnras},
         year = 2016,
        month = nov,
       volume = {462},
       number = {3},
        pages = {3265-3284},
          doi = {10.1093/mnras/stw1862},
archivePrefix = {arXiv},
       eprint = {1604.01418},
 primaryClass = {astro-ph.GA},
       adsurl = {https://ui.adsabs.harvard.edu/abs/2016MNRAS.462.3265D}
}

@ARTICLE{Crain2015,
       author = {{Crain}, Robert A. and {Schaye}, Joop and {Bower}, Richard G. and {Furlong}, Michelle and {Schaller}, Matthieu and {Theuns}, Tom and {Dalla Vecchia}, Claudio and {Frenk}, Carlos S. and {McCarthy}, Ian G. and {Helly}, John C. and {Jenkins}, Adrian and {Rosas-Guevara}, Yetli M. and {White}, Simon D.~M. and {Trayford}, James W.},
        title = "{The EAGLE simulations of galaxy formation: calibration of subgrid physics and model variations}",
      journal = {\mnras},
         year = 2015,
        month = jun,
       volume = {450},
       number = {2},
        pages = {1937-1961},
          doi = {10.1093/mnras/stv725},
archivePrefix = {arXiv},
       eprint = {1501.01311},
 primaryClass = {astro-ph.GA},
       adsurl = {https://ui.adsabs.harvard.edu/abs/2015MNRAS.450.1937C}
}

@ARTICLE{Choi2012,
       author = {{Choi}, Ena and {Ostriker}, Jeremiah P. and {Naab}, Thorsten and {Johansson}, Peter H.},
        title = "{Radiative and Momentum-based Mechanical Active Galactic Nucleus Feedback in a Three-dimensional Galaxy Evolution Code}",
      journal = {\apj},
         year = 2012,
        month = aug,
       volume = {754},
       number = {2},
          eid = {125},
        pages = {125},
          doi = {10.1088/0004-637X/754/2/125},
archivePrefix = {arXiv},
       eprint = {1205.2082},
 primaryClass = {astro-ph.GA},
       adsurl = {https://ui.adsabs.harvard.edu/abs/2012ApJ...754..125C}
}

@ARTICLE{Hopkins2011,
       author = {{Hopkins}, Philip F. and {Quataert}, Eliot},
        title = "{An analytic model of angular momentum transport by gravitational torques: from galaxies to massive black holes}",
      journal = {\mnras},
         year = 2011,
        month = aug,
       volume = {415},
       number = {2},
        pages = {1027-1050},
          doi = {10.1111/j.1365-2966.2011.18542.x},
archivePrefix = {arXiv},
       eprint = {1007.2647},
 primaryClass = {astro-ph.CO},
       adsurl = {https://ui.adsabs.harvard.edu/abs/2011MNRAS.415.1027H}
}

@ARTICLE{Choi2015,
       author = {{Choi}, Ena and {Ostriker}, Jeremiah P. and {Naab}, Thorsten and {Oser}, Ludwig and {Moster}, Benjamin P.},
        title = "{The impact of mechanical AGN feedback on the formation of massive early-type galaxies}",
      journal = {\mnras},
         year = 2015,
        month = jun,
       volume = {449},
       number = {4},
        pages = {4105-4116},
          doi = {10.1093/mnras/stv575},
archivePrefix = {arXiv},
       eprint = {1403.1257},
 primaryClass = {astro-ph.CO},
       adsurl = {https://ui.adsabs.harvard.edu/abs/2015MNRAS.449.4105C}
}

@ARTICLE{Hopkins2015,
       author = {{Hopkins}, Philip F.},
        title = "{A new class of accurate, mesh-free hydrodynamic simulation methods}",
      journal = {\mnras},
         year = 2015,
        month = jun,
       volume = {450},
       number = {1},
        pages = {53-110},
          doi = {10.1093/mnras/stv195},
archivePrefix = {arXiv},
       eprint = {1409.7395},
 primaryClass = {astro-ph.CO},
       adsurl = {https://ui.adsabs.harvard.edu/abs/2015MNRAS.450...53H}
}

@ARTICLE{Hopkins2014,
       author = {{Hopkins}, Philip F. and {Kere{\v{s}}}, Du{\v{s}}an and {O{\~n}orbe}, Jos{\'e} and {Faucher-Gigu{\`e}re}, Claude-Andr{\'e} and {Quataert}, Eliot and {Murray}, Norman and {Bullock}, James S.},
        title = "{Galaxies on FIRE (Feedback In Realistic Environments): stellar feedback explains cosmologically inefficient star formation}",
      journal = {\mnras},
         year = 2014,
        month = nov,
       volume = {445},
       number = {1},
        pages = {581-603},
          doi = {10.1093/mnras/stu1738},
archivePrefix = {arXiv},
       eprint = {1311.2073},
 primaryClass = {astro-ph.CO},
       adsurl = {https://ui.adsabs.harvard.edu/abs/2014MNRAS.445..581H}
}

@ARTICLE{Birnboim2003,
       author = {{Birnboim}, Yuval and {Dekel}, Avishai},
        title = "{Virial shocks in galactic haloes?}",
      journal = {\mnras},
         year = 2003,
        month = oct,
       volume = {345},
       number = {1},
        pages = {349-364},
          doi = {10.1046/j.1365-8711.2003.06955.x},
archivePrefix = {arXiv},
       eprint = {astro-ph/0302161},
 primaryClass = {astro-ph},
       adsurl = {https://ui.adsabs.harvard.edu/abs/2003MNRAS.345..349B}
}

@ARTICLE{Muratov2015,
       author = {{Muratov}, Alexander L. and {Kere{\v{s}}}, Du{\v{s}}an and {Faucher-Gigu{\`e}re}, Claude-Andr{\'e} and {Hopkins}, Philip F. and {Quataert}, Eliot and {Murray}, Norman},
        title = "{Gusty, gaseous flows of FIRE: galactic winds in cosmological simulations with explicit stellar feedback}",
      journal = {\mnras},
         year = 2015,
        month = dec,
       volume = {454},
       number = {3},
        pages = {2691-2713},
          doi = {10.1093/mnras/stv2126},
archivePrefix = {arXiv},
       eprint = {1501.03155},
 primaryClass = {astro-ph.GA},
       adsurl = {https://ui.adsabs.harvard.edu/abs/2015MNRAS.454.2691M}
}

@ARTICLE{Broderick25,
       author = {{Broderick}, Ariel O. and {Roberts}, Ian D. and {Hudson}, Michael J.},
        title = "{Truncated Star Formation and Ram Pressure Stripping in the Coma Cluster}",
      journal = {\apj},
         year = 2025,
        month = jun,
       volume = {986},
       number = {2},
          eid = {132},
        pages = {132},
          doi = {10.3847/1538-4357/add739},
archivePrefix = {arXiv},
       eprint = {2505.10633},
 primaryClass = {astro-ph.GA},
       adsurl = {https://ui.adsabs.harvard.edu/abs/2025ApJ...986..132B}
}

@ARTICLE{Wright22,
       author = {{Wright}, Ruby J. and {Lagos}, Claudia del P. and {Power}, Chris and {Stevens}, Adam R.~H. and {Cortese}, Luca and {Poulton}, Rhys J.~J.},
        title = "{An orbital perspective on the starvation, stripping, and quenching of satellite galaxies in the EAGLE simulations}",
      journal = {\mnras},
         year = 2022,
        month = oct,
       volume = {516},
       number = {2},
        pages = {2891-2912},
          doi = {10.1093/mnras/stac2042},
archivePrefix = {arXiv},
       eprint = {2205.08414},
 primaryClass = {astro-ph.GA},
       adsurl = {https://ui.adsabs.harvard.edu/abs/2022MNRAS.516.2891W}
}

@ARTICLE{Stevens21,
       author = {{Stevens}, Adam R.~H. and {Lagos}, Claudia del P. and {Cortese}, Luca and {Catinella}, Barbara and {Diemer}, Benedikt and {Nelson}, Dylan and {Pillepich}, Annalisa and {Hernquist}, Lars and {Marinacci}, Federico and {Vogelsberger}, Mark},
        title = "{Molecular hydrogen in IllustrisTNG galaxies: carefully comparing signatures of environment with local CO and SFR data}",
      journal = {\mnras},
         year = 2021,
        month = apr,
       volume = {502},
       number = {3},
        pages = {3158-3178},
          doi = {10.1093/mnras/staa3662},
archivePrefix = {arXiv},
       eprint = {2011.03226},
 primaryClass = {astro-ph.GA},
       adsurl = {https://ui.adsabs.harvard.edu/abs/2021MNRAS.502.3158S}
}

@ARTICLE{Wells22,
       author = {{Wells}, Natalie K. and {Prescott}, Moire K.~M. and {Finlator}, Kristian M.},
        title = "{Brighter and More Massive Galaxies in the Vicinity of Ly{\ensuremath{\alpha}} Nebulae}",
      journal = {\apj},
         year = 2022,
        month = dec,
       volume = {941},
       number = {2},
          eid = {180},
        pages = {180},
          doi = {10.3847/1538-4357/aca3ac},
archivePrefix = {arXiv},
       eprint = {2301.02755},
 primaryClass = {astro-ph.GA},
       adsurl = {https://ui.adsabs.harvard.edu/abs/2022ApJ...941..180W}
}

@ARTICLE{Etherington17,
       author = {{Etherington}, J. and {Thomas}, D. and {Maraston}, C. and {Sevilla-Noarbe}, I. and {Bechtol}, K. and {Pforr}, J. and {Pellegrini}, P. and {Gschwend}, J. and {Carnero Rosell}, A. and {Maia}, M.~A.~G. and {da Costa}, L.~N. and {Benoit-L{\'e}vy}, A. and {Swanson}, M.~E.~C. and {Hartley}, W.~G. and {Abbott}, T.~M.~C. and {Abdalla}, F.~B. and {Allam}, S. and {Bernstein}, R.~A. and {Bertin}, E. and {Brooks}, D. and {Buckley-Geer}, E. and {Carrasco Kind}, M. and {Carretero}, J. and {Castander}, F.~J. and {Crocce}, M. and {Cunha}, C.~E. and {Desai}, S. and {Doel}, P. and {Eifler}, T.~F. and {Evrard}, A.~E. and {Fausti Neto}, A. and {Finley}, D.~A. and {Flaugher}, B. and {Fosalba}, P. and {Frieman}, J. and {Gerdes}, D.~W. and {Gruen}, D. and {Gruendl}, R.~A. and {Gutierrez}, G. and {Honscheid}, K. and {James}, D.~J. and {Kuehn}, K. and {Kuropatkin}, N. and {Lahav}, O. and {Lima}, M. and {Martini}, P. and {Melchior}, P. and {Miquel}, R. and {Mohr}, J.~J. and {Nord}, B. and {Ogando}, R. and {Plazas}, A.~A. and {Romer}, A.~K. and {Rykoff}, E.~S. and {Sanchez}, E. and {Scarpine}, V. and {Schubnell}, M. and {Smith}, R.~C. and {Soares-Santos}, M. and {Sobreira}, F. and {Tarle}, G. and {Vikram}, V. and {Walker}, A.~R. and {Zhang}, Y.},
        title = "{Environmental dependence of the galaxy stellar mass function in the Dark Energy Survey Science Verification Data}",
      journal = {\mnras},
         year = 2017,
        month = apr,
       volume = {466},
       number = {1},
        pages = {228-247},
          doi = {10.1093/mnras/stw3069},
archivePrefix = {arXiv},
       eprint = {1701.06066},
 primaryClass = {astro-ph.GA},
       adsurl = {https://ui.adsabs.harvard.edu/abs/2017MNRAS.466..228E}
}

@ARTICLE{Xu25,
       author = {{Xu}, Ke and {Wang}, Tao and {Daddi}, Emanuele and {Elbaz}, David and {Sun}, Hanwen and {Chen}, Longyue and {Gobat}, Raphael and {Zanella}, Anita and {Liu}, Daizhong and {Xiao}, Mengyuan and {Cen}, Renyue and {Kodama}, Tadayuki and {Kohno}, Kotaro and {Yang}, Tiancheng and {Zhang}, Zhi-Yu and {Zhou}, Luwenjia and {Valentino}, Francesco},
        title = "{Ram-pressure stripping caught in action in a forming galaxy cluster 3 billion years after the Big Bang}",
      journal = {arXiv e-prints},
         year = 2025,
        month = mar,
          eid = {arXiv:2503.21724},
        pages = {arXiv:2503.21724},
          doi = {10.48550/arXiv.2503.21724},
archivePrefix = {arXiv},
       eprint = {2503.21724},
 primaryClass = {astro-ph.GA},
       adsurl = {https://ui.adsabs.harvard.edu/abs/2025arXiv250321724X}
}

@ARTICLE{Overzier2016,
       author = {{Overzier}, Roderik A.},
        title = "{The realm of the galaxy protoclusters. A review}",
      journal = {\aapr},
         year = 2016,
        month = nov,
       volume = {24},
       number = {1},
          eid = {14},
        pages = {14},
          doi = {10.1007/s00159-016-0100-3},
archivePrefix = {arXiv},
       eprint = {1610.05201},
 primaryClass = {astro-ph.GA},
       adsurl = {https://ui.adsabs.harvard.edu/abs/2016A&ARv..24...14O}
}

@ARTICLE{Traina2025,
       author = {{Traina}, A. and {Vito}, F. and {Arrigoni-Battaia}, F. and {Chen}, C.-C. and {Vignali}, C. and {Prochaska}, X. and {Cantalupo}, S. and {Pensabene}, A. and {Tozzi}, P. and {Travascio}, A. and {Gilli}, R. and {Isla Llave}, M.~N. and {Marchesi}, S. and {Mazzolari}, G.},
        title = "{The properties of X-ray-selected active galactic nuclei in protoclusters pinpointed by enormous Lyman alpha nebulae}",
      journal = {\aap},
         year = 2025,
        month = sep,
       volume = {701},
          eid = {A158},
        pages = {A158},
          doi = {10.1051/0004-6361/202555443},
       adsurl = {https://ui.adsabs.harvard.edu/abs/2025A&A...701A.158T}
}

@ARTICLE{Vito2024,
       author = {{Vito}, F. and {Brandt}, W.~N. and {Comastri}, A. and {Gilli}, R. and {Ivison}, R.~J. and {Lanzuisi}, G. and {Lehmer}, B.~D. and {Lopez}, I.~E. and {Tozzi}, P. and {Vignali}, C.},
        title = "{Fast supermassive black hole growth in the SPT2349─56 protocluster at z = 4.3}",
      journal = {\aap},
         year = 2024,
        month = sep,
       volume = {689},
          eid = {A130},
        pages = {A130},
          doi = {10.1051/0004-6361/202450225},
archivePrefix = {arXiv},
       eprint = {2406.13005},
 primaryClass = {astro-ph.GA},
       adsurl = {https://ui.adsabs.harvard.edu/abs/2024A&A...689A.130V}
}

@ARTICLE{Schaye2015,
       author = {{Schaye}, Joop and {Crain}, Robert A. and {Bower}, Richard G. and {Furlong}, Michelle and {Schaller}, Matthieu and {Theuns}, Tom and {Dalla Vecchia}, Claudio and {Frenk}, Carlos S. and {McCarthy}, I.~G. and {Helly}, John C. and {Jenkins}, Adrian and {Rosas-Guevara}, Y.~M. and {White}, Simon D.~M. and {Baes}, Maarten and {Booth}, C.~M. and {Camps}, Peter and {Navarro}, Julio F. and {Qu}, Yan and {Rahmati}, Alireza and {Sawala}, Till and {Thomas}, Peter A. and {Trayford}, James},
        title = "{The EAGLE project: simulating the evolution and assembly of galaxies and their environments}",
      journal = {\mnras},
         year = 2015,
        month = jan,
       volume = {446},
       number = {1},
        pages = {521-554},
          doi = {10.1093/mnras/stu2058},
archivePrefix = {arXiv},
       eprint = {1407.7040},
 primaryClass = {astro-ph.GA},
       adsurl = {https://ui.adsabs.harvard.edu/abs/2015MNRAS.446..521S}
}

@ARTICLE{somerville2015,
       author = {{Somerville}, Rachel S. and {Dav{\'e}}, Romeel},
        title = "{Physical Models of Galaxy Formation in a Cosmological Framework}",
      journal = {\araa},
         year = 2015,
        month = aug,
       volume = {53},
        pages = {51-113},
          doi = {10.1146/annurev-astro-082812-140951},
archivePrefix = {arXiv},
       eprint = {1412.2712},
 primaryClass = {astro-ph.GA},
       adsurl = {https://ui.adsabs.harvard.edu/abs/2015ARA&A..53...51S}
}

@ARTICLE{Genel2014,
       author = {{Genel}, Shy and {Vogelsberger}, Mark and {Springel}, Volker and {Sijacki}, Debora and {Nelson}, Dylan and {Snyder}, Greg and {Rodriguez-Gomez}, Vicente and {Torrey}, Paul and {Hernquist}, Lars},
        title = "{Introducing the Illustris project: the evolution of galaxy populations across cosmic time}",
      journal = {\mnras},
         year = 2014,
        month = nov,
       volume = {445},
       number = {1},
        pages = {175-200},
          doi = {10.1093/mnras/stu1654},
archivePrefix = {arXiv},
       eprint = {1405.3749},
 primaryClass = {astro-ph.CO},
       adsurl = {https://ui.adsabs.harvard.edu/abs/2014MNRAS.445..175G}
}

@ARTICLE{Tillman23b,
       author = {{Tillman}, Megan Taylor and {Burkhart}, Blakesley and {Tonnesen}, Stephanie and {Bird}, Simeon and {Bryan}, Greg L. and {Angl{\'e}s-Alc{\'a}zar}, Daniel and {Dav{\'e}}, Romeel and {Genel}, Shy},
        title = "{Efficient Long-range Active Galactic Nuclei (AGNs) Feedback Affects the Low-redshift Ly{\ensuremath{\alpha}} Forest}",
      journal = {\apjl},
         year = 2023,
        month = mar,
       volume = {945},
       number = {1},
          eid = {L17},
        pages = {L17},
          doi = {10.3847/2041-8213/acb7f1},
archivePrefix = {arXiv},
       eprint = {2210.02467},
 primaryClass = {astro-ph.GA},
       adsurl = {https://ui.adsabs.harvard.edu/abs/2023ApJ...945L..17T}
}

@ARTICLE{Booth&Schaye09,
       author = {{Booth}, C.~M. and {Schaye}, Joop},
        title = "{Cosmological simulations of the growth of supermassive black holes and feedback from active galactic nuclei: method and tests}",
      journal = {\mnras},
         year = 2009,
        month = sep,
       volume = {398},
       number = {1},
        pages = {53-74},
          doi = {10.1111/j.1365-2966.2009.15043.x},
archivePrefix = {arXiv},
       eprint = {0904.2572},
 primaryClass = {astro-ph.CO},
       adsurl = {https://ui.adsabs.harvard.edu/abs/2009MNRAS.398...53B}
}

@ARTICLE{Borrow20,
       author = {{Borrow}, Josh and {Angl{\'e}s-Alc{\'a}zar}, Daniel and {Dav{\'e}}, Romeel},
        title = "{Cosmological baryon transfer in the SIMBA simulations}",
      journal = {\mnras},
         year = 2020,
        month = feb,
       volume = {491},
       number = {4},
        pages = {6102-6119},
          doi = {10.1093/mnras/stz3428},
archivePrefix = {arXiv},
       eprint = {1910.00594},
 primaryClass = {astro-ph.GA},
       adsurl = {https://ui.adsabs.harvard.edu/abs/2020MNRAS.491.6102B}
}

@ARTICLE{Tillman2023,
       author = {{Tillman}, Megan Taylor and {Burkhart}, Blakesley and {Tonnesen}, Stephanie and {Bird}, Simeon and {Bryan}, Greg L. and {Angl{\'e}s-Alc{\'a}zar}, Daniel and {Hassan}, Sultan and {Somerville}, Rachel S. and {Dav{\'e}}, Romeel and {Marinacci}, Federico and {Hernquist}, Lars and {Vogelsberger}, Mark},
        title = "{An Exploration of AGN and Stellar Feedback Effects in the Intergalactic Medium via the Low-redshift Ly{\ensuremath{\alpha}} Forest}",
      journal = {\aj},
         year = 2023,
        month = dec,
       volume = {166},
       number = {6},
          eid = {228},
        pages = {228},
          doi = {10.3847/1538-3881/ad02f5},
archivePrefix = {arXiv},
       eprint = {2307.06360},
 primaryClass = {astro-ph.GA},
       adsurl = {https://ui.adsabs.harvard.edu/abs/2023AJ....166..228T}
}

@ARTICLE{Pathak2025,
       author = {{Pathak}, Debosmita and {Christensen}, Charlotte R. and {Brooks}, Alyson M. and {Munshi}, Ferah and {Wright}, Anna C. and {Carter}, Courtney},
        title = "{Survivors and Zombies: The Quenching and Disruption of Satellites Around Milky Way Analogs}",
      journal = {\apj},
         year = 2025,
        month = aug,
       volume = {989},
       number = {2},
          eid = {178},
        pages = {178},
          doi = {10.3847/1538-4357/adec94},
archivePrefix = {arXiv},
       eprint = {2505.22742},
 primaryClass = {astro-ph.GA},
       adsurl = {https://ui.adsabs.harvard.edu/abs/2025ApJ...989..178P}
}

@ARTICLE{davies2020,
       author = {{Davies}, Jonathan J. and {Crain}, Robert A. and {Oppenheimer}, Benjamin D. and {Schaye}, Joop},
        title = "{The quenching and morphological evolution of central galaxies is facilitated by the feedback-driven expulsion of circumgalactic gas}",
      journal = {\mnras},
         year = 2020,
        month = jan,
       volume = {491},
       number = {3},
        pages = {4462-4480},
          doi = {10.1093/mnras/stz3201},
archivePrefix = {arXiv},
       eprint = {1908.11380},
 primaryClass = {astro-ph.GA},
       adsurl = {https://ui.adsabs.harvard.edu/abs/2020MNRAS.491.4462D}
}

@ARTICLE{Wellons2023,
       author = {{Wellons}, Sarah and {Faucher-Gigu{\`e}re}, Claude-Andr{\'e} and {Hopkins}, Philip F. and {Quataert}, Eliot and {Angl{\'e}s-Alc{\'a}zar}, Daniel and {Feldmann}, Robert and {Hayward}, Christopher C. and {Kere{\v{s}}}, Du{\v{s}}an and {Su}, Kung-Yi and {Wetzel}, Andrew},
        title = "{Exploring supermassive black hole physics and galaxy quenching across halo mass in FIRE cosmological zoom simulations}",
      journal = {\mnras},
         year = 2023,
        month = apr,
       volume = {520},
       number = {4},
        pages = {5394-5412},
          doi = {10.1093/mnras/stad511},
archivePrefix = {arXiv},
       eprint = {2203.06201},
 primaryClass = {astro-ph.GA},
       adsurl = {https://ui.adsabs.harvard.edu/abs/2023MNRAS.520.5394W}
}

@ARTICLE{Koudmani2022,
       author = {{Koudmani}, Sophie and {Sijacki}, Debora and {Smith}, Matthew C.},
        title = "{Two can play at that game: constraining the role of supernova and AGN feedback in dwarf galaxies with cosmological zoom-in simulations}",
      journal = {\mnras},
         year = 2022,
        month = oct,
       volume = {516},
       number = {2},
        pages = {2112-2141},
          doi = {10.1093/mnras/stac2252},
archivePrefix = {arXiv},
       eprint = {2206.11274},
 primaryClass = {astro-ph.GA},
       adsurl = {https://ui.adsabs.harvard.edu/abs/2022MNRAS.516.2112K}
}

@ARTICLE{Vogelsberger2014,
       author = {{Vogelsberger}, Mark and {Genel}, Shy and {Springel}, Volker and {Torrey}, Paul and {Sijacki}, Debora and {Xu}, Dandan and {Snyder}, Greg and {Nelson}, Dylan and {Hernquist}, Lars},
        title = "{Introducing the Illustris Project: simulating the coevolution of dark and visible matter in the Universe}",
      journal = {\mnras},
         year = 2014,
        month = oct,
       volume = {444},
       number = {2},
        pages = {1518-1547},
          doi = {10.1093/mnras/stu1536},
archivePrefix = {arXiv},
       eprint = {1405.2921},
 primaryClass = {astro-ph.CO},
       adsurl = {https://ui.adsabs.harvard.edu/abs/2014MNRAS.444.1518V}
}

@ARTICLE{Fabian2012,
       author = {{Fabian}, A.~C.},
        title = "{Observational Evidence of Active Galactic Nuclei Feedback}",
      journal = {\araa},
         year = 2012,
        month = sep,
       volume = {50},
        pages = {455-489},
          doi = {10.1146/annurev-astro-081811-125521},
archivePrefix = {arXiv},
       eprint = {1204.4114},
 primaryClass = {astro-ph.CO},
       adsurl = {https://ui.adsabs.harvard.edu/abs/2012ARA&A..50..455F}
}

@ARTICLE{Peng2012,
       author = {{Peng}, Ying-jie and {Lilly}, Simon J. and {Renzini}, Alvio and {Carollo}, Marcella},
        title = "{Mass and Environment as Drivers of Galaxy Evolution. II. The Quenching of Satellite Galaxies as the Origin of Environmental Effects}",
      journal = {\apj},
         year = 2012,
        month = sep,
       volume = {757},
       number = {1},
          eid = {4},
        pages = {4},
          doi = {10.1088/0004-637X/757/1/4},
archivePrefix = {arXiv},
       eprint = {1106.2546},
 primaryClass = {astro-ph.CO},
       adsurl = {https://ui.adsabs.harvard.edu/abs/2012ApJ...757....4P}
}

@ARTICLE{Faltenbacher2010,
       author = {{Faltenbacher}, A.},
        title = "{The impact of environment on the dynamical structure of satellite systems}",
      journal = {\mnras},
         year = 2010,
        month = oct,
       volume = {408},
       number = {2},
        pages = {1113-1119},
          doi = {10.1111/j.1365-2966.2010.17185.x},
archivePrefix = {arXiv},
       eprint = {0912.0013},
 primaryClass = {astro-ph.CO},
       adsurl = {https://ui.adsabs.harvard.edu/abs/2010MNRAS.408.1113F}
}

@ARTICLE{Springel2010,
       author = {{Springel}, Volker},
        title = "{E pur si muove: Galilean-invariant cosmological hydrodynamical simulations on a moving mesh}",
      journal = {\mnras},
         year = 2010,
        month = jan,
       volume = {401},
       number = {2},
        pages = {791-851},
          doi = {10.1111/j.1365-2966.2009.15715.x},
archivePrefix = {arXiv},
       eprint = {0901.4107},
 primaryClass = {astro-ph.CO},
       adsurl = {https://ui.adsabs.harvard.edu/abs/2010MNRAS.401..791S}
}

@ARTICLE{Blanton2009,
       author = {{Blanton}, Michael R. and {Moustakas}, John},
        title = "{Physical Properties and Environments of Nearby Galaxies}",
      journal = {\araa},
         year = 2009,
        month = sep,
       volume = {47},
       number = {1},
        pages = {159-210},
          doi = {10.1146/annurev-astro-082708-101734},
archivePrefix = {arXiv},
       eprint = {0908.3017},
 primaryClass = {astro-ph.GA},
       adsurl = {https://ui.adsabs.harvard.edu/abs/2009ARA&A..47..159B}
}

@ARTICLE{Shreeram25,
       author = {{Shreeram}, Soumya and {Gal{\'a}rraga-Espinosa}, Daniela and {Comparat}, Johan and {Merloni}, Andrea and {Nagai}, Daisuke and {Peroux}, C{\'e}line and {Marini}, Ilaria and {Gouin}, C{\'e}line and {Nandra}, Kirpal and {Zhang}, Yi and {Ponti}, Gabriele and {Olechowska}, Anna},
        title = "{Impact of the large-scale cosmic web on the X-ray emitting circumgalactic medium}",
      journal = {arXiv e-prints},
         year = 2025,
        month = jun,
          eid = {arXiv:2506.17222},
        pages = {arXiv:2506.17222},
          doi = {10.48550/arXiv.2506.17222},
archivePrefix = {arXiv},
       eprint = {2506.17222},
 primaryClass = {astro-ph.GA},
       adsurl = {https://ui.adsabs.harvard.edu/abs/2025arXiv250617222S}
}

@ARTICLE{Tonnesen2009,
       author = {{Tonnesen}, Stephanie and {Bryan}, Greg L.},
        title = "{Gas Stripping in Simulated Galaxies with a Multiphase Interstellar Medium}",
      journal = {\apj},
         year = 2009,
        month = apr,
       volume = {694},
       number = {2},
        pages = {789-804},
          doi = {10.1088/0004-637X/694/2/789},
archivePrefix = {arXiv},
       eprint = {0901.2115},
 primaryClass = {astro-ph.GA},
       adsurl = {https://ui.adsabs.harvard.edu/abs/2009ApJ...694..789T}
}

@ARTICLE{Boselli2006,
       author = {{Boselli}, Alessandro and {Gavazzi}, Giuseppe},
        title = "{Environmental Effects on Late-Type Galaxies in Nearby Clusters}",
      journal = {\pasp},
         year = 2006,
        month = apr,
       volume = {118},
       number = {842},
        pages = {517-559},
          doi = {10.1086/500691},
archivePrefix = {arXiv},
       eprint = {astro-ph/0601108},
 primaryClass = {astro-ph},
       adsurl = {https://ui.adsabs.harvard.edu/abs/2006PASP..118..517B}
}

@ARTICLE{Blanton2005,
       author = {{Blanton}, Michael R. and {Eisenstein}, Daniel and {Hogg}, David W. and {Schlegel}, David J. and {Brinkmann}, J.},
        title = "{Relationship between Environment and the Broadband Optical Properties of Galaxies in the Sloan Digital Sky Survey}",
      journal = {\apj},
         year = 2005,
        month = aug,
       volume = {629},
       number = {1},
        pages = {143-157},
          doi = {10.1086/422897},
archivePrefix = {arXiv},
       eprint = {astro-ph/0310453},
 primaryClass = {astro-ph},
       adsurl = {https://ui.adsabs.harvard.edu/abs/2005ApJ...629..143B}
}

@ARTICLE{Springel2005,
       author = {{Springel}, Volker},
        title = "{The cosmological simulation code GADGET-2}",
      journal = {\mnras},
         year = 2005,
        month = dec,
       volume = {364},
       number = {4},
        pages = {1105-1134},
          doi = {10.1111/j.1365-2966.2005.09655.x},
archivePrefix = {arXiv},
       eprint = {astro-ph/0505010},
 primaryClass = {astro-ph},
       adsurl = {https://ui.adsabs.harvard.edu/abs/2005MNRAS.364.1105S}
}

@ARTICLE{Hogg2004,
       author = {{Hogg}, David W. and {Blanton}, Michael R. and {Brinchmann}, Jarle and {Eisenstein}, Daniel J. and {Schlegel}, David J. and {Gunn}, James E. and {McKay}, Timothy A. and {Rix}, Hans-Walter and {Bahcall}, Neta A. and {Brinkmann}, J. and {Meiksin}, Avery},
        title = "{The Dependence on Environment of the Color-Magnitude Relation of Galaxies}",
      journal = {\apjl},
         year = 2004,
        month = jan,
       volume = {601},
       number = {1},
        pages = {L29-L32},
          doi = {10.1086/381749},
archivePrefix = {arXiv},
       eprint = {astro-ph/0307336},
 primaryClass = {astro-ph},
       adsurl = {https://ui.adsabs.harvard.edu/abs/2004ApJ...601L..29H}
}

@ARTICLE{Kauffman2004,
       author = {{Kauffmann}, Guinevere and {White}, Simon D.~M. and {Heckman}, Timothy M. and {M{\'e}nard}, Brice and {Brinchmann}, Jarle and {Charlot}, St{\'e}phane and {Tremonti}, Christy and {Brinkmann}, Jon},
        title = "{The environmental dependence of the relations between stellar mass, structure, star formation and nuclear activity in galaxies}",
      journal = {\mnras},
         year = 2004,
        month = sep,
       volume = {353},
       number = {3},
        pages = {713-731},
          doi = {10.1111/j.1365-2966.2004.08117.x},
archivePrefix = {arXiv},
       eprint = {astro-ph/0402030},
 primaryClass = {astro-ph},
       adsurl = {https://ui.adsabs.harvard.edu/abs/2004MNRAS.353..713K}
}

@ARTICLE{Rojas2004,
       author = {{Rojas}, Randall R. and {Vogeley}, Michael S. and {Hoyle}, Fiona and {Brinkmann}, Jon},
        title = "{Photometric Properties of Void Galaxies in the Sloan Digital Sky Survey}",
      journal = {\apj},
         year = 2004,
        month = dec,
       volume = {617},
       number = {1},
        pages = {50-63},
          doi = {10.1086/425225},
archivePrefix = {arXiv},
       eprint = {astro-ph/0307274},
 primaryClass = {astro-ph},
       adsurl = {https://ui.adsabs.harvard.edu/abs/2004ApJ...617...50R}
}

@ARTICLE{Lizhi25,
       author = {{Xie}, Lizhi and {De Lucia}, Gabriella and {Fossati}, Matteo and {Fontanot}, Fabio and {Hirschmann}, Michaela},
        title = "{The impact of ram pressure on cluster galaxies, insights from GAEA and TNG}",
      journal = {\aap},
         year = 2025,
        month = jun,
       volume = {698},
          eid = {A73},
        pages = {A73},
          doi = {10.1051/0004-6361/202553915},
archivePrefix = {arXiv},
       eprint = {2504.12863},
 primaryClass = {astro-ph.GA},
       adsurl = {https://ui.adsabs.harvard.edu/abs/2025A&A...698A..73X}
}

@ARTICLE{Artale2018,
       author = {{Artale}, M. Celeste and {Zehavi}, Idit and {Contreras}, Sergio and {Norberg}, Peder},
        title = "{The impact of assembly bias on the halo occupation in hydrodynamical simulations}",
      journal = {\mnras},
         year = 2018,
        month = nov,
       volume = {480},
       number = {3},
        pages = {3978-3992},
          doi = {10.1093/mnras/sty2110},
archivePrefix = {arXiv},
       eprint = {1805.06938},
 primaryClass = {astro-ph.GA},
       adsurl = {https://ui.adsabs.harvard.edu/abs/2018MNRAS.480.3978A}
}

@ARTICLE{dekel2009,
       author = {{Dekel}, A. and {Birnboim}, Y. and {Engel}, G. and {Freundlich}, J. and {Goerdt}, T. and {Mumcuoglu}, M. and {Neistein}, E. and {Pichon}, C. and {Teyssier}, R. and {Zinger}, E.},
        title = "{Cold streams in early massive hot haloes as the main mode of galaxy formation}",
      journal = {\nat},
         year = 2009,
        month = jan,
       volume = {457},
       number = {7228},
        pages = {451-454},
          doi = {10.1038/nature07648},
archivePrefix = {arXiv},
       eprint = {0808.0553},
 primaryClass = {astro-ph},
       adsurl = {https://ui.adsabs.harvard.edu/abs/2009Natur.457..451D}
}

@ARTICLE{dekel2003,
       author = {{Dekel}, Avishai and {Devor}, Jonathan and {Hetzroni}, Guy},
        title = "{Galactic halo cusp-core: tidal compression in mergers}",
      journal = {\mnras},
         year = 2003,
        month = may,
       volume = {341},
       number = {1},
        pages = {326-342},
          doi = {10.1046/j.1365-8711.2003.06432.x},
archivePrefix = {arXiv},
       eprint = {astro-ph/0204452},
 primaryClass = {astro-ph},
       adsurl = {https://ui.adsabs.harvard.edu/abs/2003MNRAS.341..326D}
}

@ARTICLE{Gomez2003,
       author = {{G{\'o}mez}, Percy L. and {Nichol}, Robert C. and {Miller}, Christopher J. and {Balogh}, Michael L. and {Goto}, Tomotsugu and {Zabludoff}, Ann I. and {Romer}, A. Kathy and {Bernardi}, Mariangela and {Sheth}, Ravi and {Hopkins}, Andrew M. and {Castander}, Francisco J. and {Connolly}, Andrew J. and {Schneider}, Donald P. and {Brinkmann}, Jon and {Lamb}, Don Q. and {SubbaRao}, Mark and {York}, Donald G.},
        title = "{Galaxy Star Formation as a Function of Environment in the Early Data Release of the Sloan Digital Sky Survey}",
      journal = {\apj},
         year = 2003,
        month = feb,
       volume = {584},
       number = {1},
        pages = {210-227},
          doi = {10.1086/345593},
archivePrefix = {arXiv},
       eprint = {astro-ph/0210193},
 primaryClass = {astro-ph},
       adsurl = {https://ui.adsabs.harvard.edu/abs/2003ApJ...584..210G}
}

@ARTICLE{Abadi1999,
       author = {{Abadi}, Mario G. and {Moore}, Ben and {Bower}, Richard G.},
        title = "{Ram pressure stripping of spiral galaxies in clusters}",
      journal = {\mnras},
         year = 1999,
        month = oct,
       volume = {308},
       number = {4},
        pages = {947-954},
          doi = {10.1046/j.1365-8711.1999.02715.x},
archivePrefix = {arXiv},
       eprint = {astro-ph/9903436},
 primaryClass = {astro-ph},
       adsurl = {https://ui.adsabs.harvard.edu/abs/1999MNRAS.308..947A}
}

@ARTICLE{Wang2025,
       author = {{Wang}, Yu-Jan and {Chen}, Chian-Chou and {Arrigoni Battaia}, Fabrizio and {Decarli}, Roberto and {Dannerbauer}, Helmut and {Wu}, Po-Feng},
        title = "{A Multiwavelength Study of ELAN Environments (AMUSE$^{2}$): The Impact of Dense Environment on Massive Dusty Star-forming Galaxies at Cosmic Noon}",
      journal = {\apj},
         year = 2025,
        month = nov,
       volume = {993},
       number = {1},
          eid = {111},
        pages = {111},
          doi = {10.3847/1538-4357/adff7d},
archivePrefix = {arXiv},
       eprint = {2509.03027},
 primaryClass = {astro-ph.GA},
       adsurl = {https://ui.adsabs.harvard.edu/abs/2025ApJ...993..111W}
}

@ARTICLE{Dressler1980,
       author = {{Dressler}, A.},
        title = "{Galaxy morphology in rich clusters: implications for the formation and evolution of galaxies.}",
      journal = {\apj},
         year = 1980,
        month = mar,
       volume = {236},
        pages = {351-365},
          doi = {10.1086/157753},
       adsurl = {https://ui.adsabs.harvard.edu/abs/1980ApJ...236..351D}
}

@ARTICLE{Gunn&Gott1972,
       author = {{Gunn}, James E. and {Gott}, J. Richard, III},
        title = "{On the Infall of Matter Into Clusters of Galaxies and Some Effects on Their Evolution}",
      journal = {\apj},
         year = 1972,
        month = aug,
       volume = {176},
        pages = {1},
          doi = {10.1086/151605},
       adsurl = {https://ui.adsabs.harvard.edu/abs/1972ApJ...176....1G}
}

@ARTICLE{Bondi1952,
       author = {{Bondi}, H.},
        title = "{On spherically symmetrical accretion}",
      journal = {\mnras},
         year = 1952,
        month = jan,
       volume = {112},
        pages = {195},
          doi = {10.1093/mnras/112.2.195},
       adsurl = {https://ui.adsabs.harvard.edu/abs/1952MNRAS.112..195B}
}

%%%%%%%%%%%%%%%%%%%%%%%%%%%%%%%%%%%%%%%%%%%%%%%%%%

%%%%%%%%%%%%%%%%% APPENDICES %%%%%%%%%%%%%%%%%%%%%

\appendix

\section{Definition of local galaxy density}
\label{sec: Appendix}

\begin{figure}
    \centering
    \includegraphics[width=\columnwidth]{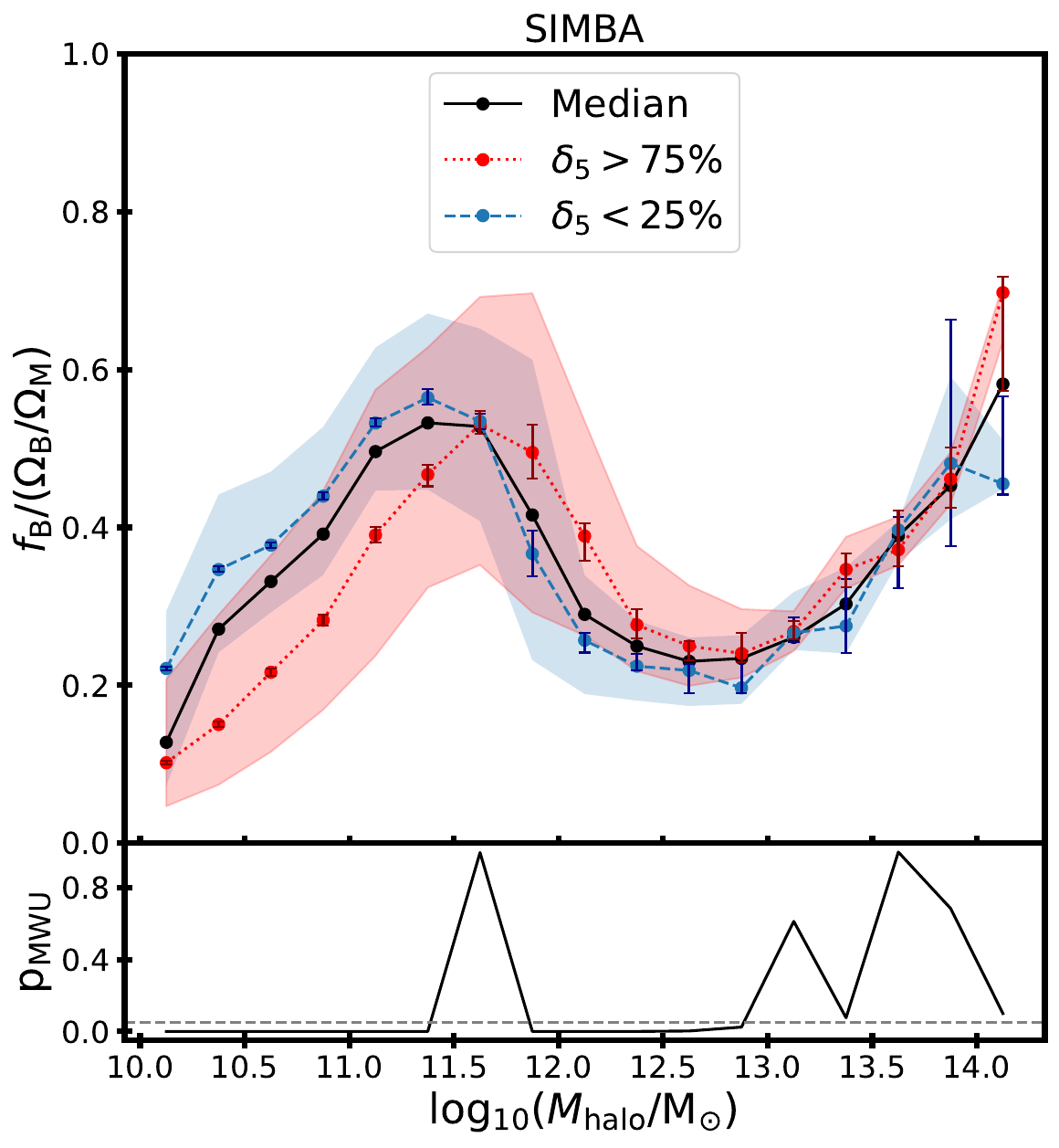}
    \vspace{-0.2in}
    \caption{Baryon fraction, $f_{\rm B}$, as a function of $M_{\rm halo}$ and environment for the SIMBA CV simulations, similar to Figure~\ref{fig:f_b} but quantifying overdensity as $\delta_5$ (based on distance to the five nearest galaxies) instead of $\delta_{10}$. Low-mass haloes ($M_{\rm halo} < 10^{11.5}$\,M$_{\odot}$) in overdense regions have lower baryon fractions, with the impact of environment reverting at higher masses, in qualitative agreement with the SIMBA results in Figure~\ref{fig:f_b} based on $\delta_{10}$.}
    \label{fig:d5}
\end{figure}

Throughout this paper, we have used $\delta_{\rm N}$ with $N=10$ to quantify the environment of galaxies and haloes, representing the galaxy number density relative to average in a sphere containing the ten nearest galaxy neighbors. Figure~\ref{fig:d5} explores the dependence of results on the specific number $N$ of neighbors used, showing halo baryon fraction as a function of halo mass for haloes in overdense and underdense regions in the SIMBA CV simulations at $z=0$. This is similar to Figure~\ref{fig:f_b} but now using $\delta_{5}$ (based on $N=5$ nearest neighbors) instead of $\delta_{10}$ to quantify environment. We find that low-mass haloes have lower baryon fraction in overdense regions as quantified by $\delta_{5}$, with this trend reversing (and weakening) at $M_{\rm halo} > 10^{11.5}$\,M$_\odot$, in agreement with our previous results based on $\delta_{10}$. Overall, the main trends identified for all galaxy formation models based on $\delta_{\rm N}$ are robust relative to changes in the number $N$ of neighbors used. These results are in agreement with \citet{2017MNRAS.466.3460V}, who stated that using $\delta_{\rm N}$ as a measure of environment yields similar results between $N=5$ and $N=10$ in the original EAGLE large-volume simulation.

\section{Physical scale of environment measurements}
\label{sec:Appendix_B}

\begin{figure*}
    \includegraphics[width=0.45\textwidth]{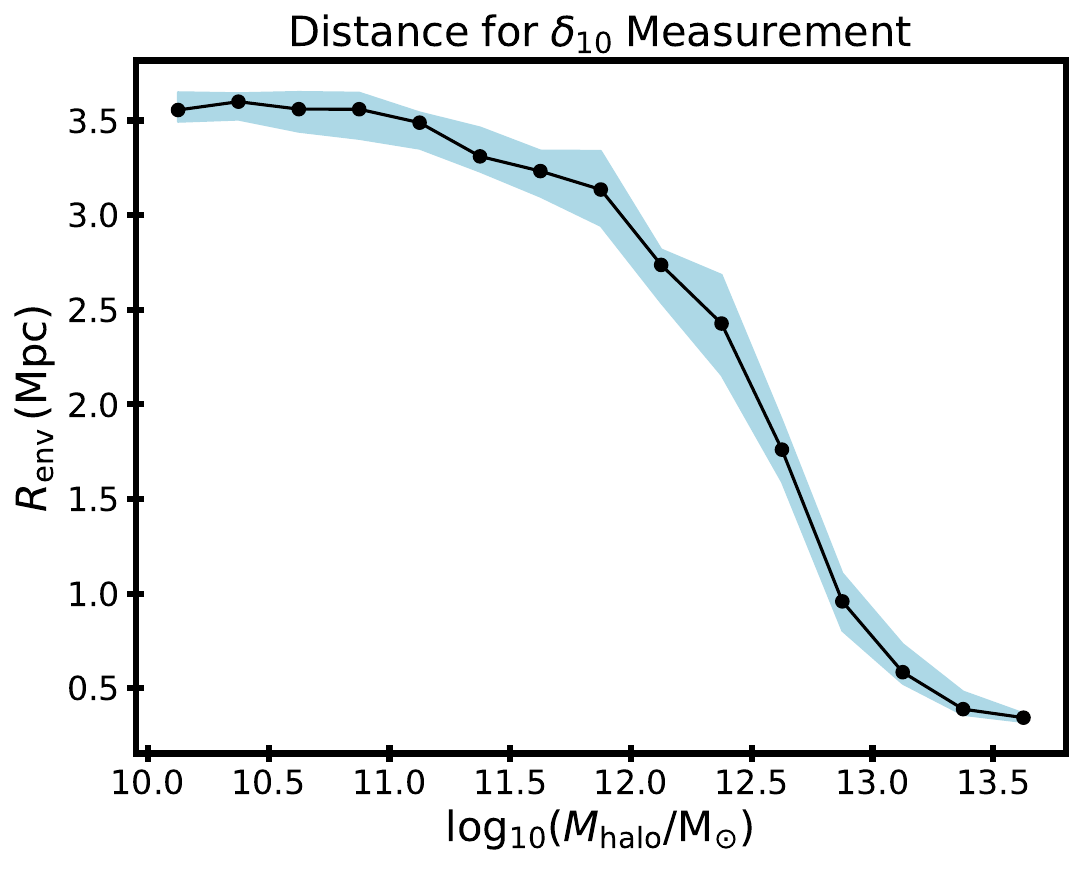}
    \includegraphics[width=0.45\textwidth]{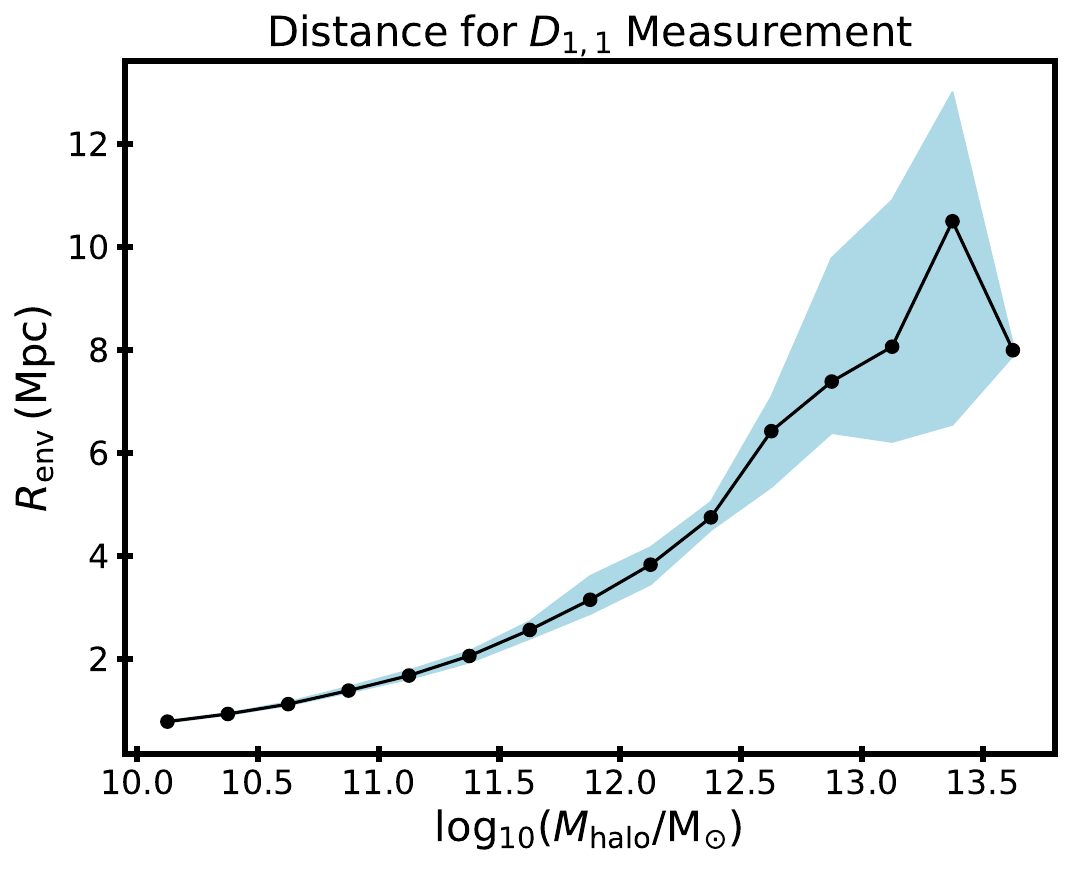}
    \caption{Physical distance corresponding to our measures of environment at $z=0$ for the $\delta_{10}$ (left) and $D_{1,1}$ (right) environment definitions (see Section~\ref{sec:env_def} for details) as a function of halo mass in the SIMBA CV simulations. The black points represent the median radial distance in each mass bin, and the blue region represents the 25$^{\rm th}$ to 75$^{\rm th}$ percentile range quantifying cosmic variance across the CV set. 
    The $\delta_{10}$ and $D_{1,1}$ overdensity definitions correspond to different physical scales over which local environment is measured and show opposite trends as a function of halo mass. }
    \label{fig:scale}
\end{figure*}

Figure~\ref{fig:scale} quantifies the physical scale over which local environment is measured ($R_{\rm env}$) as a function of halo mass for the $\delta_{10}$ (left) and $D_{1,1}$ (right) environment definitions, complementing the analysis of the overdensity--$M_{\rm halo}$ connection presented in Section~\ref{subsec:local environment}.  
We find that there is a clear negative correlation between the distance to a given halo's $10^{\rm th}$ nearest neighbour galaxy and the corresponding halo mass (left panel), with the physical scale corresponding to $\delta_{10}$ decreasing from $R_{\rm env} \sim 3.5$\,Mpc for $M_{\rm halo} \lesssim 10^{11}$\,M$_{\odot}$ to $R_{\rm env} \lesssim 0.5$\,Mpc for $M_{\rm halo} > 10^{13}$\,M$_{\odot}$. This is expected, since lower-mass haloes tend to populate lower-density regions (Figure~\ref{fig:delta-DNF}), resulting in larger $R_{\rm env}$ to find the required 10 nearest galaxies for the $\delta_{10}$ measurement.
In contrast, there is a positive correlation between halo mass and the distance to its nearest neighbour halo of at least equivalent mass (right panel), with the corresponding physical scale for $D_{1,1}$ measurements increasing from $R_{\rm env} \sim 1 \rightarrow 10$\,Mpc across the halo mass range. This trend can be explained by the lower abundance of higher-mass haloes, resulting in larger $R_{\rm env}$ to find the nearest similar-mass halo required for the measurement of $D_{1,1}$.  
Our local environment definitions $\delta_{10}$ and $D_{1,1}$ are thus probing different physical distances as a function of halo mass, providing further support for the systematic environmental impacts identified in this work.   

%%%%%%%%%%%%%%%%%%%%%%%%%%%%%%%%%%%%%%%%%%%%%%%%%%

\bsp	
\label{lastpage}
\end{document}